\documentclass[a4paper,11pt]{article}
\pdfoutput=1 % if your are submitting a pdflatex (i.e. if you have
\usepackage{jheppub}

\usepackage{amsmath,amssymb,mathtools,dsfont,subcaption,graphicx}
\usepackage{mathrsfs}
\usepackage{enumitem}
\usepackage[utf8]{inputenc}
\usepackage{amsfonts}
\usepackage[normalem]{ulem}
\usepackage{tensor}
\usepackage{array}

\newcommand{\cTFD}{{\mathrm{cTFD}}}
\newcommand{\TFD}{{\mathrm{TFD}}}
\newcommand{\lR}{{\bar{L}R}}
\newcommand{\Lr}{{L\bar{R}}}
\newcommand{\rf}{{\mathrm{ref}}}
\newcommand{\vac}{{\mathrm{vac}}}
\newcommand{\Sp}{{\mathrm{Sp}}}
\newcommand{\ssp}{{\mathfrak{sp}}}
\newcommand{\SO}{{\mathrm{SO}}}
\newcommand{\vol}{{\mathrm{vol}}}

\newcommand{\LL}{{\mathscr{L}}}
\newcommand{\CC}{{\mathscr{C}}}
\newcommand{\Cbb}{{\mathbb{C}}}
\newcommand{\Rbb}{{\mathbb{R}}}
\newcommand{\HH}{{\mathscr{H}}}
\newcommand{\cmplx}{{\mathcal{C}}}
\newcommand{\mC}{{\mathcal{C}}}
\newcommand{\Oplus}{{\tilde{\oplus}}}

\newcommand{\T}{{\mathrm{T}}}

\newcommand{\UB}{{\mathrm{UB}}}
\newcommand{\crit}{{\mathrm{crit}}}

\makeatletter
\renewcommand*\env@matrix[1][\arraystretch]{%
  \edef\arraystretch{#1}%
  \hskip -\arraycolsep
  \let\@ifnextchar\new@ifnextchar
  \array{*\c@MaxMatrixCols c}}
\makeatother

\DeclareMathOperator{\diag}{diag}
\DeclareMathOperator{\tr}{tr}
\newcommand{\eg}{{\it e.g.,}\ }
\newcommand{\ie}{{\it i.e.,}\ }

\title{Charged Complexity and the Thermofield Double State}

\author[a,b]{Shira Chapman}
\author[c]{and Hong Zhe (Vincent) Chen}

\affiliation[a]{Department of Physics, Ben-Gurion University of the Negev, Beer Sheva 84105, Israel}
\affiliation[b]{Institute for Theoretical Physics, University of Amsterdam\\
Science Park 904, Postbus 94485, 1090 GL Amsterdam, The Netherlands}
\affiliation[c]{Perimeter Institute for Theoretical Physics\\
31 Caroline Street North, Waterloo, ON N2L 2Y5, Canada}

\emailAdd{schapman@bgu.ac.il}
\emailAdd{hchen2@perimeterinstitute.ca}

\usepackage{cancel}
\usepackage[dvipsnames]{xcolor}
\abstract{We establish a systematic framework for studying quantum computational complexity of Gaussian states of charged systems based on Nielsen's geometric approach. We use this framework to examine the effect of a chemical potential on the dynamics of complexity. As an example, we consider the complexity of a charged thermofield double state constructed from two free massive complex scalar fields in the presence of a chemical potential.
We show that this state factorizes between positively and negatively charged modes and demonstrate that this fact can be used to relate it, for each momentum mode separately, to two uncharged thermofield double states with shifted temperatures and times. We evaluate the complexity of formation for the charged thermofield double state, both numerically and in certain analytic expansions. We further present numerical results for the time dependence of complexity. We compare various aspects of these results to those obtained in holography for charged black holes.}

\begin{document}
\maketitle
\flushbottom

\section{Introduction}\label{sec:intro}
Quantum computational complexity provides an estimate of the difficulty of
constructing quantum states for the purpose of performing quantum computations
\cite{Aaronson:2016vto,nielsen2002quantum}. Traditionally, complexity is defined
for spin/qubit chains, using a universal set of unitary gates chosen such that
each gate acts only on a small number of spins/qubits. Appropriately chosen
sequences of these gates, \ie circuits, are able to reproduce, to a given
precision, arbitrary target states of the spin/qubit chain, starting from a
simple unentangled reference state. The complexity of a given target state is
then defined as the minimal number of gates required for such a circuit. The
problem of finding the shortest circuits is challenging if one naively attempts
an exhaustive check of all possible circuits. An alternative geometric approach,
proposed by Nielsen, translates the problem of finding the minimal circuits to a
geometric problem of finding geodesics in the manifold of unitaries equipped
with the metric that naturally arises from the algebra on its tangent space
\cite{nielsen2005geometric,Nielsen1133,dowling2006geometry}.

Nielsen's approach was extended to Gaussian states of quantum field theories
(QFTs) in \cite{QFT1}, where the authors studied the complexity of the vacuum
state of a free scalar QFT (see \cite{QFT2} for an alternative approach based on
the Fubini-Study (FS) metric). Extensions for various classes of Gaussian states
were also studied, see, \eg
\cite{Khan:2018rzm,Hackl:2018ptj,Reynolds:2017jfs,TFD,Jiang:2018nzg,Camargo:2018eof,Jiang:2018gft,Caceres:2019pgf}.
Despite this progress, little is known beyond the free field theory regime, see
however some recent progress in
\cite{Bhattacharyya:2018bbv,Caputa:2018kdj,Caputa:2017yrh,Belin:2018bpg}.

In holography, quantum complexity is proposed to be related to certain geometric
bulk observables by means of the complexity=volume (CV) proposal
\cite{Volume1,Volume2}, relating the complexity to the volume of a maximal bulk
slice anchored at the boundary time slice where the state is defined, and the
complexity=action (CA) proposal \cite{Brown1,Brown2}, relating the complexity to
the action of a certain region in the bulk --- the Wheeler-DeWitt (WDW) patch
--- bounded by light sheets and anchored at that same boundary time slice.
Complexity comes about as part of a wider line of research studying the way in
which quantum information notions are encoded in gravity, which dates back to
the relation between entanglement entropy and Ryu-Takayanagi surfaces, see, \eg
\cite{Ryu:2006bv,Nishioka:2009un}.

The properties of the complexity of Gaussian states of free QFTs turn out to
have surprising qualitative similarities to the properties of holographic
complexity.\footnote{As is standard in the literature, we use ``holographic
  complexity'' here and elsewhere to mean the bulk quantities (volume or action)
  dual to the CFT complexity in the CV and CA conjectures.} This is despite the
fact that these free systems are significantly simpler than the strongly coupled
theories which are of interest in holography. One similarity has to do with the
structure of divergences in complexity \cite{QFT1,QFT2}. Complexity in QFT is
divergent due to the necessity of establishing short range correlations in the
QFT state. This is similar to what happens for entanglement entropy. The
divergences have to be regulated by the introduction of a UV cutoff, for
example, a lattice spacing $\delta$. It turns out that the structure of
divergences of complexity is similar when comparing the free QFT results with
those found in holography, see \cite{QFT1,QFT2,Comments,Reynolds:2016rvl} and
the discussion of \cite{Vaidya2}, with a leading divergence of the form of a
volume law. For subregions of the vacuum state, one obtains in the complexity,
in addition to the leading volume law divergence, also a subleading area law
divergence proportional to the entanglement entropy, both in holography and QFT
\cite{Caceres:2019pgf}. It is expected that the qualitative comparison of free
field theory results to those obtained from gravity will provide hints toward a
good definition of complexity on the field theory side applicable to the dual
field theory of holographic systems.

Despite being very well studied on the holographic side of the correspondence
(as we will explain below), the complexity of charged systems received very
little attention on the quantum field theory side and in particular no models
for complexity of Gaussian states in charged systems have been examined. The
goal of this work is to fill this gap by establishing a systematic computational
framework for evaluating the complexity in charged systems and for studying the
effect of chemical potential on the dynamics of complexity in these systems. Our
formalism utilizes complex phase space operators which turn out to provide a
simpler description of Gaussian states of charged systems. We will use our newly
developed formalism to study the complexity of the Gaussian charged version of
the thermofield double (TFD) state constructed from two copies of a free complex
scalar QFT.

This ``charged thermofield double state'' (cTFD) is a particularly symmetric
purification of a mixed state in the grand canonical ensemble at finite
temperature and chemical potential. The purification is obtained by constructing
a pure state on two identical copies of the system on which the mixed state is
defined and is given by the following expression, see \eg
\cite{Andrade:2013rra}\footnote{We have traded the traditional symbol $Q$ for
  the charge in favour of $C$, in anticipation of using $Q$ as the position
  operator for the harmonic oscillators in the Nielsen construction.
  Consequently, we will use $c$ to denote eigenvalues of charge.}
\begin{equation}\label{eq:intro:cTFD}
  |\cTFD(t_L,t_R)\rangle = \frac{1}{\sqrt{Z_{\beta,\mu}}} \sum_{n,\sigma} e^{-\beta (E_n+\mu c_\sigma)/2-i (E_n+\mu c_\sigma) (t_L+t_R)} |E_n,c_\sigma\rangle_L |E_n,-c_\sigma\rangle_R.
\end{equation}
In the above expression, the two copies of the system are denote L (the ``left''
copy) and R (the ``right'' copy) and are both equipped with an identical
Hamiltonian. $t_{L,R}$ are the times on the left and right copies, $\beta=1/T$
is the inverse temperature, $\mu$ is the chemical potential and $|E_n,
c_\sigma\rangle$ are a basis of energy and charge eigenstates where $E_n$ is the
energy eigenvalue and $c_\sigma$ is the eigenstate of a conserved global $U(1)$
charge. Finally, $Z_{\beta,\mu}$ is a normalization constant. The two states in
each term were chosen to have opposite charges such that they are CPT conjugates
\cite{Andrade:2013rra}. It is apparent from this expression that the time
evolution of the cTFD is governed by the deformed Hamiltonians $H_{L,R} \pm \mu
C_{R,L}$, where $C_{R,L}$ are the $U(1)$ charges on the left/right copies,
respectively. The neutral thermofield double (TFD) state is simply obtained in
the $\mu=0$ limit of the above expression.

TFD and cTFD states received special attention in holographic studies due to
their duality to (neutral and charged) eternal black holes of the corresponding
temperature \cite{Maldacena:2001kr} and chemical potential
\cite{Chamblin:1999tk,Hartnoll:2009sz}. Outside the context of holography, a
number of works have been written on building the TFD state in the laboratory
\cite{Cottrell:2018ash,Wu:2018nrn,Zhu:2019bri,Martyn:2018wli,Maldacena:2018lmt}
and these further motivate focusing on the TFD for simple quantum mechanical
systems. Since charge is very natural in an experimental setup, we expect that
the cTFD state is also a natural state to study in the laboratory.

The complexity of formation of the cTFD and TFD states
\cite{Formation,Brown2,Carmi:2017jqz} is defined as the additional complexity
required in order to prepare the entangled (c)TFD state compared to preparing
both copies of the field theory in their vacuum state. This quantity is UV
finite. For the uncharged TFD state, it was found that the complexity of
formation is proportional to the entropy with positive proportionality
coefficient, both in free QFT and in holography\footnote{The holographic result
  is valid for planar black holes in $d>2$ using both CV and CA, where $d$ is
  the number of boundary spacetime dimensions.} \cite{Formation,TFD}. This
provides another point of similarity between holography and free field theory.
The complexity of formation in charged black hole backgrounds was also studied
in holography in \cite{Brown2,Carmi:2017jqz} and was found to diverge in the
extremal (zero-temperature) limit. This is a result of an IR divergence due to
the infinitely long throat of the wormhole in the extremal limit. The
interpretation suggested in \cite{Carmi:2017jqz} was that of a third law of
complexity, namely that cTFD states at finite chemical potential and zero
temperature are infinitely more complex compared to their finite temperature
counterparts and cannot be formed by any physical process in a finite amount of
time.\footnote{It is worth mentioning that the gravitational setup with charged
  black holes is somewhat advantageous compared to the neutral setup especially
  with regards to evaluating the complexity with the CA proposal since this
  result is influenced by regions of the WDW patch which go arbitrarily near the
  singularity of the neutral black hole. This does not happen for charged black
  holes due to the different causal structure including two horizons.} We will
use our formalism for charged complexity to verify the third law of complexity
for simple Gaussian states and check how universal it is.

The time dependence of complexity of the TFD and cTFD states using the two
holographic proposals was also studied. It was found that the complexity
increases (approximately) linearly as a function of the time $t_L=t_R=t/2$, for
a very long time, much longer than the typical time it takes for other
observables, \eg the entanglement entropy \cite{Hartman:2013qma}, to saturate.
In fact, it was suggested that holographic complexity keeps growing for an
exponential amount of time, until the semi-classical gravity approximation
breaks down \cite{Brown2}. This unusual time dependence captures the growth of
the volume of the wormhole/Einstein-Rosen bridge connecting the two sides of the
geometry via the behind horizon region \cite{Susskind:2014moa} and is also
typical of complexity in spin chains of fast scrambling systems, \eg
\cite{Brown:2016wib,Balasubramanian:2019wgd}.\footnote{Fast scramblers are
  systems which spread the effects of localized perturbations in a time which is
  logarithmic in the entropy \cite{Sekino:2008he}. Holographic systems are
  expected to be fast scrambling \cite{Shenker:2013pqa,Roberts:2014isa}.} The
linear growth of complexity (for long times) served as one of the original
motivations for proposing the holographic conjectures for complexity (CV and
CA). Additional evidence for the validity of these conjectures follows from the
effects of perturbations on the complexity, and in particular from the
manifestation of chaos and scrambling
\cite{Volume2,Bridges,Brown:2016wib,Vaidya1,Vaidya2}. The simple Gaussian states
in free QFTs fail to capture the properties of complexity typical to fast
scrambling systems such as the time dependence of complexity and its reaction to
perturbations.

For charged black holes, the rate of change in complexity over time was found to
vanish in the extremal (vanishing temperature) limit at all times using both the
CV and the CA complexity proposals \cite{Carmi:2017jqz}. It is interesting to
verify how generic this property is --- for example, we would like to understand
whether this property would hold also for simpler Gaussian cTFD states in
charged systems.

To test the effect of charge on properties of the complexity in a simple setup
and as a specific example to our charged complexity formalism, we explore the
complexity of Gaussian cTFD states constructed from two copies of a free complex
scalar QFT.\footnote{A previous attempt to study the complexity of cTFDs was
  made in \cite{Mann1} using the Fubini-Study approach. However, in that work,
  the authors found that the complexity grows linearly for a long time for the
  free complex scalar QFT. We find that this is a gross overestimate of
  complexity since they have not identified correctly the optimal circuit. This
  is easy to see --- while their control function in eq.~(38) starts and stops
  at the right points, it periodically tracks back on itself over the course of
  the circuit when $t$ is large. Considering the $k$-th mode for concreteness,
  when $(\omega_k + \mu q) t$ is a multiple of $2\pi$, the reference and target
  states coincide for that mode (see their eqs.~(12) and (37)), so its
  complexity contribution should vanish --- eq.~(38) fails this basic test as is
  clear from the non-vanishing integrand in the third line of (A-1) when $t$ is
  a multiple of $\frac{2\pi}{\omega_k + \mu q}$. Hence the circuit trajectory is
  clearly not the optimal one.} Our motivation to focus on the free cTFD state
is twofold. First, we would like to qualitatively compare its complexity to the
holographic results. Second, since charges are of experimental relevance in
various quantum mechanical systems, we expect that those results will come to
use also in the context of quantum information physics.

The complex scalar field factorizes to a set of complex harmonic oscillators
with frequencies $\omega_k=\sqrt{k^2+m^2}$ where $m$ is the mass of the complex
scalar and $k$ is the spatial momentum. Furthermore, each harmonic oscillator
can be decomposed using particle and anti-particle modes. A very simple relation
then connects the energy and charge to the number operators for the particle and
anti-particle modes, \ie the energy is proportional to the sum of the numbers of
particles and anti-particles and the charge to their difference. It then becomes
obvious that we can regard the cTFD state of each mode as a product of two
independent TFD states --- one which includes particles on the left copy and
anti-particles on the right copy and another one which includes particles on the
right copy and anti-particles on the left copy. These two TFD states are
associated with effective shifted temperatures and times given by\footnote{We
  have set the elementary unit of charge to one in eq.~\eqref{eqintro:factor}.
  This amounts to the replacement $\mu \leftrightarrow \mu c$ where $c$ is the
  elementary unit of charge in our results below.}
\begin{equation}\label{eqintro:factor}
  \beta_{k,\text{eff}}
  \equiv \beta\left(1\pm\frac{\mu}{\omega_k}\right), \qquad
  t_{k,\text{eff}}
  \equiv t\left(1\pm\frac{\mu}{\omega_k}\right),
  % \omega_{\text{eff}} = \omega \pm \mu,
\end{equation}
respectively, see eq.~\eqref{eq:peppermint}-\eqref{eq:earlgrey}, where
$\omega_k$ are the frequencies of the oscillators. Note that we have a different
effective temperature and time for each mode. It is worth emphasizing that while
the cTFD state (for each momentum mode) factorizes to two uncharged TFD states,
the complexity of the cTFD state cannot be directly related to the complexity of
the two uncharged TFD states, see footnote \ref{foot:complexityNotTrivial}. The
effective temperatures in eq.~\eqref{eqintro:factor} imply that the cTFD state
is only well defined so long as $|\mu|<\omega_k$. As $\mu\rightarrow \omega_k$,
one of the effective temperatures approaches infinity and this means that the
associated mode will be infinitely populated.
% , \ie a condensate develops.
This forces us to consider only\footnote{In any dimension, taking $|\mu|>m$
    leads to a non-normalizable state as all modes in the range $0\le |k| \le
    \sqrt{\mu^2-m^2}$ become infinitely populated. In $d\le 2$, just taking the
    limit $|\mu|\to m$ leads to a non-normalizable state, as can be seen from
    the divergence of the (vacuum-subtracted) grand canonical potential (which
    is proportional to the logarithm of the normalization factor). In $d \ge 3$,
    when naively neglecting possible condensation in the zero mode $k=0$, the
    $|\mu|\to m$ state is normalizable, but perhaps still physically
    questionable at $d=3$ as it gives an infinite particle number density.
    Finally, in $d\ge 4$, states with $|\mu|=m$ correspond to a phase with
    Bose-Einstein condensation where the zero mode $k=0$ makes up a nontrivial
    fraction of the total (finite) particle number density. (The derivation of
    these facts follows from a straightforward extension of the standard
    statistical mechanics treatment of a Bose gas, except here we consider the
    relativistic spectrum $\omega_k = \sqrt{k^2+m^2}$ rather than the
    non-relativistic spectrum $\frac{k^2}{2m}$.) In this paper, however,
    whenever we take the $|\mu|\to m$ limit, we shall assume that the
    contribution from the condensate is negligible. We expect this to be the
    case, for example, when the $|\mu|\to m$ limit and the limit of infinite
    spatial extent $L$ are taken such that we are on the boundary of (or just
    inside the) uncondensed phase, where the number density
    $\frac{1}{L^{d-1}} \frac{1}{e^{\beta (m-|\mu|)} - 1}$ of the condensate
    vanishes. \label{foot:blahblahblah}} $|\mu|<m$ and prevents us from taking
the conformal limit $m \rightarrow 0$ while holding the chemical potential fixed
when studying the complexity of the full fledged complex scalar QFT. It would be
interesting to verify if this problem can be avoided in a different
setup.\footnote{For example, by studying the theory in a (sufficiently small)
  finite volume with Dirichlet boundary conditions, one should be able to take
  the $m\rightarrow 0$ limit without encountering a condensate.} Furthermore, it
would be interesting to include the condensate explicitly in evaluating the
complexity. We leave these problems for the future.

In evaluating the complexity of the charged thermofield double state we find
that the third law of complexity is not reproduced in the free charged scalar
example. We do find however that the computation slows down as we decrease the
temperature which is similar to what happens in holography.

This paper is organized as follows. In section \ref{sec:covprel}, we present our
framework for studying charged complexity in the context of quantum mechanical
systems. We explain how to generalize Nielsen's complexity approach to circuits
between Gaussian states using covariance matrices defined with respect to
complex phase space operators which are useful in the presence of charge. In
section \ref{sec:praline}, we review the various necessary ingredients for
studying the complexity of states in a QFT of a complex scalar. In section
\ref{sec:George}, we explain how to evaluate the complexity of the cTFD state.
We present our numerical results for the complexity of formation, as well as a
number of analytic expansions for low and high temperatures. We then present
numerical results for the time dependence of the complexity of the cTFD state.
In section \ref{sec:holo}, we compare our results to those obtained in
holography. We end with a summary and outlook in section
\ref{sec:summaryoutlook}. We have left a number of technical details for
appendices. In appendix \ref{sec:deadline}, we present the generators used to
evaluate the complexity as explained in section \ref{sec:covprel}. In appendix
\ref{app:timeevol}, we present an explanation for the form of the time evolution
of the cTFD state. Finally, in appendices \ref{app:lowTlowT} and \ref{sec:cv},
we present additional details required for evaluating the low temperature limits
in section \ref{lowtemplim}.

%%% Local Variables:
%%% mode: latex
%%% TeX-master: "../ChargedComplexityV7"
%%% End:

\section{Charged Complexity from Complex Covariance Matrices}\label{sec:covprel}

We start by generalizing several aspects of Nielsen's geometric approach to the
quantum complexity of Gaussian states using covariance matrices \cite{TFD} to
account for the possibility of complex phase space
operators.
Complex phase space operators are a simple rotation of real phase space
operators as we demonstrate below, but as we will see, they provide a much more
natural framework for studying charged systems. In particular, they are very
commonly used when studying charged (complex) scalar field theory.

Gaussian states are completely characterized by the one and two-point functions
of a set of phase space operators. When working with complex phase space
operators, we start by listing the coordinates and their conjugates as
follows\footnote{Although we do not put hats on top of our phase space
  operators, here we always mean operators in the quantum mechanical system.}
\begin{equation}\label{complex:phasespace}
  \xi_{\mathbb{C}} = (q_1,q_1^\dagger,\ldots,q_{N/2},q_{N/2}^\dagger,p_1^\dagger,p_1,\ldots,p_{N/2}^\dagger,p_{N/2}),
\end{equation}
where we have specifically chosen to invert the order of the momentum operators
$p_i^\dagger$ and $p_i$ compared to the order of the corresponding position
operators $q_i$ and $q_i^\dagger$ and where $N$, the number of (real) degrees of
freedom, is an even integer. The complex phase space operators can also be
re-expressed in terms of real phase space operators
\begin{align}
  \xi_{\mathbb{R}} = (q_1,\ldots,q_N,p_1,\ldots,p_N),
  \label{eq:oink}
\end{align}
where here, $N$ is the same (even integer) as in eq.~\eqref{complex:phasespace}.
From now on we will use subscripts $\mathbb{C}$ and $\Rbb$ to refer to the cases
of complex and real phase space operators, respectively. Statements written
without a subscript are correct for both types of operators. Note that $\xi$ and
$\xi^\dagger$ are not independent from each other but are rather related
according to
\begin{equation}\label{xi:xidagger:real}
  (\xi^{a})^{\dagger}=A^{a}{}_{b}\, \xi^{b}
\end{equation}
where $A$ is the following matrix
\begin{equation}
  A_{\mathbb{R}}=\mathds{1}_{2N \times 2N}, \qquad A_{\mathbb{C}}=\bigoplus_{I=1\ldots N} \begin{bmatrix}
    0 && 1 \\
    1 && 0
  \end{bmatrix},
  \label{eq:purr}
\end{equation}
for the cases of real and complex operators, respectively. $\bigoplus$ stands
for a direct sum of 2 by 2 matrices acting on each pair of operators by their
order of appearance in eq.~\eqref{complex:phasespace}. Note that both these
transformations satisfy $A^2=1$ and $A^{\mathrm{T}}=A$.

The phase space operators satisfy canonical commutation relations given in terms
of the symplectic form $\Omega$ as follows\footnote{In the first equality of
  eq.~\eqref{comrel:complex}, $\Omega^{ab}$ is implicitly multiplying a unit
  operator in the Hilbert space.}
\begin{equation}\label{comrel:complex} [\xi^a, (\xi^{b})^{\dagger}] = i
  \Omega^{ab}, \qquad \Omega
  = \begin{bmatrix}
    0 & \mathds{1}_{N \times N} \\
    -\mathds{1}_{N \times N} & 0
  \end{bmatrix}.
\end{equation}
As mentioned earlier, we will focus on Gaussian states which are completely
characterized by the two point functions of the phase space operators
\begin{align}\label{covcov0}
  \langle \psi|\xi^a (\xi^b)^{\dagger}|\psi\rangle
  \equiv \frac{1}{2}(G^{ab}+i\Omega^{ab}).
\end{align}
In the above expression, the anti-symmetric part is the matrix $\Omega^{ab}$, and
the symmetric part
\begin{align}\label{covcov}
  G^{ab}=G^{(ab)}=\langle \psi|\xi^a (\xi^b)^{\dagger}+ (\xi^b)^{\dagger}\xi^a|\psi\rangle
\end{align}
is known as the {\it covariance matrix}. Note that the matrix representation of
the covariance matrix, \ie $G$, is Hermitian, and therefore satisfies
$(G^{\dagger})^{ab} =(G^{ba})^{\dagger}=G^{ab}$. Everywhere in this paper, we
will only focus on Gaussian states with vanishing one point functions $\langle
\psi|\xi^a |\psi\rangle=0$ which will be completely characterized by their
covariance matrices. By restricting our attention to Gaussian states, \ie
selecting a reference and target state which are Gaussian and only considering
circuits moving through the space of Gaussian states, we are able to make some
progress in solving for the optimal circuits in the complexity geometry. The
treatment of complexity using covariance matrices was proposed in \cite{TFD} as
an alternative to the wavefunction approach of \cite{QFT1}. This approach proves
simpler in cases where the circuit utilizes gates which are quadratic in both
the position and the momentum operators such as when constructing the TFD and
cTFD states.

The action of a circuit moving between Gaussian states can be parameterized by
the action of Hermitian generators which are quadratic in the canonical
operators $\xi$,
\begin{equation}\label{unitop}
  |\tilde \psi \rangle =\hat U |\psi\rangle, \qquad \hat U = e^{-i \hat K},
\end{equation}
\begin{equation}\label{Qgenerators}
  \qquad \hat K \equiv \frac{1}{2} (\xi^{a})^{\dagger}\, k_{ab} \, \xi^{b},
\end{equation}
where $k_{ab}$ is the matrix characterization of a given Hermitian generator
$\hat K$. Note that in order for $\hat K$ to be Hermitian, the matrix $k_{ab}$
should satisfy
\begin{equation}\label{constraint2}
  k=k^\dagger.
\end{equation}
Furthermore, due to the relation \eqref{xi:xidagger:real}, we can assume,
without loss of generality, that $A \cdot k$ is symmetric, namely that the $k$
matrices satisfy
\begin{equation}\label{constraint1}
  k^{\mathrm{T}} = A \cdot k \cdot A\,.
\end{equation}

The action of the unitary in eq.~\eqref{unitop} on the state can be represented
directly as an operation on the covariance matrix. To see this, we start by
exploring the effect of conjugating the canonical operators with the unitary
operation \eqref{unitop}
\begin{equation}\label{tosum}
  \tilde \xi^a   \equiv \hat U^{\dagger}\, \xi^a \, \hat U = \sum_{n=0}^{\infty} \frac{1}{n!} [i \hat K, \xi^a]_{(n)}
\end{equation}
where the Taylor expansion is given in terms of the nested commutator defined
recursively by $[i \hat K, \xi^a]_{(n)}\equiv [i \hat K ,[i \hat K,
\xi^a]_{(n-1)}]$, where $[i \hat K, \xi^a]_{(0)}=\xi^a$. Using the commutation
relations in eq.~\eqref{comrel:complex}, we find
\begin{equation} [i \hat K, \xi^a] = \frac{1}{2} \left[\Omega \cdot (A \cdot k^T
    \cdot A + k)\right]^{a}{}_{b}\, \xi^b = (\Omega\cdot k)^{a}{}_{b}\, \xi^b
  \equiv K^a{}_b\,\xi^b \;,
\end{equation}
where we have defined
\begin{equation}\label{Kuseful}
  K \equiv \Omega\cdot k, \qquad K^\dagger = - k \cdot \Omega \;.
\end{equation}
Resumming eq.~\eqref{tosum}, we obtain
\begin{equation}\label{allS}
  \tilde \xi^a = S^a{}_b \, \xi^b, \qquad  S \equiv e^{K}.
\end{equation}
It is then straightforward to check that this induces the following
transformations on the covariance matrix \eqref{covcov}
\begin{equation}
  \tilde G = S \cdot G \cdot  S^\dagger,
\end{equation}
where $\tilde G$ is the covariance matrix associated with the state $|\tilde
\psi\rangle$. The unitary conjugation \eqref{tosum} does not modify the
commutation relations since
\begin{equation} [\tilde \xi^a, \tilde \xi^b] = S (i \Omega) S^\dagger =
  (i\Omega) \;,
\end{equation}
which is satisfied automatically by virtue of the identity $K \Omega + \Omega
K^\dagger=0$, see eq.~\eqref{Kuseful}. Note however that this last condition is
less restrictive than requiring $K=\Omega \cdot k$ for some $k$ satisfying
\eqref{constraint2}-\eqref{constraint1}.

The next step is to identify a basis of independent generators which span the
space of possible Hermitian generators $\hat K$ in eq.~\eqref{Qgenerators}.
Naively, we could think that taking a basis for general Hermitian matrices
$k_{ab}$ in eq.~\eqref{Qgenerators} would naturally induce a basis for the
independent Hermitian generators. However, due to the relation
\eqref{xi:xidagger:real}, this would be over-counting and it is straightforward
to see that the independent generators are instead given by matrices $k$ which
satisfy the conditions \eqref{constraint2}-\eqref{constraint1}.

Whether the operators are real or complex, the number of independent generators
for these transformations between Gaussian states can be counted by counting the
number of constraints in eqs.~\eqref{constraint2}-\eqref{constraint1} and is
equal to $N(2N+1)$. This is due to the fact that the group of all $S$ in
eq.~\eqref{allS}, where $K=\Omega \cdot k$ and $k$ satisfies the conditions
\eqref{constraint2}-\eqref{constraint1}, is simply isomorphic to
$\Sp(2N,\mathbb{R})=\{e^{\bar K}\in M_{2N\times 2N}(\mathbb{R}) \text{ with }
\bar K\cdot \Omega+\Omega \cdot \bar K^T=0\}$. To see this, first note that the
algebra of generators of $\Sp(2N,\mathbb{R})$ can be recast as the algebra of
real symmetric matrices $\bar{k}$ defined by $\bar K=\Omega\cdot \bar k$, since
the condition that $\bar k$ be symmetric is satisfied if and only if $\bar
K\cdot \Omega+\Omega \cdot \bar K^T=0$. Further note that we may then relate the
generators $\bar k$ to the independent generators $k$ satisfying
eqs.~\eqref{constraint2}-\eqref{constraint1} according to $\bar{k}=k$ for real
phase space operators and $\bar k = R \cdot k \cdot R^\dagger$, where
\begin{equation}\label{Strans0}
  R=\bigoplus_{I=1\ldots N} \frac{1}{\sqrt 2}\begin{bmatrix}
    1 && 1 \\
    -i && i
  \end{bmatrix},
\end{equation}
for complex phase space operators, and these $\bar{k}$ will automatically be
real and symmetric using eqs.~\eqref{constraint2}-\eqref{constraint1} due to the
identity $A_\Cbb \cdot R^\mathrm{T} = R^{\dagger}$.

We will label by $K_I$ a complete basis of independent generators for the
transformations above and will generally assume that they are orthonormal with
respect to the inner product
\begin{equation}
  \frac{1}{2}\tr(K_I K_J^\dagger)=\delta_{IJ}.
  \label{eq:application}
\end{equation}
Explicit expressions for those generators using both real and complex operators
can be found in appendix \ref{sec:deadline}.

At the next step, we will want to construct circuits through the space of
Gaussian states.
The circuits will act on the covariance matrices while leaving the symplectic
form invariant, \ie{}
\begin{align}\label{circuitgens}
  G(\sigma)
  = S(\sigma) \, G_\rf \, S^{\dagger}(\sigma), \qquad S(\sigma)\, \Omega \, S^\dagger(\sigma)=\Omega ,
\end{align}
where $\sigma\in[0,1]$ is a trajectory parameter. Boundary conditions are
imposed such that the circuit $G(\sigma)$ moves between the covariance matrix of
the \emph{reference state} $G_\rf$ at $\sigma=0$ and ends at the covariance matrix of
the \emph{target state} $G_\mathrm{target}$ at $\sigma=1$. Explicitly, this means that
$S(\sigma=0)$ acts trivially on $G_\rf$, while
\begin{align}\label{cond}
  S(\sigma=1) \, G_\rf \, S^\dagger(\sigma=1)
  = G_\mathrm{target}.
\end{align}
The symplectic transformation $S(\sigma)$ can be decomposed according to
\begin{equation}\label{pathorder}
  S(\sigma)
  \equiv \overleftarrow{\mathcal{P}} \exp \int_0^\sigma d\sigma'\; Y^I(\sigma') K_I,
\end{equation}
where $K_I$ are the generators of the symplectic group, assumed to be
orthonormal, as in eq.~\eqref{eq:application}, and $\overleftarrow{\mathcal{P}}$
denotes right to left path ordering, \ie the product integral of infinitesimal exponentials re-ordered from right to left in ascending order in $\sigma'$.
Eq.~\eqref{pathorder} may be
interpreted as decomposing the path $S(\sigma)$ through the space of unitary
operations on Gaussian states as a sequence of infinitesimal operations
generated by $K_I$ turned on by the control functions $Y^I(\sigma)$. This is
completely analogous to the usual quantum computation picture mentioned in
section \ref{sec:intro}, where a circuit is composed of a sequential application
of gates chosen from some set of elementary gates --- here $S(\sigma)$ is the
circuit and infinitesimal exponentiation of the generators $K_I$ produces the
elementary gates (or rather, their action on the covariance matrices).

This geometric approach was implemented in \cite{TFD} using several different cost
functions $F$ to evaluate the length
\begin{align}
  d[S(\sigma)]
  =& \int_0^1 d\sigma\; F(S(\sigma),Y^I(\sigma))
     \label{eq:meow}
\end{align}
of a given circuit. The cost functions considered were:
\begin{align}
  F_1
  =& \sum_I |Y^I|,
  &
    F_2
    =& \sqrt{\sum_I (Y^I)^2},
  &
    D_\kappa
    =& \sum_I |Y^I|^\kappa.
\end{align}
The complexity of the target state is given by the length of the shortest
circuit, \ie path $S(\sigma)$ through $\Sp(2N,\mathbb{R})$, satisfying the
boundary condition \eqref{cond} (as well as acting trivially on the reference
state at $\sigma=0$), for a given choice of cost function,
\begin{align}
  \cmplx
  \equiv \min_{S(\sigma)} d[S(\sigma)].
  \label{eq:pupper}
\end{align}
Note that the result for the complexity is basis dependent, \ie in general the
choice of basis $K_I$ influences this result. However, the $F_2$ and $\kappa=2$
cost functions remain unchanged when the two bases are related by an orthogonal
transformation on the position operators and an identical orthogonal
transformation on the momentum operators.

Before proceeding, let us comment on the dimensions of various quantities
introduced thus far. In order for Nielsen's notion of complexity, defined in
eqs.~\eqref{eq:meow}-\eqref{eq:pupper}, to be sensible generalizations of gate
counting, it must be dimensionless.
This is achieved by considering dimensionless phase space operators $\xi$ (and
hence dimensionless $F$, $Y^I$, $K_I$, $K$, and $k$). In order to absorb the
dimensions intrinsic to the usual phase space operators of physical systems, it
will therefore be necessary to introduce a dimensionful scale $\omega_g$, see
\eg{}section 2.2.3 of \cite{TFD}. In section \ref{sec:woof}, we shall use
$\omega_g$ to translate between dimensionful physical operators and the
dimensionless operators with respect to which complexity is defined.

A particularly simple circuit between the reference state and the target state
is the `straight line' circuit, obtained by exponentiating a constant Lie
algebra element multiplied by $\sigma$. The straight line circuit can be
obtained as follows. Given a target state covariance matrix $G_\mathrm{target}$
and a reference state covariance matrix $G_\rf$, the relative covariance matrix
is defined to be their ratio:
\begin{align}
  \Delta_\mathrm{target}
  \equiv& G_\mathrm{target} G_\rf^{-1}\, .
          \label{eq:piano}
\end{align}
The straight line circuit is then given by
\begin{align} \label{eq:piano2} S(\sigma) =& e^{\sigma K}, \qquad \text{where}
  \qquad K=\frac{1}{2}\log\Delta_\mathrm{target}.
\end{align}

It was proved in \cite{TFD} that, for the case of the $F_2$ and $D_{\kappa=2}$
cost functions, when the reference state scale and the gate scale (to be
introduced below in section \ref{sec:woof}) are taken to be equal, the path of
minimal cost is indeed the straight line circuit. With these scales equal to each other, it was further suggested in \cite{TFD} that the straight line circuit provides a good approximation for the $F_1$ complexity. In other cases, \cite{TFD} have suggested that
  the straight line circuit yields a non-trivial upper bound for the
  complexity.

Of the cost functions considered in \cite{TFD}, it was found that the $F_1$ cost
produces results most similar to those obtained in holography using the
holographic complexity proposals, for the structure of UV divergences in the
complexity of the full boundary state \cite{QFT1} as well as that of mixed
states \cite{Caceres:2019pgf}. Hence, in this note, we too will focus on the
$F_1$ complexity, evaluating it on the straight line
circuit. Furthermore, since the $F_1$ cost function depends on the choice of
generators, two choices were proposed in \cite{TFD}. The first choice retains
the left-right coordinate split between the two sides of the TFD, while the
second choice mixes the two into a ``diagonal'' basis. The first choice was
found to yield properties more similar to those of holographic complexity, in
particular in reproducing the proportionality between the complexity of
formation and the entropy. We will therefore focus on this choice in this paper.

In order to obtain the $F_1$ complexity, we have to decompose the trajectory
\eqref{eq:piano2} according to the expression \eqref{pathorder} and extract the
scalar coefficients $Y^I$, which appear in the cost function. Since we have
assumed that our basis of generators $K_I$ is orthonormal, see eq.
\eqref{eq:application}, we can do this by using the inner product
\begin{align}
  Y^I
  =& \frac{1}{2} \tr \left( K K_I^\dagger\right).
     \label{eq:Topaz}
\end{align}
Finally, an upper bound $\cmplx_1^\UB$ for the complexity $\cmplx_1$ associated
with the $F_1$ cost function is given by integrating $F_1$ along the straight
line circuit. Since the generator of the trajectory is simply constant along the
path, this yields
\begin{align}
  \cmplx_1\le&~ \cmplx_1^\UB
               = \sum_I |Y^I| = \frac{1}{2} \sum_I \left|\tr \left( K K_I^\dagger\right)\right| = \frac{1}{4} \sum_I \left|\tr \left( \log(\Delta_{\text{target}}) \cdot K_I^\dagger\right)\right|.   \label{eq:ciaccino}
\end{align}

Before we end this section, let us mention that, in order to move between
different bases of generators, it is possible to use a coordinate transformation
\begin{equation}
  \tilde \xi^a = R^a{}_b \,\xi^b,
  \label{eq:fluvoxamine}
\end{equation}
where $R$ is a general complex matrix preserving the commutation relations
\begin{equation}\label{cond:trans}
  \Omega=R \cdot \Omega \cdot R^\dagger.
\end{equation}
The covariance matrices, circuit \eqref{circuitgens} and generators \eqref{allS}
get rotated according to
\begin{equation}
  \tilde G = R \cdot G \cdot R^{\dagger}, \qquad
  \tilde S = R\cdot  S \cdot R^{-1}, \qquad
  \tilde K = R \cdot  K\cdot  R^{-1}\,.
  \label{eq:fox}
\end{equation}
One such useful transformation mentioned earlier is the one moving between the
real and complex operators, \ie $\xi_\mathbb{R} = R_{\mathbb{C}\to
  \mathbb{R}}\xi_\mathbb{C}$ where $R_{\mathbb{C}\to \mathbb{R}}$ is the same
transformation as given in eq.~\eqref{Strans0}. Note that not all such basis
transformations can be represented as unitary transformations acting on the
state $|\psi\rangle$ used to define the covariance matrix. Finally, let us point
out that the inner product \eqref{eq:Topaz} has to be evaluated in the basis in
which we want to compute the complexity. Alternatively, we may compute it in a
different basis by using the rotated inner product
\begin{align}\label{modified}
  Y^I
  =& \frac{1}{2} \tr \left( \tilde K \tilde{G}_{\mathds{1}} \tilde K_I^\dagger \tilde{G}_{\mathds{1}}^{-1}\right),
\end{align}
defined in terms of the positive symmetric matrix $\tilde{G}_{\mathds{1}} = R
R^{\dagger}$, where $R$ is the transformation matrix between the two bases.

%%% Local Variables:
%%% mode: latex
%%% TeX-master: "../ChargedComplexityV6"
%%% End:

\section{Complex Scalar Field Theory}\label{sec:praline}
In this section, we establish our notation for the complex scalar QFT and
explain how to put it on a lattice. This translates the problem of studying the
complexity of charged states in this QFT to a problem of studying the complexity
of charged states of a set of coupled harmonic oscillators, which can then be
addressed using the tools of section \ref{sec:covprel}. In subsection
\ref{decpartapart}, we present a useful choice of basis which decouples the
contributions of the particles and anti-particles to the Hamiltonian and charge.
This basis is generally useful for studying the complexity of charged states,
which are typically defined in terms of eigenstates of these two operators. We
will demonstrate how to use all this machinery to study the specific example of
the complexity of the cTFD state in the next section.

\subsection{Preliminaries}\label{sec:mocca}

We will focus on a theory consisting of a complex scalar field in $d$ spacetime
dimensions. The Hamiltonian of the system is given in terms of the fields $\phi$, $\phi^{\dagger}$ and
conjugate momenta $\pi$, $\pi^\dagger$ according to
\begin{align}
  H
  =& \int d^{d-1}x\;
     \left(\pi^\dagger \pi + \vec \nabla \phi^\dagger \cdot \vec \nabla \phi + m^2 \phi^\dagger \phi \right) . \label{eq:green}
\end{align}
The field and momentum operators obey the equal time commutation relations
$[\phi(\vec x),\pi(\vec y)] = [\phi^\dagger(\vec x),\pi^\dagger(\vec y)] = i
\delta^{d-1} (\vec x-\vec y)$. The charge is given by
\begin{align}
  C
  =&\, i\int d^{d-1}x\; (\phi^\dagger\pi^\dagger-\phi \pi ).
     \label{eq:maize}
\end{align}
In the above expression, we chose our convention such that the fundamental unit
of charge is set to one, but of course, this dependence can be recovered later
by redefining the chemical potential appropriately. The complex scalar field can
be decomposed in terms of the following Fourier modes
\begin{equation}
  \begin{split}
    \phi(x)&=\int \frac{d^{d-1} p}{(2\pi)^{d-1}}
    \frac{1}{\sqrt{2\omega_p}}\left(a_{\vec p} \,e^{-i\omega_p t+i\vec p
        \cdot\vec x} + \bar a_{\vec p}^{\dagger}\,e^{i\omega_p t-i\vec
        p\cdot\vec x}\right),
    \\
    \pi(x)&=-i \int \frac{d^{d-1} p}{(2\pi)^{d-1}}
    \sqrt{\frac{\omega_p}{2}}\left(\bar a_{\vec p}\, e^{-i\omega_p t+i\vec p
        \cdot\vec x} - a_{\vec p}^{\dagger}\,e^{i\omega_p t-i\vec p\cdot\vec
        x}\right),
  \end{split}
\end{equation}
where $a_{\vec p}$, $a_{\vec p}^{\dagger}$ and $\bar a_{\vec p}$, $\bar a_{\vec
  p}^\dagger$ are annihilation and creation operators for particle and
antiparticle modes, respectively, satisfying the following commutation relations
\begin{equation} [a_{\vec p},a_{\vec p\,'}^\dagger]=[\bar a_{\vec p},\bar
  a_{\vec p\,'}^\dagger]= (2\pi)^{d-1} \delta(\vec p-\vec p\,'),
\end{equation}
and where $\omega_p\equiv\sqrt{{\vec p}\,^2 + m^2}$. In terms of those creation
and annihilation operators, the Hamiltonian and charge are given by
\begin{equation}\label{sunnysideup}
  H = \int  \frac{d^{d-1} p}{(2\pi)^{d-1}} \, \omega_p \left(a_{\vec p}^\dagger\, a_{\vec p}+\bar a_{\vec p}\, \bar a_{\vec p}^\dagger  \right), ~~~
  C = \int \frac{d^{d-1} p}{(2\pi)^{d-1}}\left(a_{\vec p}^\dagger \,a_{\vec p} - \bar a_{\vec p}^\dagger \, \bar a_{\vec p}\right).
\end{equation}
This reflects the fact that particles and antiparticles contribute to the energy
of a given state according to the sum of their number operators while
contributing to the charge with opposite signs.

\subsection{Normal Mode Decomposition on the Lattice}
\label{sec:woof}
As explained in section \ref{sec:intro}, the complexity is divergent and has to
be regularized. This was done in \cite{QFT1,TFD} by placing the theory on a
spatial periodic lattice. Hence, we will start by briefly reviewing how to place
the free complex scalar on such a lattice. The resulting theory will be a sum of
harmonic oscillators for the different momentum modes.

We will use a periodic lattice of size $L$ in each space direction with
$N^{d-1}$ sites and lattice spacing $\delta=L/N$. For convenience, we assume
that $N$ is odd. The different sites will be labelled by indices
\begin{equation}
  \vec a\equiv(a_1,\ldots,a_{d-1}) \in \left\{-\tilde N,\ldots,\tilde N\right\}^{d-1}, \quad\text{where} \quad \tilde N \equiv\frac{N-1}{2}. \label{eq:cucumber}
\end{equation}
The discretized versions of eqs.~\eqref{eq:green} and \eqref{eq:maize} take the
form
\begin{align}
  \begin{split}
    H =& \sum_{\vec a} \left[ \delta \hat{P}_{\vec a}^\dagger \hat{P}_{\vec a} +
      m^2 \delta^{-1} \hat{Q}_{\vec a}^\dagger \hat{Q}_{\vec a} + \delta^{-3}
      \sum_j (\hat{Q}_{\vec a+\vec e_j}-\hat{Q}_{\vec a})^\dagger(\hat{Q}_{\vec
        a+\vec e_j}-\hat{Q}_{\vec a})
    \right] \label{eq:cornetto},\\
    C =& i\sum_{\vec a}(\hat{Q}_{\vec a}^\dagger \hat{P}_{\vec a}^\dagger
    -\hat{Q}_{\vec a} \hat{P}_{\vec a}),
  \end{split}
\end{align}
where we have defined
\begin{align}\label{deltarels}
  \hat{Q}_{\vec a}
  \equiv \delta^{d/2} \phi(\delta \cdot \vec a), \quad
  \hat{P}_{\vec a}
  \equiv \delta^{d/2-1}\pi(\delta \cdot  \vec a),
\end{align}
$\vec e_j$ denotes the unit vector in the $j$-th direction and we have written
the position and momentum operators with hats in order to keep the symbols $P,Q$
free for later use. These coordinates and momentum operators satisfy the
commutation relations
\begin{equation} [\hat{Q}_{\vec a},\hat{P}_{\vec b}] = i \delta_{\vec a ,\vec
    b}.
\end{equation}
We see that, on the lattice, the field theory reduces to a theory of coupled
harmonic oscillators. To decouple these oscillators, we move into Fourier space
by defining
\begin{align}
  \begin{split}
    & \tilde{Q}_{\vec n} \equiv N^{-\frac{d-1}{2}} \sum_{\vec a} e^{-\frac{2\pi
        i \vec n\cdot \vec a}{N}} \hat{Q}_{\vec a}\;,\qquad \tilde{P}_{\vec n}
    \equiv N^{-\frac{d-1}{2}} \sum_{\vec a} e^{\frac{2\pi i \vec n\cdot \vec a}{N}} \hat{P}_{\vec a}\;,\\
    &~~~~~~~~~~~~~~~~~~~~~ n\equiv (n_1,\ldots,n_{d-1})
    \in\{-\tilde{N},\ldots,\tilde{N}\}^{d-1}.
    \label{eq:sandwich}
  \end{split}
\end{align}
Note that we have chosen opposite signs for the phases in the Fourier transforms
of the position and momentum operators. These operators satisfy the commutation
relations
\begin{equation}\label{comrelnk} [\tilde{Q}_{\vec n},\tilde{P}_{\vec k}] = i
  \delta_{\vec n ,\vec k}.
\end{equation}
The Hamiltonian and charge then read\footnote{Throughout the following, we will
  stick to the convention where $\tilde{Q}_{\vec n}^\dagger$ is the complex
  conjugate of $\tilde{Q}_{\vec n}$, rather than being the Fourier transform of
  the coordinate $Q_{\vec a}^\dagger$, with conventions as in
  \eqref{eq:sandwich}.}
\begin{equation}
  \begin{split}\label{eq:omelette}
    & H = \sum_{\vec n} \left( \delta \tilde{P}_{\vec n}^\dagger \tilde{P}_{\vec
        n} + \omega_n^2 \delta^{-1} \tilde{Q}_{\vec n}^\dagger \tilde{Q}_{\vec
        n} \right),\qquad \omega_n^2 \equiv m^2 + \frac{4}{\delta^{2}} \sum_j
    \sin^2\left(\frac{n_j\pi}{N}\right),
    \\
    & ~~~~~~~~~~~~~~~~~~~~~~~~~~~~~~ C = i\sum_{\vec n}(\tilde{Q}_{\vec
      n}^\dagger \tilde{P}_{\vec n}^\dagger - \tilde{Q}_{\vec n}\tilde{P}_{\vec
      n})\, .
  \end{split}
\end{equation}

In order to gain physical intuition, it is also instructive to consider the
decomposition of the scalar field in terms of creation and annihilation
operators. On the lattice, the complex scalar field and its conjugate momentum
have mode expansions\footnote{The creation and annihilation operators in this
  section are dimensionless and are related to the ones in the previous section
  according to $a_{\vec
    p}^{\text{continuous}}=L^{\frac{d-1}{2}}a^{\text{lattice}}_{\vec n}$ where
  $\vec{p}=\frac{2\pi}{L}\vec{n}$.\label{bigfoot11}}
\begin{align}\label{discrete1}
  \begin{split}
    \phi(\delta\cdot \vec a,t) =& L^{-\frac{d-1}{2}} \sum_{\vec n}
    \frac{1}{\sqrt{2\omega_n}} \left(a_{\vec n} e^{-i\left(\omega_n t-\frac{2\pi
            \vec n\cdot \vec a}{ N }\right)} + \bar{ a}_{\vec n}^\dagger
      e^{i\left(\omega_n t-\frac{2\pi \vec n\cdot \vec a}{N}\right)}\right),
    \\
    \pi(\delta\cdot \vec a,t) =& -i L^{-\frac{d-1}{2}} \sum_{\vec n}
    \sqrt{\frac{\omega_n}{2}} \left(\bar{a}_{\vec n} e^{-i\left(\omega_n
          t-\frac{2\pi \vec n\cdot \vec a}{N}\right)} - a_{\vec n}^\dagger
      e^{i\left(\omega_n t-\frac{2\pi \vec n\cdot \vec a }{N}\right)}\right),
  \end{split}
\end{align}
where $\vec a,\vec n$ take values as indicated by eqs.~\eqref{eq:cucumber} and
\eqref{eq:sandwich}, $\omega_n$ is defined in eq.~\eqref{eq:omelette}, and
\begin{align}
  [a_{\vec n},a_{\vec n'}^\dagger]
  =& [\bar{a}_{\vec n},\bar{a}_{\vec n'}^\dagger]
     = \delta_{\vec n\vec n'}
\end{align}
with other creation and annihilation commutators vanishing. Using
eqs.~\eqref{discrete1}, \eqref{eq:sandwich} and \eqref{deltarels} we can deduce
\begin{equation}\label{theRealexpansionofchops}
  \tilde{Q}_{\vec n}=\sqrt{\frac{\delta}{2\omega_n}} \left(a_{\vec n}+\bar{ a}_{-\vec n}^\dagger\right), \qquad
  \tilde{P}_{\vec n}=-i \sqrt{\frac{\omega_n}{2\delta}}\left(\bar a_{-\vec n} - a_{\vec n}^\dagger\right).
\end{equation}
The $a_{\vec n},\bar{a}_{\vec n}$ can be regarded as annihilation operators for
particles and anti-particles respectively since the Hamiltonian and charge of
the field are given by
\begin{align}
  H
  =& \sum_{\vec n} \omega_n (N_{\vec n} + \bar{N}_{\vec n}+1),
     \label{eq:dragonFruit}
  &
    C
    =& \sum_{\vec{n}} (N_{\vec n} - \bar{N}_{\vec n}),
\end{align}
where the number operators are defined to be
\begin{align}\label{numberops1}
  N_{\vec n} \equiv& a_{\vec n}^\dagger a_{\vec n},
  & \bar{N}_{\vec n} \equiv& \bar{a}_{\vec n}^\dagger \bar{a}_{\vec n}.
\end{align}
Note that one usually normal-orders the Hamiltonian, removing the last term of
the summand in the first equation of \eqref{eq:dragonFruit} representing the
zero point energy.

As already mentioned in the discussion below eq.~\eqref{eq:pupper}, the gates
used in constructing quantum circuits for studying complexity in this paper and
also those used to construct the TFD state in \cite{TFD} consist of quadratic
combinations of the coordinate and momentum operators. Since these are
dimensionful operators we will have to introduce an additional scale $\omega_g$
(with inverse length dimensions) in our complexity model. This scale is used to
rescale the position and momentum operators in such a way that they become
dimensionless
\begin{align}
  \tilde{q}_{\vec n} \equiv& \omega_g \, \tilde{Q}_{\vec n} =\sqrt{\frac{1}{2\lambda_n}} \left(a_{\vec n}+\bar{ a}_{-\vec n}^\dagger\right) ,
  &
    \tilde{p}_{\vec n}
    \equiv& \frac{\tilde{P}_{\vec n}}{\omega_g}=-i \sqrt{\frac{\lambda_n}{2}}\left(\bar a_{-\vec n} - a_{\vec n}^\dagger\right), \label{eq:lemonade2}
\end{align}
where we have defined
\begin{equation}\label{gatescale1}
  \lambda_n
  \equiv \frac{\omega_n}{\mu_g}, \qquad \text{where} \qquad \mu_g \equiv \delta \omega_g^2,
\end{equation}
and we refer to $\mu_g$ as the \emph{gate scale}.

Using these new dimensionless operators, we can express the Hamiltonian and
charge in eq.~\eqref{eq:omelette} as
\begin{align}\label{dimlessHamiltonian1}
  H
  = \sum_{\vec n} \omega_n \left(
  \lambda_n^{-1} \tilde{p}_{\vec n}^\dagger \tilde{p}_{\vec n}
  + \lambda_n \tilde{q}_{\vec n}^\dagger \tilde{q}_{\vec n}
  \right),
  \quad
  C
  = i\sum_{\vec n}(\tilde{q}_{\vec n}^\dagger \tilde{p}_{\vec n}^\dagger - \tilde{q}_{\vec n}\tilde{p}_{\vec n}).
\end{align}
For later reference, we denote this basis of complex operators by:
\begin{align}
  \xi_{\vec n}^{\Cbb} %_\Cbb
  =& \begin{bmatrix}
    \tilde q_{{\vec n}}, &
    \tilde q_{{\vec n}}^\dagger , &
    \tilde p_{{\vec n}}^\dagger, &
    \tilde p_{{\vec n}}&
  \end{bmatrix}^\T. %, \qquad s\in[L,R].
                         \label{eq:guitar}
\end{align}
Recall from eq.~\eqref{comrel:complex} that the unusual ordering of the
operators and their conjugates was chosen such that $[\xi_{\vec
  n}^{\Cbb},(\xi_{\vec n}^{\Cbb})^{\dagger}]=i\Omega$.

\subsection{Decoupling the Particles and Anti-Particles}\label{decpartapart}
So far, we have reframed the theory of the complex scalar field as a theory of
decoupled harmonic oscillators in the complex ``$\Cbb$''
% $LR_$
basis defined in eq.~\eqref{eq:guitar}. Since many states in this field theory
are defined in terms of eigenvalues of the Hamiltonian and charge, it will be
useful to perform an additional transformation to identify degrees of freedom
associated with positive and negative charges. This transformation decouples the contributions of particles and anti-particles to the Hamiltonian
and charge operators simultaneously. This transformation is given by the explicit expression
\begin{align}
  \begin{split}
    & \xi_{\vec n}^{\Rbb} \equiv R_{\Cbb\to \Rbb} \xi_{\vec n}^{\Cbb}, \qquad
    \xi_{\vec n}^{\Cbb} \equiv
    \begin{bmatrix}
      \tilde q_{{\vec n}} \\ \tilde q_{{\vec n}}^\dagger \\
      \tilde p_{{\vec n}}^\dagger \\ \tilde p_{{\vec n}}
    \end{bmatrix}, \qquad \xi_{\vec n}^\Rbb \equiv \begin{bmatrix}
      q_{{\vec n}} \\ \bar{q}_{{\vec n}} \\
      p_{{\vec n}} \\ \bar{p}_{{\vec n}}
    \end{bmatrix},
    \\
    & R_{\Cbb \to \Rbb} \equiv \frac{1}{2}\begin{bmatrix}
      1 & 1 & i \lambda_n^{-1} & -i \lambda_n^{-1} \\
      1 & 1 & -i \lambda_n^{-1} & i \lambda_n^{-1} \\
      -i \lambda_n & i \lambda_n & 1 & 1 \\
      i \lambda_n & -i \lambda_n & 1 & 1
    \end{bmatrix}.%, \qquad s\in [L,R].
    \label{eq:maccu}
  \end{split}
\end{align}
This transformation of phase space operators does not modify the commutation
relations \eqref{comrel:complex} since it satisfies the condition
\eqref{cond:trans}. It is also easy to check explicitly that this transformation
generates real operators $\xi_n^\Rbb= (\xi_n^{\Rbb})^{\dagger}$.\footnote{Here,
  when we write $\xi^\dagger$, we mean that we simply Hermitian conjugate the
  contents of $\xi$ without additionally transposing $\xi$ between being a row
  vector and a column vector.} By substituting this coordinate transformation
into eq.~\eqref{dimlessHamiltonian1}, we obtain
\begin{align}
  \begin{split}
    H =& \frac{1}{2} \sum_{\vec{n}} \omega_n \left[ \lambda_n^{-1} (p_{{\vec n}}^2 +
      \bar{p}_{{\vec n}}^2) + \lambda_n (q_{{\vec n}}^2 + \bar{q}_{{\vec n}}^2)
    \right],\\
    C =& \frac{1}{2}\sum_{\vec{n}} \left[ \lambda_n^{-1}(p_{{\vec n}}^2-\bar{p}_{{\vec
          n}}^2) +\lambda_n (q_{{\vec n}}^2-\bar{q}_{{\vec n}}^2) \right].
  \end{split}
         \label{eq:crescentina}
\end{align}
Here, we see that the oscillators remain decoupled in the expressions for the
Hamiltonian and the charge, where the oscillators $(q_{{\vec n}},p_{{\vec n}})$
have positive charge while $(\bar{q}_{{\vec n}},\bar{p}_{{\vec n}})$ have
negative charge, cf.~eq.~\eqref{eq:dragonFruit}.

In order to gain physical intuition for this decomposition, it is instructive to
consider the above transformations in terms of creation and annihilation
operators. Written in terms of creation and annihilation operators, the phase
space operators introduced previously read
\begin{align}\label{phasebarnotbar}
  q_{{\vec n}}
  =& \sqrt{\frac{1}{2\lambda_n}} (a_{{\vec n}} + a_{{\vec n}}^\dagger),
  &
    p_{{\vec n}}
    =& -i \sqrt{\frac{\lambda_n}{2}} (a_{{\vec n}}-a_{{\vec n}}^\dagger), \\
  \bar{q}_{-{\vec n}}
  =& \sqrt{\frac{1}{2\lambda_n}} (\bar{a}_{{\vec n}} + \bar{a}_{{\vec n}}^\dagger),
  &
    \bar{p}_{-{\vec n}}
    =& -i \sqrt{\frac{\lambda_n}{2}} (\bar{a}_{{\vec n}}-\bar{a}_{{\vec n}}^\dagger).
\end{align}
Thus, $(q_{{\vec n}},p_{{\vec n}})$ and $(\bar{q}_{{\vec n}},\bar{p}_{{\vec
    n}})$ correspond to the real phase space operators for particles and
anti-particles of the field theory, respectively. This explains the signs of the
charge contributions in \eqref{eq:crescentina} from $(q_{{\vec n}},p_{{\vec
    n}})$ and $(\bar{q}_{{\vec n}},\bar{p}_{{\vec n}})$. Furthermore, the
transformation in eq.~\eqref{eq:maccu} can be decomposed into a rotation of the
complex phase space operators to real phase space operators as in
\eqref{Strans0}, followed by a symplectic transformation which separates the
particles' and anti-particles' creation and annihilation operators.
The set of operators $\xi^{\Rbb}$ in eq.~\eqref{eq:maccu} will allow us to
directly utilize certain covariance matrices from \cite{TFD} which we review in
section \ref{sec:bagel} below, while the use of $\xi^{\Cbb}$ will allow us to
more naturally evaluate the covariance matrix for the reference state, see
eq.~\eqref{eq:cello} below. Finally, the transformations between these two bases
will allow us to freely move between the different bases as we evaluate the
complexity.

Before moving on, let us take a moment to compare the transformations introduced
in this section and those introduced for the uncharged TFD problem for a real
scalar field \cite{TFD}. Due to the lack of charge and the reality of their
field, the authors of \cite{TFD} were content to stop at
\eqref{dimlessHamiltonian1} after performing the Fourier transform ---
note that, modulo Hermitian conjugation, $H$ is decoupled in the tilde phase
space operators there.\footnote{Actually, the Hamiltonian of \cite{TFD} had a
  very similar form to \eqref{dimlessHamiltonian1}, but with $n$ and $-n$ modes
  mixed. This is because for the case of a real field we have $q_{\vec
    n}^\dagger=q_{-{\vec n}}$ and $p_{\vec n}^\dagger=p_{-{\vec n}}$. These can
  be decoupled as in appendix D of \cite{TFD} by performing a coordinate
  transformation of the type \eqref{Strans0}.} It is the second part of the
symplectic transformation \eqref{eq:maccu},
which splits the particles and anti-particles degrees of freedom, which was not
needed for the case of the uncharged TFD. We will find this transformation crucial later to the
decomposition of charged oscillator TFDs to uncharged TFDs.

\subsection{The Reference State}

The complexity problem involves a reference state, see
discussion around eq.~\eqref{cond}. The reference state serves as the starting
point for our quantum circuits and is usually chosen to be simple in the sense
that it is unentangled. It was proposed in \cite{QFT1,QFT2} that a natural
reference state for complexity in QFTs is the ground state of some Hamiltonian
\begin{align}
  % H_\rf
  % =& H_{R,\rf} + H_{L,\rf}; \qquad
       H_{\rf}
       \equiv \int d^{d-1} x\; \left(\pi^\dagger \pi + m_\rf^2 \phi^\dagger \phi\right),%, \quad s\in[L,R].
       \label{eq:mozzarella}
\end{align}
given in terms of a different frequency $m_{\rf}$, known as the {\it reference
  state scale}.\footnote{$m_\rf$ was denoted $\mu$ in \cite{TFD}, which we have
  changed in order to reserve $\mu$ for the chemical
  potential.}
Note that, due to the lack of the derivative term compared to
eq.~\eqref{eq:green}, this Hamiltonian only couples the various position and
momentum degrees of freedom -- $\phi(x),\pi(x)$ -- at the same spatial point $x$
rather than introducing a more complicated structure where the degrees of
freedom at different points are coupled to each other.
Hence, the vacuum state for the Hamiltonian \eqref{eq:mozzarella} is spatially
unentangled --- a desirable property for a `simple' reference state.

Note that, in the absence of a UV regulator, \eg a lattice, the energy of the reference state
of the QFT diverges in the UV (with respect to the physical Hamiltonian) and therefore
resides outside the `physical' Hilbert space of states whose energy is UV-finite.
The motivation behind this definition of the reference state in \cite{QFT1,QFT2}
stems from the fact that physical states in QFT have correlations down to
arbitrary short distance scales, and therefore it is expected that those states
are infinitely more complex than the unentangled reference state. It is
therefore sensible to have a reference state which is ``infinitely far'' from
our target states. A finite notion of complexity can only be defined in the
presence of a UV-regulator (\eg on the lattice). This is similar to what happens
for the entanglement entropy which is also a UV-divergent quantity in the continuum
QFT. Once a lattice regulator is introduced, our reference and target states
both live in the same Hilbert space. As discussed in the introduction, UV-divergences in the complexity can be regarded as a constant overhead which
cancels when comparing the complexity of different states in the physical Hilbert
space (\eg when evaluating the complexity of formation).

The Hamiltonian in eq.~\eqref{eq:mozzarella} can be discretized and expressed in
terms of the operators \eqref{eq:lemonade2} yielding
\begin{align}
  H_{\rf}
  =& \sum_{\vec n} m_\rf \left[
     \lambda_\rf^{-1} \tilde{p}_{\vec n}^\dagger \tilde{p}_{\vec n} + \lambda_\rf \tilde{q}_{\vec n}^\dagger \tilde{q}_{\vec n}
     \right] \label{eq:feta}
\end{align}
where above we have defined
\begin{align}
  \lambda_\rf
  \equiv& \frac{m_\rf}{\mu_g}.
          \label{eq:tedious}
\end{align}
Recall from the discussion around eq.~\eqref{eq:piano2} that the straight line
trajectory is optimal with respect to the $F_2$ norm and provides a good approximation for the optimal trajectory with respect to the $F_1$ norm when the reference state
scale and the gate scale are equal, \ie when $\lambda_\rf=1$ \cite{TFD}.
Applying the extra coordinate transformation \eqref{eq:maccu} decoupling the
particles and anti-particles unfortunately does not preserve the simple form of
the reference Hamiltonian but rather yields:
\begin{align}
  \begin{split}
    \label{eq:cheddar}
    H_{\rf} =& \sum_{\vec n} m_\rf \left[ \lambda_\rf^{-1} \tilde{p}_{{\vec
          n}}^\dagger \tilde{p}_{s,{\vec n}} + \lambda_\rf \tilde{q}_{{\vec
          n}}^\dagger \tilde{q}_{{\vec n}} \right]\\ %\label{eq:feta2}\\
    =& \frac{1}{4}\sum_{\vec n} m_\rf \Bigg\{ \lambda_\rf^{-1} (p_{{\vec n}} +
    \bar{p}_{{\vec n}})^2
    + \frac{\lambda_\rf}{\lambda_n^2} (p_{{\vec n}}-\bar{p}_{{\vec n}})^2 \\
    &~~~~~~~~~~~~~~~~+ \lambda_\rf (q_{{\vec n}}+\bar{q}_{{\vec n}})^2 +
    \frac{\lambda_n^2}{\lambda_\rf} (q_{{\vec n}}-\bar{q}_{{\vec n}})^2 \Bigg\}.
  \end{split}
\end{align}
We will therefore be mostly using the complex coordinates $\xi^\Cbb$ when
discussing the reference state.

%%% Local Variables:
%%% mode: latex
%%% TeX-master: "../ChargedComplexityV7"
%%% End:

\section{Complexity of the cTFD State}\label{sec:George}

Having separately discussed complexity and the complex scalar field theory in
sections \ref{sec:covprel} and \ref{sec:praline}, respectively, in this section,
we combine the machinery developed thus far, in order to compute the complexity
of the cTFD state of two copies of a complex scalar field theory.
We begin in subsection \ref{sec:bagel} with a review of the properties of the
uncharged TFD state of two harmonic oscillators \cite{TFD}, specifically noting
the expression for its covariance matrix. In subsection \ref{sec:argh}, we next
consider the cTFD state of two complex harmonic oscillators, finding that
this state factorizes to two uncharged TFDs. The upshot is that we may reuse the
covariance matrix of the uncharged TFD for the cTFD, by merely shifting
the temperatures and times according to the chemical potential --- we explain
this in subsection \ref{sec:George2}.\footnote{Note however that while we
    are able to adapt the TFD covariance matrices from \cite{QFT1} to the case
    of the cTFD by using the particle anti-particle factorization, the result
    for the complexity cannot be adapted directly from the uncharged case due to
    the non-trivial structure of the reference state of the charged system in
    the particle anti-particle basis, cf. eq.~\eqref{eq:cheddar}.\label{foot:complexityNotTrivial}} In
subsection \ref{sec:paroxetine}, we use the decomposition of the complex scalar
field into harmonic oscillators, as discussed in subsections
\ref{sec:woof}-\ref{decpartapart}, to argue that the complexity of the complex
scalar cTFD is simply the harmonic oscillator answer summed over the modes of
the scalar. Finally, in subsections \ref{sec:ouch} and \ref{sec:byeah}, we use
these results to compute the complexity of formation and the time evolution of
complexity for the complex scalar, respectively.

\subsection{Properties of the Uncharged Thermofield Double State}
\label{sec:bagel}
Before we consider the cTFD state, let us briefly summarize some
useful results from \cite{TFD} about the uncharged TFD state of two harmonic
oscillators. The state is a purification of a mixed thermal state on a system
with two copies, just like eq.~\eqref{eq:intro:cTFD}, but without chemical
potential and charge. More explicitly, it is defined as
\begin{equation}\label{NTFD}
  |\text{TFD}(t_L,t_R)\rangle = \frac{1}{\sqrt{Z_\beta}} \sum_n e^{-\beta E_n/2} e^{-i E_n (t_L+t_R)} |E_n\rangle_L |E_n\rangle_R,
\end{equation}
where $t_L$ and $t_R$ are the times on the two identical copies, $\beta$ is the
inverse temperature and $|E_n\rangle_L$ and $|E_n\rangle_R$ are energy
eigenstates of the left and right harmonic oscillators respectively, with energy
eigenvalues $E_n$.\footnote{Here again, we follow the common nomenclature of
  referring to the two harmonic oscillators as the ``left'' and ``right''
  copies.} We will consider a single mode/oscillator in the left system with
(dimensionless) position and momentum denoted $(q_L,p_L)$ and a single harmonic
oscillator in the right system with position and momentum denoted $(q_R,p_R)$,
both taken to have the same frequency $\omega$.
The Hamiltonian for this system is given by
\begin{equation}\label{oboe}
  H=\sum_{s\in[L,R]} \frac{\omega}{2} \left( \lambda^{-1} \, p_s^2  + \lambda \, q_s^2 \right)
  = \sum_{s\in[L,R]}\omega \left(a_s^\dagger a_s+\frac{1}{2}\right)
\end{equation}
where $\lambda$ is a parameter encoding the mass of the oscillators and
where here the phase space operators are {\it real} and are related to the
creation and annihilation operators according to
\begin{equation}
  q_s=\frac{1}{\sqrt{2\lambda}}(a_s^\dagger+a_s), \quad p_s = i \sqrt{\frac{\lambda}{2}}(a_s^\dagger-a_s), \qquad s\in [L,R].
\end{equation}
From now on, we will use a subscript $s\in [L,R]$ on our variables, indicating
which copy we are referring to. Note that eq.~\eqref{NTFD} depends on the
combination of times $t_L+t_R$ and so the full time dependence can be captured
by setting, \eg $t_L=t_R=t/2$ as we do in the following. The time dependent TFD
state of the two harmonic oscillators is given by the following explicit
expression
\begin{equation}\label{neutralTFD}
  \begin{split}
    |\text{TFD}(t)\rangle = &Z_{\beta}^{-1/2} \sum_{n=0}^\infty e^{-\frac{\beta\omega}{2}(n+\frac{1}{2})}e^{-i\omega (n+\frac{1}{2})t} |n\rangle_L |n\rangle_R\\
    = &Z_{\beta}^{-1/2} e^{-\frac{\beta\omega}{4}} \,e^{-\frac{i}{2}\omega t}
    \sum_{n=0}^\infty \exp \left[e^{-\beta\omega/2}e^{-i\omega t}a_L^\dagger
      a_R^\dagger\right] |0\rangle_L |0\rangle_R
  \end{split}
\end{equation}
where the normalization factor is defined as $Z_{\beta}\equiv
e^{-\beta\omega/2}\, (1-e^{-\beta \omega})^{-1}$.

The TFD state takes a simpler form in the ``diagonal'' $\pm$ basis mentioned
below eq.~\eqref{eq:piano2}, mixing the two sides, defined according to
\begin{align}
  q_\pm
  \equiv& \frac{1}{\sqrt{2}}(q_L\pm q_R),
  &
    p_\pm
    \equiv& \frac{1}{\sqrt{2}} (p_L\pm p_R) \;,
            \label{eq:harpsichord}
\end{align}
\ie{}
\begin{align}\label{Strans}
  \begin{bmatrix}
    q_+ \\
    q_-
  \end{bmatrix}=& R_{LR\to \pm}
                  \begin{bmatrix}
                    q_L \\
                    q_R
                  \end{bmatrix}, &
                                   \begin{bmatrix}
                                     p_+ \\
                                     p_-
                                   \end{bmatrix}=& R_{LR\to \pm}
                                                   \begin{bmatrix}
                                                     p_L \\
                                                     p_R
                                                   \end{bmatrix}, & R_{LR\to
                                                     \pm}\equiv& \frac{1}{\sqrt
                                                     2}\begin{bmatrix}
                                                     1 & 1 \\
                                                     1& -1
                                                   \end{bmatrix}.
\end{align}
In this basis, the TFD state can be written as follows (see eqs.~(35), (36) and (77) in \cite{TFD})
\begin{align}
  \begin{split}
    & |\TFD(t)\rangle = e^{-i\alpha \hat{\mathcal{O}}_+(t)} |0\rangle_+
    \otimes e^{i\alpha \hat{\mathcal{O}}_-(t)} |0\rangle_- \\
    & \hat{\mathcal{O}}_\pm(t) \equiv \frac{1}{2}\cos(\omega t) (q_\pm p_\pm +
    p_\pm q_\pm) + \frac{1}{2}\sin(\omega t) (\lambda q_\pm^2-\lambda^{-1}
    p_\pm^2)
  \end{split}
\end{align}
where we have defined
\begin{equation}
  \alpha             \label{eq:bass}
  \equiv
  \frac{1}{2}\log\left(
    \frac{1+e^{-\beta\omega/2}}{1-e^{-\beta\omega/2}}
  \right).
\end{equation}
The covariance matrix of the TFD state in the $\pm$ basis is given by eq.~(76)
of \cite{TFD}, \ie
\begin{equation}
  G_\TFD^\pm(t,\alpha)
  \equiv \begin{bmatrix}
    \lambda^{-1}[\cosh(2\alpha)\pm\sinh(2\alpha)\cos(\omega t)]
    & \mp\sinh(2\alpha)\sin(\omega t)
    \\
    \mp\sinh(2\alpha) \sin(\omega t)
    & \lambda[\cosh(2\alpha)\mp\sinh(2\alpha)\cos(\omega t)]
  \end{bmatrix}.
  \label{eq:viola}
\end{equation}
This TFD state consists of a single Harmonic oscillator of frequency $\omega$.

When studying the uncharged TFD state of a real scalar QFT, \cite{TFD} have
shown that the problem factorizes to evaluating the complexity of a product of
different one-mode TFD states, each with a different frequency
$\omega_k=\sqrt{k^2+m^2}$ where $k$ is the spatial momentum of the different
modes and $m$ is the QFT mass.

\subsection{cTFD of Two Complex Harmonic Oscillators}
\label{sec:argh}
Next, we consider the charged thermofield double consisting of two complex or
four real harmonic oscillators. We will label each oscillator as right or left
($R$ or $L$) and particle or anti-particle (no overbar or overbar). The complete
Hamiltonian is given by
\begin{align}
  H
  =& H_R + H_L; \qquad
     H_s
     =\frac{\omega}{2} \left[\lambda^{-1}(p_s^2+\bar{p}_s^2) + \lambda (q_s^2+\bar{q}_s^2)\right],\quad s\in[L,R].
     \label{eq:coffee}
\end{align}
This is to be representative of a single mode in eq.~\eqref{eq:crescentina}.
Alternatively, in terms of complex phase space operators we have $H_s = \omega
\left(\lambda^{-1} \tilde{p}_{s}^\dagger \tilde{p}_{s} + \lambda
  \tilde{q}_{s}^\dagger \tilde{q}_{s} \right)$. We may use similar creation and
annihilation operators to those in eq.~\eqref{phasebarnotbar} to expand the
Hamiltonian and charge, \ie{} we take\footnote{Here $a_s$ is identified with
  $a_{s,\vec n}$ while $\bar a_s$ is identified with $\bar a_{s,-\vec n}$ from
  the previous discussion around eq.~\eqref{phasebarnotbar}.}
\begin{align}\label{phasebarnotbar2}
  q_{s}
  =& \sqrt{\frac{1}{2\lambda}} (a_{s} + a_{s}^\dagger),
  &
    p_{s}
    =& -i \sqrt{\frac{\lambda}{2}} (a_{s}-a_{s}^\dagger), \\
  \bar{q}_{s}
  =& \sqrt{\frac{1}{2\lambda}} (\bar{a}_{s} + \bar{a}_{s}^\dagger),
  &
    \bar{p}_{s}
    =& -i \sqrt{\frac{\lambda}{2}} (\bar{a}_{s}-\bar{a}_{s}^\dagger),
\end{align}
and define the number operators as in \eqref{numberops1}
\begin{equation}
  N_{s} \equiv a_{s}^\dagger a_{s},\qquad
  \bar{N}_{s} \equiv \bar{a}_{s}^\dagger \bar{a}_{s}.
\end{equation}
In terms of creation $a_s^\dagger,\bar{a}_s^\dagger$, annihilation
$a_s,\bar{a}_s$, and number $N_s,\bar{N}_s$, operators, we
have
\begin{align}
  H_s
  =& \omega\left(N_s +\bar{N}_s+ 1\right).  \label{eq:cabbage}
\end{align}
Similarly, the total charge of a given side is given by
\begin{align}
  C_s
  =& \frac{1}{2}\left[\lambda^{-1}(q_s^2-\bar{q}_s^2) + \lambda(p_s^2-\bar{p}_s^2)\right]
     = N_s-\bar{N}_s. \label{eq:pineapple}
\end{align}
Alternatively, in terms of complex phase space operators we have $C_s =
i(\tilde{q}_{s}^\dagger \tilde{p}_{s}^\dagger - \tilde{q}_{s}\tilde{p}_{s})$. We
denote eigenstates of the number operators by $|n,\bar{n}\rangle_s$, satisfying
\begin{align}
  N_s|n,\bar{n}\rangle_s
  =& n|n,\bar{n}\rangle_s,
  &
    \bar{N}_s|n,\bar{n}\rangle_s
    =& \bar{n}|n,\bar{n}\rangle_s.
\end{align}
The creation and annihilation operators raise and lower number eigenvalues
according to
\begin{align}
  |n,\bar{n}\rangle_s
  =& \frac{(a_s^\dagger)^n (\bar{a}_s^\dagger)^{\bar{n}}}{\sqrt{n! \, \bar{n}!}} |0,0\rangle_s,
     \label{eq:muffin}
\end{align}
which have energy and charge eigenvalues given by
\begin{equation}
  \begin{split}
    &H_s |n,\bar{n}\rangle_s= E_{n,\bar n}|n,\bar{n}\rangle_s,  \qquad E_{n,\bar n} = \omega (n+\bar n+1),\\
    &C_s |n,\bar{n}\rangle_s= c_{n,\bar n}|n,\bar{n}\rangle_s, \qquad ~~
    c_{n,\bar n} = n-\bar n.
  \end{split}
  \label{eq:dufus}
\end{equation}

The cTFD state, is defined in general by eq.~\eqref{eq:intro:cTFD}. As in the
uncharged case of subsection \ref{sec:bagel}, let us again set $t_L=t_R=t/2$.
Then, specializing eq.~\eqref{eq:intro:cTFD} to the present theory of harmonic
oscillators, with energy and charge eigenstates and eigenvalues given in
eq.~\eqref{eq:dufus}, we find, in analogy to eq.~\eqref{neutralTFD},
\begin{align} \label{eq:chocolate}
  \begin{split}
    & |\cTFD(\beta,\mu;t,\omega)\rangle = Z_{\beta,\mu}^{-1/2}
    \sum_{n,\bar{n}=0}^\infty \exp\left\{ -\left(\frac{\beta}{2} + it\right)
      \left(E_{n,\bar{n}} +\mu c_{n,\bar{n}}\right) \right\}
    |n,\bar{n}\rangle_L |\bar{n},n\rangle_R \\
    &~~= Z_{\beta,\mu}^{-1/2} \, e^{-\omega\left(\frac{\beta}{2}+it\right)}
    \sum_{n,\bar{n}=0}^\infty \exp\left\{ -\left(\frac{\beta}{2} + it\right)
      \left[\omega(n+\bar{n}) +\mu (n-\bar{n})\right] \right\}
    |n,\bar{n}\rangle_L |\bar{n},n\rangle_R.
  \end{split}
\end{align}
Note the ordering of $\bar{n},n$ in writing $|\bar{n},n\rangle_R$; by this, we
mean the $\bar{n}$-th eigenstate of $N_R$ and the $n$-th eigenstate of
$\bar{N}_R$. This was done in order to recover the structure in
eq.~\eqref{eq:intro:cTFD} with opposite charges on
the left and right sides. The normalization constant $Z_{\beta,\mu}$ is given by
\begin{align}
  \begin{split}
    Z_{\beta,\mu} =& e^{-\beta\omega} \sum_{n,\bar{n}=0}^\infty \exp\left\{
      -\beta\left[ \omega(n+\bar{n}) +\mu (n-\bar{n}) \right]
    \right\} \\
    =& e^{-\beta\omega} \left[1-e^{-\beta(\omega-\mu)}\right]^{-1}
    \left[1-e^{-\beta(\omega+\mu)}\right]^{-1} \\
    =& e^{-\beta\omega}
    \left[1+e^{-2\beta\omega}-2e^{-\beta\omega}\cosh(\beta\mu)\right]^{-1}.
  \end{split}
       \label{eq:oreos}
\end{align}
Comparison of \eqref{eq:chocolate} with \eqref{neutralTFD} shows that this
cTFD is just a product of two uncharged TFD states
\begin{align}
  |\cTFD(\beta,\mu;t,\omega)\rangle
  =& |\TFD(\beta_\lR,\mu;t_\lR)\rangle_\lR
     \otimes |\TFD(\beta_\Lr,\mu;t_\Lr)\rangle_\Lr \label{eq:waffle}
\end{align}
at temperatures and times shifted by the chemical potential\footnote{Note that
  this cannot be rephrased purely as a shift in the frequencies $\omega_\Lr
  \equiv \omega+\mu$, $\omega_\lR \equiv \omega-\mu$ since the states $|n,\bar
  n\rangle_s$ used to construct the cTFD in eq.~\eqref{eq:chocolate} are created
  from the vacuum state of each side $|0,\bar 0\rangle$ with the creation
  operators defined with respect to the Hamiltonian containing the original
  frequency of the theory. Therefore, in \eqref{eq:viola}, it is the case that,
  while $\alpha$ becomes shifted due to the modified temperatures in
  eq.~\eqref{eq:peppermint}-\eqref{eq:earlgrey}, the parameter $\lambda$ is
  defined with the frequency of the original theory rather than with the
  shifted frequencies.}
\begin{align}
  \beta_\lR
  \equiv& \beta\left(1-\frac{\mu}{\omega}\right),
  &
    t_\lR
    \equiv& t\left(1-\frac{\mu}{\omega}\right),
            \label{eq:peppermint}
  \\
  \beta_\Lr
  \equiv& \beta\left(1+\frac{\mu}{\omega}\right),
  &
    t_\Lr
    \equiv& t\left(1+\frac{\mu}{\omega}\right). \label{eq:earlgrey}
\end{align}
Thus, we see that we can use the covariance matrices
of the uncharged TFD from reference \cite{TFD}, to study
the complexity of the cTFD.

From eqs.~\eqref{eq:peppermint} and \eqref{eq:earlgrey}, we already see
something interesting. As $|\mu|\to \omega$, one of the effective temperatures
$\beta_{\lR}^{-1},\beta_{\Lr}^{-1}$ blows up; further for $|\mu|>\omega$, we have an ill-defined
negative-effective-temperature state in either $\lR$ or $\Lr$. In particular,
notice that, for $|\mu|>\omega$, the cTFD state \eqref{eq:chocolate} becomes
ill-defined since the normalization factor \eqref{eq:oreos} diverges. This will force
us to take all frequencies in our field theory construction, see
eq.~\eqref{eq:omelette}, to be at least as large as the chemical potential and,
in particular, this implies that in our field theory setup we must have $m\ge
|\mu|$.\footnote{The case $m=|\mu|$ is rather singular, however as we explained in
  footnote \ref{foot:blahblahblah} and revisit below in footnote \ref{foot:woopwoopwoop}, this limit can be taken smoothly in
  the field theory in sufficiently high dimensions.} This means that
we will not be able to reach the conformal limit for fixed $\mu\ne 0$ in this
work. Moreover, note that in general replacing $\mu\leftrightarrow-\mu$ is
equivalent to swapping $\lR\leftrightarrow\Lr$, under which the
complexity is invariant.

\subsection{Complexity of the cTFD of Two Complex Harmonic Oscillators}
\label{sec:George2}
We now have the necessary ingredients to compute the complexity of the charged
thermofield double state \eqref{eq:chocolate} of the two complex harmonic
oscillator system. Recall from the paragraph above eq.~\eqref{eq:Topaz} that the
complexity depends on the choice of basis, and that it was found in \cite{TFD}
that the $F_1$ cost function with a choice of basis which does not mix the left
and right degrees of freedom was the one which reproduced best a number of
qualitative features of complexity in holography. We have chosen to focus on a similar choice of bases below. We will
consider two different bases, the complex left-right basis $LR_\Cbb$
corresponding to operators
\begin{equation}\label{eq:defLRCbasis}
  \xi^{LR_\Cbb} \equiv
  \begin{bmatrix}
    \tilde q_{L}, & \tilde q_{L}^\dagger, & \tilde q_{R}, & \tilde
    q_{R}^\dagger, & \tilde p_{L}^\dagger, & \tilde p_{L}, & \tilde
    p_{R}^\dagger, & \tilde p_{R}
  \end{bmatrix},
\end{equation}
and the real left-right basis $LR$ corresponding to the particle and
anti-particle degrees of freedom
\begin{equation}\label{eq:defLRbasis}
  \xi^{LR}  \equiv \begin{bmatrix}
    q_{L}, & \bar{q}_{L}, & q_{R}, & \bar{q}_{R}, &
    p_{L}, & \bar{p}_{L}, &p_{R}, & \bar{p}_{R}
  \end{bmatrix},
\end{equation}
where $\xi^{LR_\Cbb}$ is related to $\xi^{LR}$ by
eq.~\eqref{eq:maccu}.\footnote{Of course, here we mean that the operators are
  related by a matrix constructed as a direct sum of \eqref{eq:maccu} for the
  left and right copies.} (To reuse the results of subsection \ref{sec:bagel},
we will also need to relate the above to a ``diagonal'' $\pm$ basis
(cf. eq.~\eqref{eq:harpsichord}), as we do below; however, we will always
transform back to the $\xi^{LR_\Cbb}$ or $\xi^{LR}$ bases to compute
complexity.)

For the reference state, we take the vacuum of a single mode of the complex
scalar reference Hamiltonian \eqref{eq:feta}:
\begin{align}
  H_{R,\rf}
  =& m_\rf \left(
     \lambda_\rf^{-1} \tilde{p}_R^\dagger \tilde{p}_R + \lambda_\rf \tilde{q}_R^\dagger \tilde{q}_R
     \right).
     \label{eq:violin}
\end{align}
The covariance matrix of the reference state in the $LR_\Cbb$ basis can be
evaluated directly from the definition \eqref{covcov} by using creation and
annihilation operators adapted to the reference state as in
eq.~\eqref{eq:lemonade2} with the replacement $\lambda_n\rightarrow \lambda_\rf$
which yields
\begin{align}
  G_\rf^{L_\Cbb}
  =& G_\rf^{R_\Cbb}
     = \diag(\lambda_\rf^{-1},\lambda_\rf^{-1},\lambda_\rf,\lambda_\rf),
     \label{eq:cello}
\end{align}
where we have used the superscripts $L_\Cbb$ and $R_\Cbb$ to denote the complex
operators for the left and right copies (according to the order in
eq.~\eqref{eq:maccu}), respectively. The full covariance matrix for the
reference state will simply be the direct sum of those two 4 by 4 matrices
according to the order of operators in eq.~\eqref{eq:defLRCbasis}.

\sloppy Using the decomposition \eqref{eq:waffle} of the cTFD into
uncharged TFDs, we can deduce the (target) covariance
matrices of the cTFD. It is helpful here to define the symbol $\Oplus$ to mean direct sum,
followed by reordering rows and columns so that positions are listed before
momenta.
With this, the covariance matrix of the cTFD is given by
\begin{align}
  G_\TFD^+(t_{\lR,n},\alpha_{\lR,n})
  \Oplus
  G_\TFD^+(t_{\Lr,n},\alpha_{\Lr,n})
  \Oplus G_\TFD^-(t_{\lR,n},\alpha_{\lR,n})
  \Oplus
  G_\TFD^-(t_{\Lr,n},\alpha_{\Lr,n})
  \label{eq:tiredAndAlone}
\end{align}
where $G_\TFD^\pm$ is the uncharged TFD
covariance matrix \eqref{eq:viola}, with the shifted times
$t_{\lR,n},t_{\Lr,n}$, and $\alpha_\lR,\alpha_\Lr$ from eq.~\eqref{eq:bass}
defined in terms of the shifted temperatures $\beta_\lR,\beta_\Lr$ given in
eqs.~\eqref{eq:peppermint}-\eqref{eq:earlgrey}. The covariance matrix obtained in this way will be
given in the following basis of operators
\begin{equation}\label{xipm}
  \xi^{\pm}  \equiv \begin{bmatrix}
    q_{\bar L R}^+, q_{L \bar R}^+, q_{\bar L R}^-, q_{L \bar R}^-,
    p_{\bar L R}^+, p_{L \bar R}^+, p_{\bar L R}^-, p_{L \bar R}^-
  \end{bmatrix},
\end{equation}
obtained from eq.~\eqref{eq:defLRbasis} by the equivalent change of coordinates
to the one in eqs.~\eqref{eq:harpsichord}-\eqref{Strans}, which mixes the $\bar
LR$ operators and the $\bar RL$ operators separately, \ie
\begin{align}
  \begin{split}\label{LRbarRLbarpm}
    q^\pm_{\bar L R} \equiv \frac{1}{\sqrt{2}}(\bar q_{L}\pm q_R),\qquad
    p^\pm_{\bar L R}
    \equiv \frac{1}{\sqrt{2}} (\bar p_{L}\pm p_R),\\
    q^\pm_{L \bar R} \equiv \frac{1}{\sqrt{2}}(q_{L}\pm\bar q_R),\qquad p^\pm_{L
      \bar R} \equiv \frac{1}{\sqrt{2}} (p_{L}\pm \bar p_{R}).
  \end{split}
\end{align}

\fussy

The relevant circuits will consist of $8\times 8$ matrices acting on covariance
matrices according to \eqref{circuitgens}-\eqref{cond} where $S(\sigma)\in
\Sp(8,\Rbb)$. In particular, we shall consider the straight line circuit
described around eqs.~\eqref{eq:piano}-\eqref{eq:piano2}. We can compute the relative covariance matrix
\eqref{eq:piano} for the cTFD state in, say the $LR$ basis, as follows
\begin{align}
  \begin{split}
    \MoveEqLeft[0.5] \Delta_\cTFD^{LR}
    =\\
    & R_{\pm\to LR} \left[ G_\TFD^+(t_{\lR},\alpha_{\lR}) \Oplus
      G_\TFD^+(t_{\Lr},\alpha_{\Lr}) \Oplus G_\TFD^-(t_{\lR},\alpha_{\lR})
      \tilde{\oplus} G_\TFD^-(t_{\Lr},\alpha_{\Lr}) \right]
    R_{\pm\to LR}^\dagger \\
    &\cdot \left[(R_{L_\Cbb\to L} G_\rf^{L_\Cbb} R_{L_\Cbb \to L}^\dagger)\Oplus
      (R_{R_\Cbb\to R} G_\rf^{R_\Cbb} R_{R_\Cbb \to R}^\dagger)\right]^{-1},
  \end{split}
      \label{eq:Rose}
\end{align}
where
\begin{align}
  \xi^{LR}=&R_{\pm\to LR}\, \xi^{\pm} , & R_{\pm\to LR}
  % =& \diag(R_4,R_4),
       =& R_4 \oplus R_4,
  &
    R_4
    =& \frac{1}{\sqrt{2}} \begin{bmatrix}
      0 & 1 & 0 & 1 \\
      1 & 0 & 1 & 0 \\
      1 & 0 & -1 & 0 \\
      0 & 1 & 0 & -1
    \end{bmatrix} \in \SO(4)
                  \label{eq:Oliver}
\end{align}
inverts the transformation \eqref{LRbarRLbarpm} and $R_{L_\Cbb\to L} = R_{R_\Cbb\to R}$ is given
in eq.~\eqref{eq:maccu} applied to the left and right operators,
respectively. We can easily transform $\Delta^{LR}_\cTFD$ to another basis, say the
$LR_\Cbb$ basis as follows
\begin{align}
  \Delta_\cTFD^{LR_\Cbb}
  =& (R_{L\to L_\Cbb} \Oplus R_{R\to R_\Cbb}) \Delta_\cTFD^{LR} (R_{L\to L_\Cbb} \Oplus R_{R\to R_\Cbb})^{-1},
     \label{eq:Felix}
\end{align}
where $R_{L \to L_\Cbb} = R_{R\to R_\Cbb}$ is given by inverting eq.~\eqref{eq:maccu}.
It only remains to compute (the upper bound for) the $F_1$ complexity via
eq.~\eqref{eq:ciaccino} with an appropriate basis $K_I$ for $\ssp(8)$, see
appendix \ref{sec:deadline}. We reiterate this equation here for convenience
\begin{align}\label{finalfinal}
  \cmplx_1^\UB = \frac{1}{4} \sum_I \left|\tr \left( \log(\Delta_{\cTFD}) \cdot K_I^\dagger\right)\right|,
\end{align}
where $\UB$ stands for ``upper bound''.

We have not written the explicit expression here, since this expression
arising from the product of the $8\times 8$ matrices above is cumbersome and not particularly illuminating, but it is easily obtained by
substituting into eq.~\eqref{finalfinal} the generators in the complex basis
given in appendix \ref{sec:deadline} and the relative covariance matrix
\eqref{eq:Felix} given explicitly in terms of eqs.~\eqref{eq:maccu},
\eqref{eq:bass}-\eqref{eq:viola}, \eqref{eq:peppermint}-\eqref{eq:earlgrey}, \eqref{eq:cello} and
\eqref{eq:Rose}-\eqref{eq:Oliver}. We will study
various properties of this (upper bound on the) complexity analytically in
certain limits as well as numerically.

\subsection{cTFD States of the Complex Scalar}
\label{sec:paroxetine}
The cTFD state for the complex scalar theory is defined according to
eq.~\eqref{eq:intro:cTFD}. As in section \ref{sec:bagel} we will set
$t_L=t_R=t/2$ (but the full time dependence can easily be recovered, given that
the time dependence is simply dictated by the sum of the two times). The
relevant state for our construction is therefore given by
\begin{equation}\label{TheTFD2}
  |\cTFD(\beta,\mu;t)\rangle \equiv \frac{1}{\sqrt{Z_{\beta,\mu}}} \, \sum_{n,\sigma} e^{-\frac{\beta}{2} (E_n+\mu c_\sigma)-i (E_n+\mu c_\sigma) t } |E_n,c_\sigma\rangle_L |E_n,-c_\sigma\rangle_R,
\end{equation}
with $|E_n,c_\sigma\rangle$ denoting the Hamiltonian and charge eigenstates with
eigenvalues $E_n,c_\sigma$, respectively. The time evolution in
eq.~\eqref{TheTFD2} is motivated in appendix \ref{app:timeevol}. The goal of this section is to
calculate the complexity of this state for free complex scalar field theory.

The decomposition \eqref{eq:omelette} together with the commutation relations
\eqref{comrelnk} teach us that the cTFD state in the complex scalar field
theory factorizes into cTFD states of each of the momentum modes, \ie
\begin{align}
  |\cTFD(\beta,\mu;t)\rangle
  =& \bigotimes_{\vec n} |\cTFD_n(\beta,\mu; t,\omega_n)\rangle,
     \label{eq:vodka}
\end{align}
where the frequencies $\omega_n$ were defined in eq.~\eqref{eq:omelette}.
 Due to the mode factorization
\eqref{eq:vodka} of the complex scalar cTFD, the circuit constructing it  will similarly factor into circuits which act separately
on each mode $\vec n$. As described in sections
\ref{sec:woof}-\ref{decpartapart}, each of the momentum modes of the complex
scalar is equivalent to a complex harmonic oscillator. Therefore, we may compute
the complexity of the complex scalar cTFD by summing the complexities of
harmonic oscillator cTFDs --- the calculation of which was described in section
\ref{sec:George2} --- according to
\begin{align}
  \cmplx\left(\,|\cTFD(\beta,\mu;t)\rangle\,\right)
  =& \sum_{\vec n}
     \cmplx\left(\,|\cTFD_n(\beta,\mu;t,\omega_n)\rangle\,\right).
     \label{eq:peanut}
\end{align}

Eq.~\eqref{eq:peanut}, with the sum over discrete and finitely-many $\vec n$-s,
gives the complexity for the discretized and compactified theory of the complex
scalar, see eq.~\eqref{eq:cucumber} and the text above it. Recovering the continuum limit and decompactified space amounts to
taking $\delta \to 0$ and $L\to \infty$ respectively. The latter is implemented
by replacing sums with integrals:\footnote{There appears to be a typo in (188)
  of \cite{TFD} in that the RHS is missing the prefactor we have in
  \eqref{eq:flambe}.}
\begin{align}
  \sum_{\vec n}
  \to& \int_{-\tilde{N}}^{\tilde{N}} d^{d-1}n\;
       = \left(\frac{L}{2\pi}\right)^{d-1} \int_{-\pi/\delta}^{\pi/\delta} d^{d-1}k\;
       \label{eq:flambe}
\end{align}
where we have switched from the discrete label $\vec n$ to the continuous label $\vec k$
\begin{align}
  \vec k\equiv \frac{2\pi \vec n}{L},
\end{align}
cf.~footnote \ref{bigfoot11}, and $\tilde N$ was defined in
eq.~\eqref{eq:cucumber}. In this construction we also have to replace the
frequency by the continuous frequency $\omega_k=\sqrt{k^2+m^2}$ (see the
discussion in section 5.1 of \cite{TFD} for more details).
Similarly, the parameter $\lambda_n$ in eq.~\eqref{gatescale1} will be replaced with
\begin{equation}\label{lambdak}
\lambda_k \equiv \frac{\omega_k}{\mu_g} = \frac{\sqrt{k^2+m^2}}{\mu_g}.
\end{equation}
Furthermore, taking the continuum limit amounts to extending the domain of integration in
eq.~\eqref{eq:flambe} to be the full $\Rbb^{d-1}$. As already mentioned in the
introduction, the complexity in QFT has UV divergences and needs to be
regularized. This was the reason why we regulated the theory on the lattice in
the first place. The divergences of the complexity for the cTFD state will be
the same as those for a product state constructed from two copies of the vacuum
state since the UV structure is similar in these two states. When working in the
continuum limit, without a lattice, it is possible to regulate the UV
divergences by introducing a momentum cutoff $|\vec k|<\Lambda$ as the limit of
integration in eq.~\eqref{eq:flambe}.
For our calculations however, we will focus mostly on differences of
complexities which yield finite quantities and therefore we will be able to
simply take $\Lambda\rightarrow \infty$.

\subsection{Complexity of Formation}
\label{sec:ouch}
Following along the lines of section 5.3 of \cite{TFD}, we will now investigate
the complexity of formation \cite{Formation} of the complex scalar field in the
charged thermofield double state
\begin{align}
  \Delta\cmplx_\cTFD(t=0)
  \equiv& \, \cmplx_\cTFD(t=0)- \, \cmplx_{\vac(L\otimes R)} \;,
          \label{eq:Snowy}
\end{align}
where $\cmplx_{\vac(L\otimes R)}$ is the complexity of two copies of the vacuum
state of the same complex scalar field theory. This UV-finite quantity compares
how much harder it is to prepare the cTFD state at $t=0$ compared to preparing
two copies of the vacuum state in the same complex scalar theory. We will
evaluate this complexity difference using equation \eqref{finalfinal} which
provides an upper bound for the $F_1$ complexity. However, similarly to what was
done in \cite{TFD}, we will everywhere assume that the values of
  $\lambda_\rf$ are such that the bound \eqref{finalfinal} is approximately
  saturated (see discussion below eqs.~\eqref{eq:piano}-\eqref{eq:piano2}), in
  particular, in all our numerical analysis below, we select
  $\lambda_\rf=1$.\footnote{We keep however the general $\lambda_\rf$
    dependence in some of the analytic expressions below, since \cite{TFD} have
    shown that there exist some other values of $\lambda_\rf$ for which the
    straight line circuit provides a good approximation of the optimal circuit.
    A full analysis of the range of $\lambda_\rf\neq 1$ for which the straight
    line circuit provides a good approximation for the optimal circuit is beyond
    the scope of this work.}

By setting $t=0$ to compute the complexity of formation, it is possible to
analytically diagonalize the relative covariance matrix \eqref{eq:Rose}, or
\eqref{eq:Felix}, of each mode. We begin by noting that at $t=0$,
\eqref{eq:viola} simplifies significantly:
\begin{align}
  G_{\TFD,k}^\pm(t=0,\alpha)
  =& \diag(\lambda_k^{-1}e^{\pm 2\alpha},\lambda_k e^{\mp 2\alpha})\,,
     \label{eq:pgsd}
\end{align}
where we have appended subscripts $k$, indicating that \eqref{eq:pgsd}
  gives the covariance matrix of the $k$-th momentum mode, and $\lambda_k$ is
  given in eq.~\eqref{lambdak}. Further, transforming the covariance matrix of
the reference state \eqref{eq:cello} also to the $\pm$ basis, we find
\begin{align}
  \begin{split}\label{refpm00}
    &G_{\rf,k}^{\pm} =\, G_{\rf,k}^{LR} = (R_{L_\Cbb\to L} \Oplus R_{R_\Cbb\to
      R}) \, G_{\rf,k}^{LR_\Cbb}\, (R_{L_\Cbb\to L} \Oplus R_{R_\Cbb\to
      R})^\dagger
    % = \diag(C,C,D,D),
    = C\oplus C \oplus D \oplus D,
    \\[6pt]
    &C= \frac{1}{2}\begin{bmatrix}
      \frac{1}{\lambda_\rf}+\frac{\lambda_\rf}{\lambda_k^2} & \frac{1}{\lambda_\rf}-\frac{\lambda_\rf}{\lambda_k^2} \\
      \frac{1}{\lambda_\rf}-\frac{\lambda_\rf}{\lambda_k^2} &
      \frac{1}{\lambda_\rf}+\frac{\lambda_\rf}{\lambda_k^2}
    \end{bmatrix}, \qquad D= \frac{1}{2}\begin{bmatrix}
      \lambda_\rf+\frac{\lambda_k^2}{\lambda_\rf} & \lambda_\rf-\frac{\lambda_k^2}{\lambda_\rf} \\
      \lambda_\rf -\frac{\lambda_k^2}{\lambda_\rf} &
      \lambda_\rf+\frac{\lambda_k^2}{\lambda_\rf}
    \end{bmatrix},
  \end{split}
\end{align}
where the covariance matrix here is given in the $\xi^{\pm}$ basis from
eq.~\eqref{xipm} and $R_{L_\Cbb\to L} = R_{R_\Cbb\to R}$ is given by
eq.~\eqref{eq:maccu} with the replacement $\lambda_n \rightarrow
  \lambda_k$, applied to the left and right operators, respectively. The first
equality is due to the fact that the reference state covariance matrix is
stationary under the change of basis $R_{LR\to \pm}$, given in
\eqref{eq:Oliver}. We can then obtain the relative covariance matrix in the
$\pm$ basis as follows
\begin{align}
  \MoveEqLeft[0] \Delta_{\cTFD,k}^\pm
  \label{eq:nserc}
  = \\ &\left[
         G_{\TFD,k}^+(t_{\lR},\alpha_{\lR}) \Oplus G_{\TFD,k}^+(t_{\Lr},\alpha_{\Lr})
         \Oplus G_{\TFD,k}^-(t_{\lR},\alpha_{\lR}) \tilde{\oplus} G_{\TFD,k}^-(t_{\Lr},\alpha_{\Lr})
         \right]
         (G_{\rf,k}^\pm)^{-1}.\nonumber
\end{align}
As we show in appendix \ref{sec:cv}, we can in fact analytically diagonalize
$\Delta_{\cTFD,k}$ at $t=0$.

To evaluate the complexity of formation \eqref{eq:Snowy}, we must also evaluate
the complexity of the vacuum, which takes a relatively simple form in the
$LR_\Cbb$ basis, in which the covariance matrices of both the vacuum and the
reference state are diagonal, cf.~eq.~\eqref{eq:cello}:\footnote{The vacuum
  covariance matrix is obtained by replacing $\lambda_\rf\to\lambda_k$ in the
  reference covariance matrix, as is clear by comparison of the physical and
  reference Hamiltonians \eqref{dimlessHamiltonian1} and \eqref{eq:feta}.}
\begin{align}
  G_{\rf,k}^{LR_\Cbb} =& \diag(\lambda_\rf^{-1}\mathds{1}_{4\times 4},\lambda_\rf\mathds{1}_{4\times 4}), \\
  G_{\vac,k}^{LR_\Cbb} =& \diag(\lambda_k^{-1}\mathds{1}_{4\times 4},\lambda_k\mathds{1}_{4\times 4}), \\
  \Delta_{\vac,k}^{LR_\Cbb} =& \diag\left(
                               \frac{\lambda_\rf}{\lambda_k}\mathds{1}_{4\times 4},
                               \frac{\lambda_k}{\lambda_\rf} \mathds{1}_{4\times 4}\right).
                               \label{eq:Tigger}
\end{align}
Transforming \eqref{eq:nserc} and \eqref{eq:Tigger} to the $LR$ and $LR_\Cbb$
bases, we are then able to evaluate the complexity of formation \eqref{eq:Snowy}
for each mode using the methods described in subsection \ref{sec:George2}, then
integrate over all modes, as described in subsection \ref{sec:paroxetine}, to
obtain the complexity of formation for the full complex scalar cTFD.

In \cite{Formation,TFD}, it was found that the complexity of formation is
proportional to the entropy in several cases (see discussion in section
\ref{sec:intro}). Therefore, \cite{Formation,TFD} have argued that it is natural
to consider the ratio of complexity of formation of the TFD state over the
entanglement entropy between the two sides, or equivalently the thermal entropy
of the thermal state obtained on each side after tracing out the other. The
entropy for the uncharged TFD is obtained from the partition function with
Bose-Einstein statistics by differentiating it with respect to the temperature
(cf.~eqs.~(201)-(202) of \cite{TFD}) and reads:
\begin{align}
  S_{\TFD}=\text{vol}\int \frac{d^{d-1}k}{(2\pi)^{d-1}} \left(
  \frac{\beta\omega_k}{e^{\beta\omega_k}-1}
  -\log(1-e^{-\beta\omega_k})\right),
  \label{eq:icecream}
\end{align}
where $\omega_k=\sqrt{k^2+m^2}$ and $\text{vol}=L^{d-1}$ is an IR regulator for
the volume of the field theory. For the cTFD state, we are dealing with two sets
of modified temperatures and times, as indicated in
eqs.~\eqref{eq:peppermint}-\eqref{eq:earlgrey}. We therefore suggest that it is
natural to normalize our result with respect to the following sum of entropies
\begin{equation}
  S_{\cTFD}=S_{\TFD}(\beta\rightarrow\beta_\Lr)+S_{\TFD}(\beta\rightarrow\beta_\lR).
\end{equation}
This, of course, coincides with the entanglement entropy between the $L$ and $R$
sides of the cTFD. The entanglement entropy between the two sides of the cTFD of
the complex scalar field can be expressed as
\begin{align}
  \begin{split}
    S_{\cTFD}
    =& \,\frac{\vol}{\beta^{d-1}} [s(\beta m,\beta\mu)+s(\beta m,-\beta\mu)] \\
    s(x,y) \equiv&\, \frac{\Omega_{d-2}}{(2\pi)^{d-1}} \int_0^\infty d u\;
    u^{d-2} \Bigg[ \frac{g(u,x,y)}{e^{g(u,x,y)}-1} - \log(1-e^{-g(u,x,y)})
    \Bigg]
  \end{split}
                   \label{eq:bourbon}
\end{align}
where $g$ is given by
\begin{align}
  g(u,x,y)
  \equiv& \sqrt{u^2+x^2}-y. \label{eq:hummus}
\end{align}
In most plots below, we will consider the ratio
\begin{align}
  \frac{\Delta \cmplx_1(|\cTFD(t=0)\rangle)}{S_\cTFD}.
  \label{eq:sushi}
\end{align}
However, since at low temperatures the entropy goes to zero, in order to study
the low temperature limits we will sometimes present un-normalized plots.

\subsubsection{Vanishing Chemical Potential}
\label{sec:kaboom}
As we shall numerically verify (see figure \ref{fig:padthai} below), taking the
limit of vanishing chemical potential $\beta \mu\to 0$ in either the $LR$ or
$LR_\Cbb$ bases yields a complexity $\cmplx(t=0)$ which matches with the
complexity of the real scalar uncharged TFD \cite{TFD}, up to additional
proportionality factors. In the $LR$ basis, there is an extra overall factor of
$2$, which can be attributed to the fact that each complex field decomposes into
two real fields. For the same reason, we see from eq.~\eqref{eq:bourbon} that
entropy is similarly doubled for the complex scalar. In the $LR_\Cbb$ basis, we
find that complexity instead receives a factor of $\sqrt{2}$ compared to the
result in \cite{TFD}, which can be attributed to the fact that the straight line
circuit is better aligned with the elementary gates of this basis so that the
circuit can be generated with fewer gates, see detailed discussion in footnote
\ref{foot:doge} below.

\subsubsection{High Temperature Limit}\label{sec:zoomies}
Taking $\beta \to 0$, the dimensionless parameters of the theory $\beta m,\beta
\mu \to 0$ vanish. Hence, we expect to recover results proportional to those of
the uncharged TFD for a massless scalar in eq.~(206) of \cite{TFD} and indeed we
find
\begin{align}
  \frac{\Delta \cmplx_1(\cTFD)}{S_\cTFD}
  =& \frac{2^d-1}{d}  \times \begin{cases}
    1
    & \text{$LR$ basis}
    \\
    2^{-1/2}
    & \text{$LR_\Cbb$ basis}
  \end{cases} \qquad(\beta m=\beta\mu =0).
      \label{eq:daniels}
\end{align}
Recall that the relative factor of $2^{-1/2}$ in the $LR_\Cbb$ basis can be
attributed to the fact that the straight line circuit is better aligned with the
elementary gates of this basis, as described above in subsection
\ref{sec:kaboom}, see also footnote \ref{foot:doge} below.

\subsubsection{Low Temperature Limit}\label{lowtemplim}
Here, we consider the low
temperature (large $\beta$) limit. Focusing first on the neutral case with
$\mu=0$, and taking $\beta m \gg 1$, gives the uncharged TFD at low
temperatures, which was already treated in \cite{TFD}, see eq.~(208) there,
\begin{align}
  \frac{\Delta \cmplx_1(\cTFD)}{S_\cTFD}
  \approx\, & \frac{2^{(d+1)/2} e^{\beta m/2}}{\beta m} \times \begin{cases}
    1
    & \text{$LR$ basis}
    \\
    2^{-1/2}
    & \text{$LR_\Cbb$ basis}
  \end{cases}.
      \label{eq:Ivy}
\end{align}
As in \eqref{eq:daniels}, we have an extra factor of $2^{-1/2}$ in the $LR_\Cbb$
basis.

Next, we consider the low temperature limit with positive chemical potential,
\ie $\beta m\gg 1$ and $\beta\mu\gg 1$, where without loss of generality we have
chosen $\mu> 0$.\footnote{As noted at the end of subsection \ref{sec:argh}, our
  results are symmetric under the change $\mu \rightarrow -\mu$.} We are able to
obtain analytic expressions for the low temperature limits in the case where we
further assume $\beta(m-\mu)\gg 1$.
For the complexity of formation we obtain (see appendices \ref{app:lowTlowT} and
\ref{sec:cv} for the detailed derivation):
\begin{align}
  \begin{split}
    &\hspace{-11pt}\Delta \cmplx_1(\cTFD) \approx \vol \, \Omega_{d-2} \cdot
    \left(\frac{m}{\pi^2\beta}\right)^{\frac{d-1}{2}}
    \Gamma\left(\frac{d-1}{2}\right) e^{-\beta(m-\mu)/2}
    \\
    &\hspace{65pt}\times\begin{cases} \frac{ 2\max\left\{m^2,m_\rf^2\right\}
        \log \left(\frac{m}{m_\rf}\right) }{ m^2-m_\rf^2 } & \text{$LR$ basis}
      \\
      \frac{1}{\sqrt{2}}\left[ 1+\frac{ m m_\rf
          \left(\lambda_\rf+\lambda_\rf^{-1}\right)
          \log\left(\frac{m}{m_\rf}\right) }{ m^2-m_\rf^2 } \right] &
      \text{$LR_\Cbb$ basis}
    \end{cases}.
  \end{split}
                                                                      \label{eq:zirkova1}
\end{align}
In that same limit, the entropy \eqref{eq:bourbon} of the cTFD is given by
\begin{align}
  \begin{split}
    \hspace{-5pt} S_{\cTFD} \approx\,& \vol \, \Omega_{d-2} \cdot
    \left(\frac{m}{2\pi^2\beta}\right)^{\frac{d-1}{2}}
    \Gamma\left(\frac{d-1}{2}\right) e^{-\beta(m-\mu)} \left(\frac{\beta
        (m-\mu)}{2}\right) .
  \end{split}\label{eq:bourbon2}
\end{align}
Taking the ratio between complexity of formation \eqref{eq:zirkova1} and entropy
\eqref{eq:bourbon2}, we have
\begin{align}
  \MoveEqLeft[1]\frac{\Delta \cmplx_1(\cTFD)}{S_\cTFD}
  \approx  \frac{2^{(d-1)/2} e^{\beta(m-\mu)/2}}{\beta(m-\mu)}
  \begin{cases}
    \frac{ 4\max\{m^2,m_\rf^2\} \log \frac{m}{m_\rf} }{ m^2-m_\rf^2 } &
    \text{$LR$ basis}
    \\
    \sqrt{2}\left[ 1+\frac{ m \, m_\rf (\lambda_\rf+\lambda_\rf^{-1}) \log
        \frac{m}{m_\rf} }{ m^2-m_\rf^2 } \right] & \text{$LR_\Cbb$ basis}
  \end{cases}. \label{eq:Zooey2}
\end{align}
We will often be selecting $\lambda_\rf=m_\rf/\mu_g=1$, see comments around
eqs.\eqref{eq:piano}-\eqref{eq:piano2} and \eqref{eq:tedious}. Note that, in the
low temperature limit, both the entropy and the complexity are suppressed by
exponential factors, but the complexity goes to zero slower than the thermal
entropy. Of course, as we go away from the large $\beta$ (low temperature) limit,
this conclusion may change, as we will see in the numerics below.

\subsubsection{Numerical Results}
In this section, we present numerical plots of the complexity of formation
\eqref{eq:Snowy} of the complex scalar cTFD. In all the plots below, we have
chosen $\lambda_{\rf}=m_\rf/\mu_g=1$, see related comments around
eqs.~\eqref{eq:piano}-\eqref{eq:piano2} and \eqref{eq:tedious}. All our results
here are invariant under the symmetry $\mu \rightarrow -\mu$ of the cTFD and
will recover the neutral results (up to possible constants of proportionality)
when $\mu=0$.

\begin{figure}[h]
  \centering
  \begin{subfigure}[b]{0.49\textwidth}
    \includegraphics[width=\textwidth]{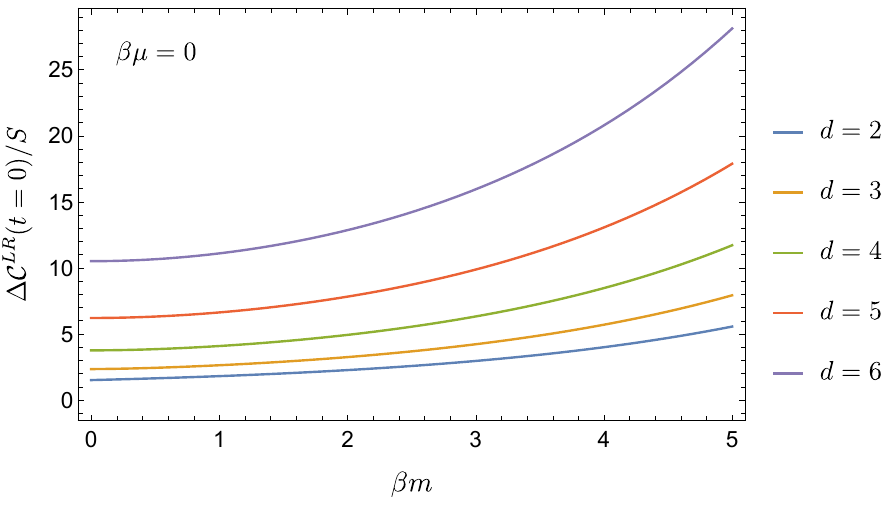}
    \caption{$LR$ basis}
    \label{fig:Leroy}
  \end{subfigure}~~~
  \begin{subfigure}[b]{0.49\textwidth}
    \includegraphics[width=\textwidth]{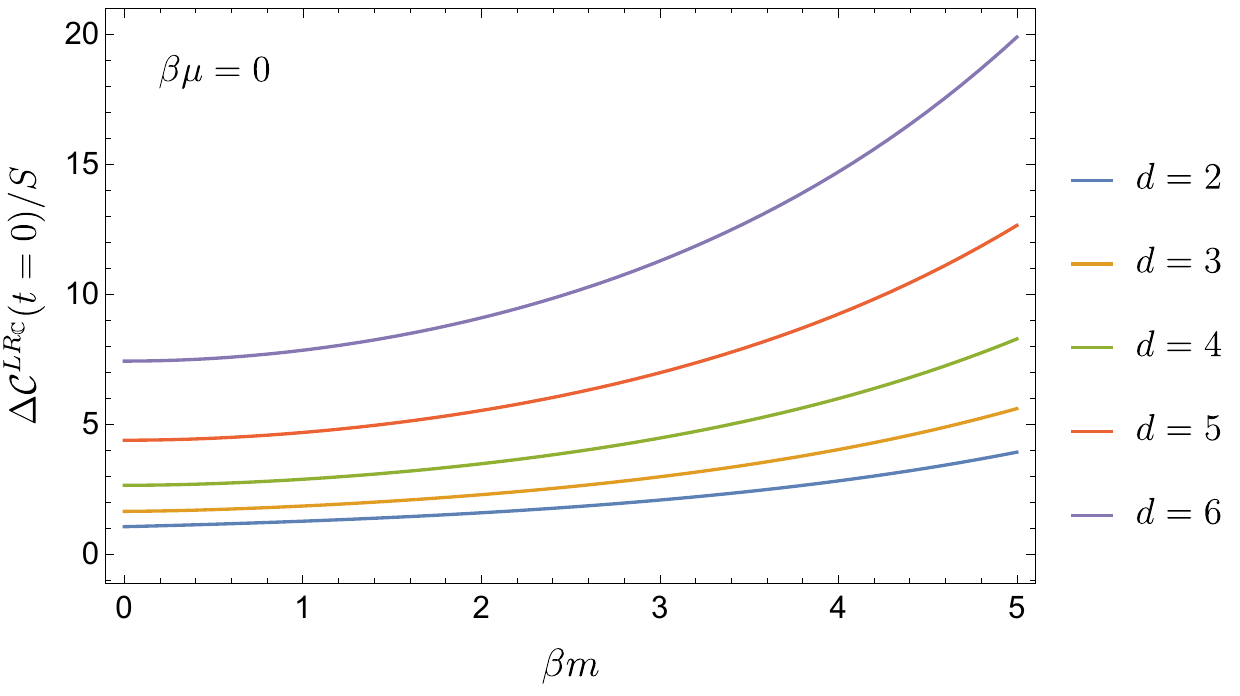}
    \caption{$LR_\Cbb$ basis}
    \label{fig:Fang}
  \end{subfigure}
  \caption{Complexity of formation, scaled by the entropy, as a function of
    $\beta m$, for the complex scalar cTFD in the special case $\mu=0$. Various
    dimensions $d$ are shown. The results for the $LR$ and $LR_\Cbb$ bases are
    proportional to each other and to figure 9 in \cite{TFD}.}
  \label{fig:padthai}
\end{figure}

We begin with figure \ref{fig:padthai}, where we plot the case of vanishing
chemical potential $\mu=0$ in both the $LR$ (figure \ref{fig:Leroy}) and
$LR_\Cbb$ (figure \ref{fig:Fang}) bases. Recall that the $LR_\Cbb$ basis was the
original basis of complex operators \eqref{eq:defLRCbasis} while the $LR$ basis
is the set of operators adapted to the particle and anti-particle degrees of
freedom \eqref{eq:defLRbasis}. Note that the vanishing of $\mu$ reduces the cTFD
to two uncharged TFDs. For this case, we see that the $LR$ and $LR_\Cbb$ bases
give proportional results, with a relative factor of $\sqrt{2}$ --- see
explanation in subsection \ref{sec:kaboom} and footnote \ref{foot:doge}. In
\cite{TFD}, the same figure (figure 9 therein) was produced for the uncharged
TFD state of a real scalar field, using a basis analogous to $LR_\Cbb$. The fact
that figure \ref{fig:Fang} is proportional to figure 9 of \cite{TFD} provides a
check of our numerics.

\begin{figure}[h]
  \centering
  \begin{subfigure}[b]{0.95\textwidth}
    \includegraphics[width=\textwidth]{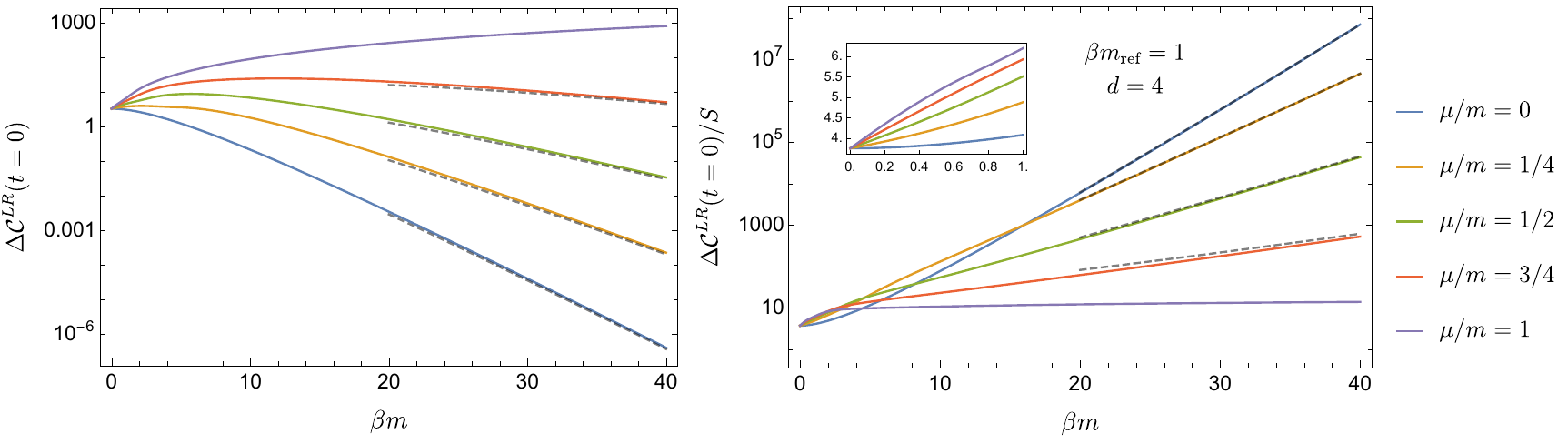}
    \caption{Fixed $\beta m_{\rf}$ and various $\mu/m$; complexity versus $\beta
      m$.}
    \label{fig:shepardspie}
  \end{subfigure}
  \begin{subfigure}[b]{0.95\textwidth}
    \includegraphics[width=\textwidth]{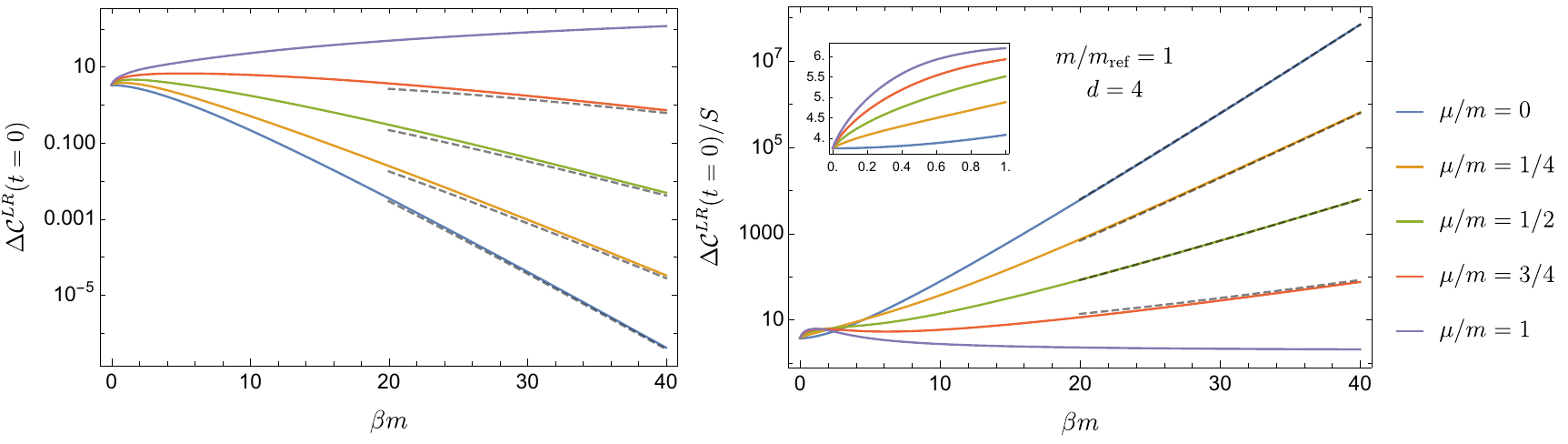}
    \caption{Fixed $m/m_{\rf}$ and various $\mu/m$; complexity versus $\beta
      m$.}\label{fig:shepardspie2}
  \end{subfigure}
  \begin{subfigure}[b]{\textwidth}
    \centering
    \includegraphics[width=0.5\textwidth]{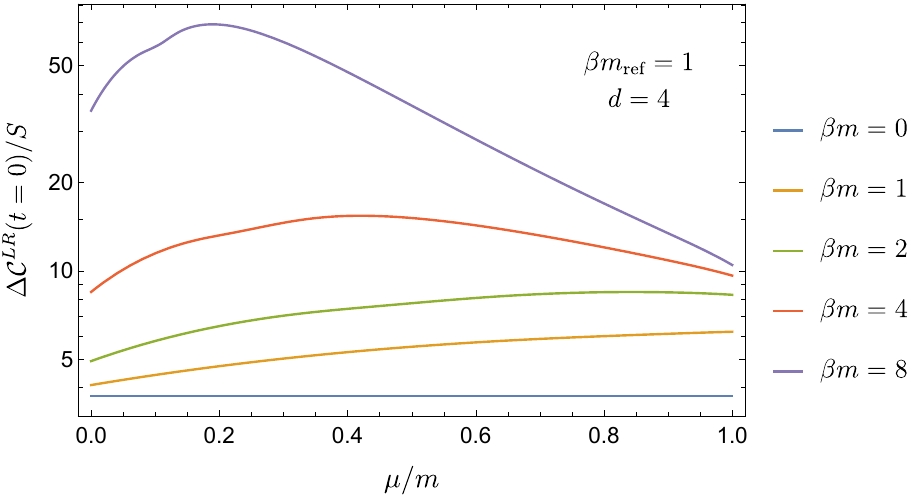}
    \caption{Fixed $\beta m_\rf$ and various $\beta m$; complexity versus
      $\mu/m$.}
    \label{fig:sundae}
  \end{subfigure}
  \caption{Complexity of formation in the $LR$ basis for the complex scalar cTFD
    in $d=4$. The curves are plotted as functions of $\beta m$ and $\mu/m$ for
    various fixed values of $\mu/m$ and $\beta m$, respectively (recall that
    $|\mu|\leq m$). The dashed curves in subfigures \ref{fig:shepardspie} and
    \ref{fig:shepardspie2} mark the low temperature limits given in subsection
    \ref{lowtemplim}. The conformal neutral limit is obtained at the left-most
    points in subfigures \ref{fig:shepardspie} and \ref{fig:shepardspie2} since
    keeping $\mu/m$ fixed and sending $m\rightarrow0$ means that we are also
    decreasing the chemical potential. We see that, in this case, the dependence
    on the ratio $\mu/m$ disappears and all the curves approach the same point
    (alternatively, this can be seen as a large temperature limit, where the
    chemical potential becomes negligible).}
  \label{fig:smokedSalmonBagel}
\end{figure}

\begin{figure}
  \centering
  \begin{subfigure}[b]{0.95\textwidth}
    \includegraphics[width=\textwidth]{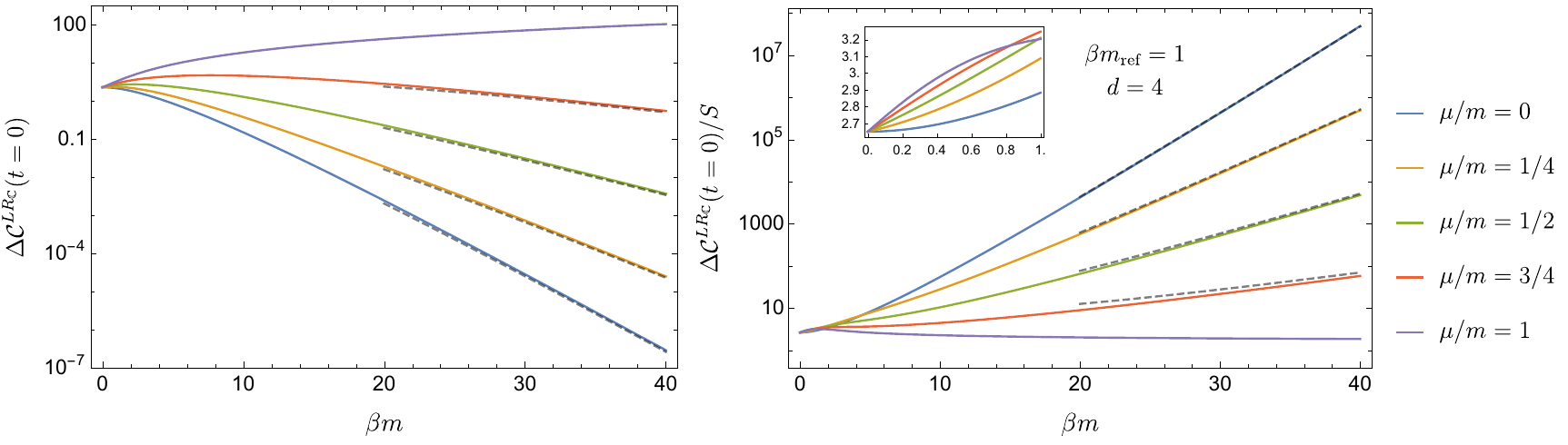}
    \caption{Fixed $\beta m_{\rf}$ and various $\mu/m$.}
    \label{fig:minestraDiCeci}
  \end{subfigure}
  \begin{subfigure}[b]{0.95\textwidth}
    \includegraphics[width=\textwidth]{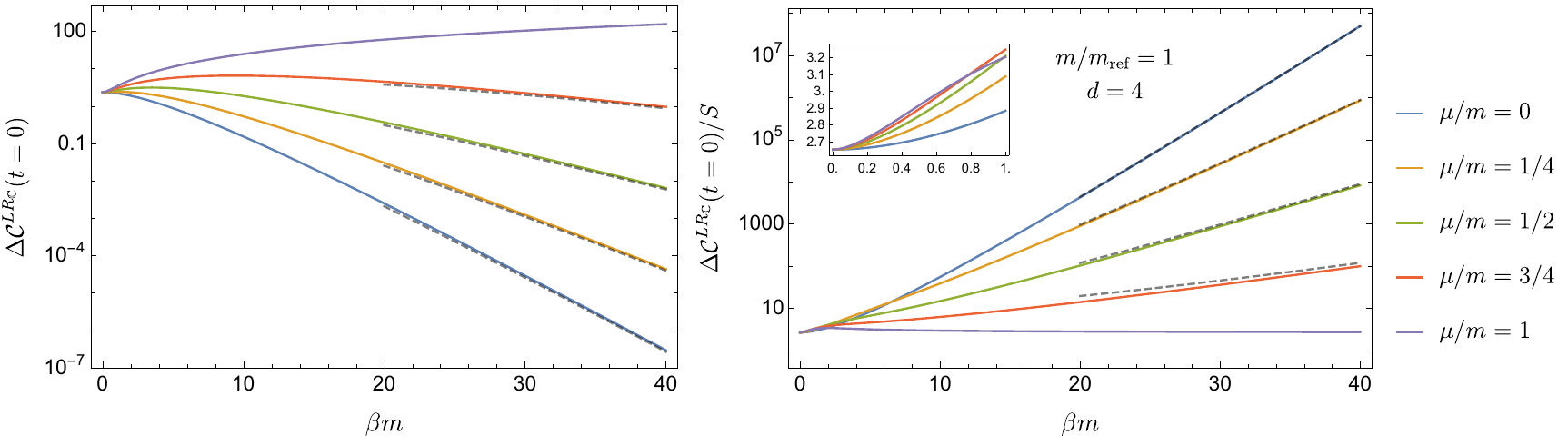}
    \caption{Fixed $m/m_{\rf}$, and various $\mu/m$.}
    \label{fig:salsiccia}
  \end{subfigure}
  \begin{subfigure}[b]{0.5\textwidth}
    \includegraphics[width=\textwidth]{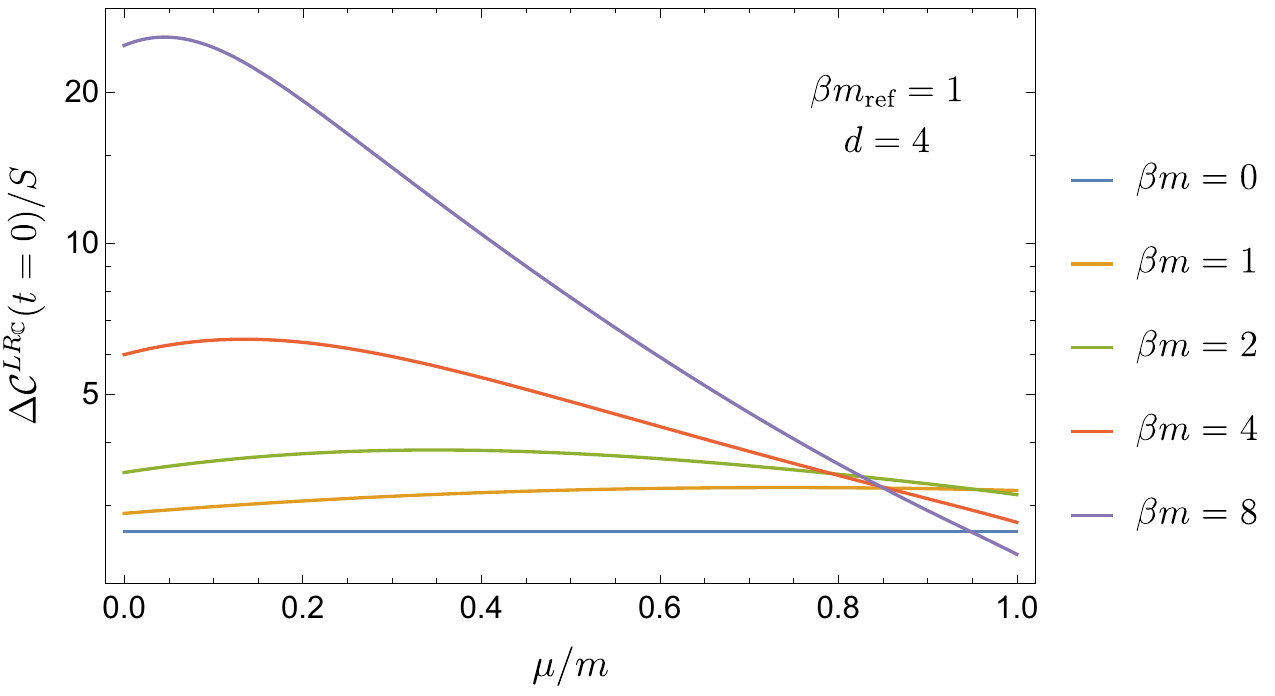}
    \caption{Fixed $\beta m_{\rf}$, and various $\beta m$.}
    \label{fig:minestrone}
  \end{subfigure}
  \caption{Complexity of formation in the $LR_\Cbb$ basis for the complex scalar
    cTFD in $d=4$. The curves are plotted as functions of $\beta m$ and $\mu/m$
    for various fixed values of $\mu/m$ and $\beta m$, respectively. The dashed
    curves in subfigures \ref{fig:minestraDiCeci} and \ref{fig:salsiccia} mark
    the low temperature limits given in subsection \ref{lowtemplim}. Note that
    although subfigure \ref{fig:minestraDiCeci} fixes $\beta m_{\rf}$ while
    subfigure \ref{fig:salsiccia} fixes $m/m_{\rf}$, the two figures are nearly
    identical --- this is because, as shown below in figure \ref{fig:Lady}, the
    dependence on $\beta m_\rf$ in the $LR_\Cbb$ basis is very weak.}
  \label{fig:bologonese}
\end{figure}

\begin{figure}
  \centering
  \begin{subfigure}[b]{0.49\textwidth}
    \includegraphics[width=\textwidth]{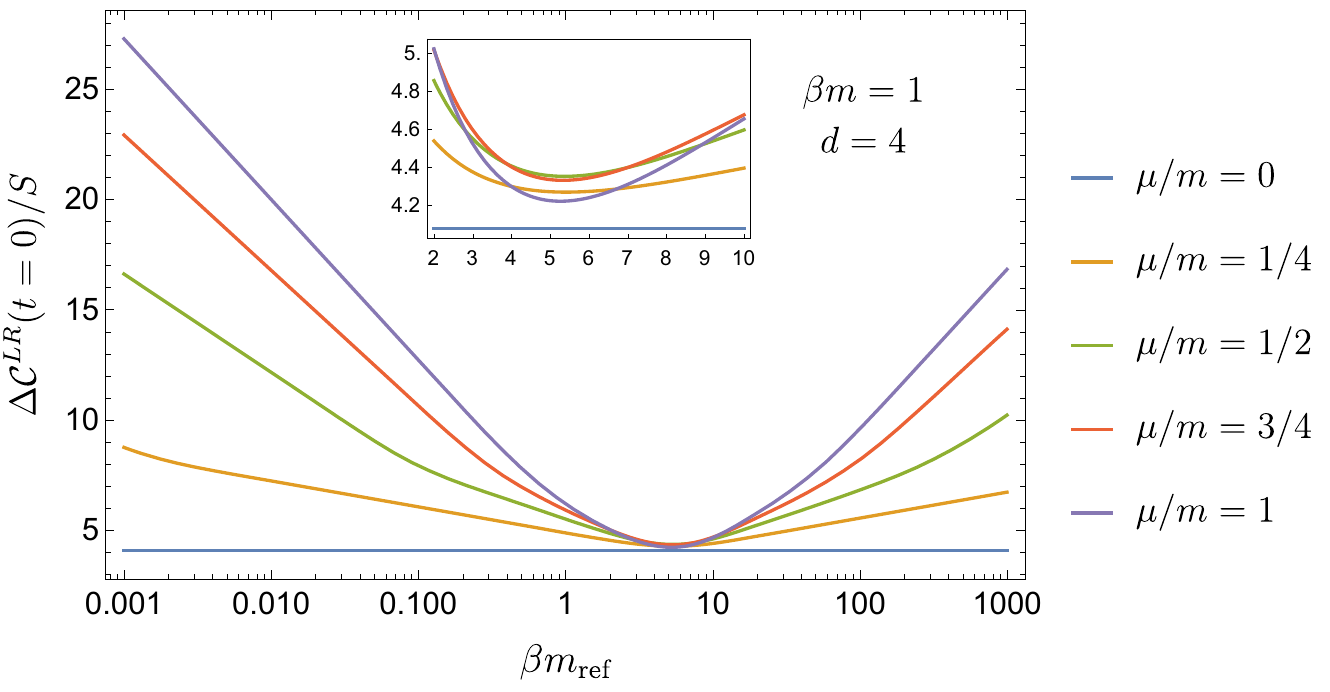}
    \caption{$LR$ basis.}
  \end{subfigure}
  \begin{subfigure}[b]{0.49\textwidth}
    \includegraphics[width=\textwidth]{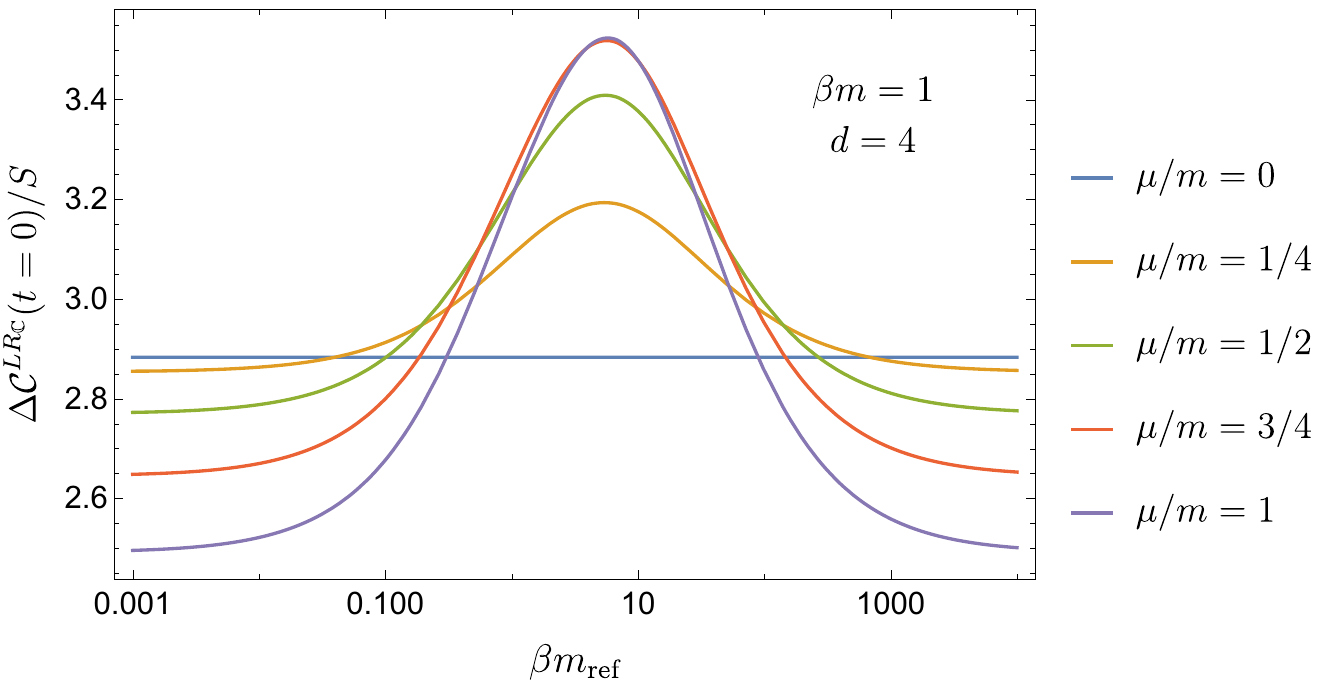}
    \caption{$LR_\Cbb$ basis.}
    \label{fig:Lady}
  \end{subfigure}
  \caption{Complexity of formation, scaled by entropy, as a function of the
    reference scale $\beta m_\rf$, for the complex scalar cTFD with $\beta m=1$
    in $d=4$. Various values of $\beta\mu$ are shown. The complexity of
    formation is independent of the reference scale when $\beta\mu=0$ as found
    in \cite{TFD}. With non-vanishing $\mu$, the dependence is weaker in the
    $LR_\Cbb$ basis than in the $LR$ basis. Note that these figures obey an
    approximate symmetry $\beta m_{\rf} \rightarrow \varkappa/(\beta m_{\rf})$
    for some constant $\varkappa$.}
  \label{fig:meatloaf}
\end{figure}

In figures \ref{fig:smokedSalmonBagel} and \ref{fig:bologonese}, we consider the
complexity of formation for $\mu\ne 0$. We also present results for the
complexity normalized by the entropy \eqref{eq:bourbon}. We have added to these
figures the low temperature approximations presented in subsection
\ref{lowtemplim}. As can be seen, moving left to right, in the right panels of
subfigures \ref{fig:shepardspie}, \ref{fig:shepardspie2},
\ref{fig:minestraDiCeci} and \ref{fig:salsiccia}, in this large $\beta$ limit,
the ratio \eqref{eq:sushi} between complexity of formation and entropy diverges
exponentially for $|\mu|<m$ and appears to vary relatively slowly in the special
case\footnote{In the limiting case $|\mu|\to m$, the zero mode in either
    $\lR$ or $\Lr$ becomes infinitely populated due to an unbounded effective
    temperature, cf.~eqs.~\eqref{eq:peppermint}-\eqref{eq:earlgrey}. However, as
    explained in footnote \ref{foot:blahblahblah}, in sufficiently high
    dimensions $d$, the full state of the scalar field may nonetheless be
    normalizable and have a finite particle number \emph{density}. Indeed, in
    $d\ge 3$, the suppression by the low frequency density of states is
    sufficient to render the entropy integrals
    \eqref{eq:icecream}-\eqref{eq:bourbon} finite as $|\mu|\to m$, which is
    reflective of our earlier claim that the grand canonical potential remains
    finite in the $|\mu|\to m$ limit for $d\ge 3$, at least when neglecting
    condensation. To have a physically reasonable $|\mu|\to m$ state with finite
    particle density would further require $d\ge 4$. In general, condensation in
    the zero mode $k=0$ may occur as $|\mu|\to m$, but we shall assume the
    condensate makes a negligible contribution such that the integral
    approximation \eqref{eq:flambe} is exact in the $L\to\infty$
    limit. \label{foot:woopwoopwoop}} $|\mu|=m$. The corresponding left panels
show that, for all $|\mu|<m$, the complexity becomes smaller in the low
temperature limit; on the other hand, for the case $|\mu|=m$, it appears to
increase. The latter case best resembles what happens in holography, where the
complexity diverges in the low temperature limit at finite chemical potential, a
phenomena known as the ``third law of complexity'' \cite{Carmi:2017jqz}. In the
scalar theory, we see that this effect is not reproduced for $|\mu|<m$, as noted
above and can also be seen from eq.~\eqref{eq:zirkova1}. In general, moving from
lower to higher curves in the left panels, we observe that the unnormalized
complexity of formation increases as $|\mu|/m$ increases for fixed values of
$\beta m$ and $\beta m_{\rf}$. Some intuition for the behaviour of the complexity of formation as a function of the chemical
potential and temperature can be developed by considering the particle number
density, which also increases with the temperature and
chemical potential. Although the detailed dependence does not precisely match
complexity (as to be expected), this perhaps suggests that states become more
complex with increasing particle density. We leave it for future work to explore
this correlation in greater detail. Figure \ref{fig:meatloaf} explores the
dependence of the complexity of formation on the reference scale $\beta m_\rf$
for general $|\mu|\le m$. This dependence is weaker in the $LR_\Cbb$ basis than
in the $LR$ basis. We observe an approximate symmetry $\beta m_{\rf} \rightarrow
\varkappa/(\beta m_{\rf})$ for some constant $\varkappa$ in these figures.

\subsection{Time Dependence}
\label{sec:byeah}
\begin{figure}
  \centering
  \begin{subfigure}[b]{0.49\textwidth}
    \centering
  \includegraphics[width=\textwidth]{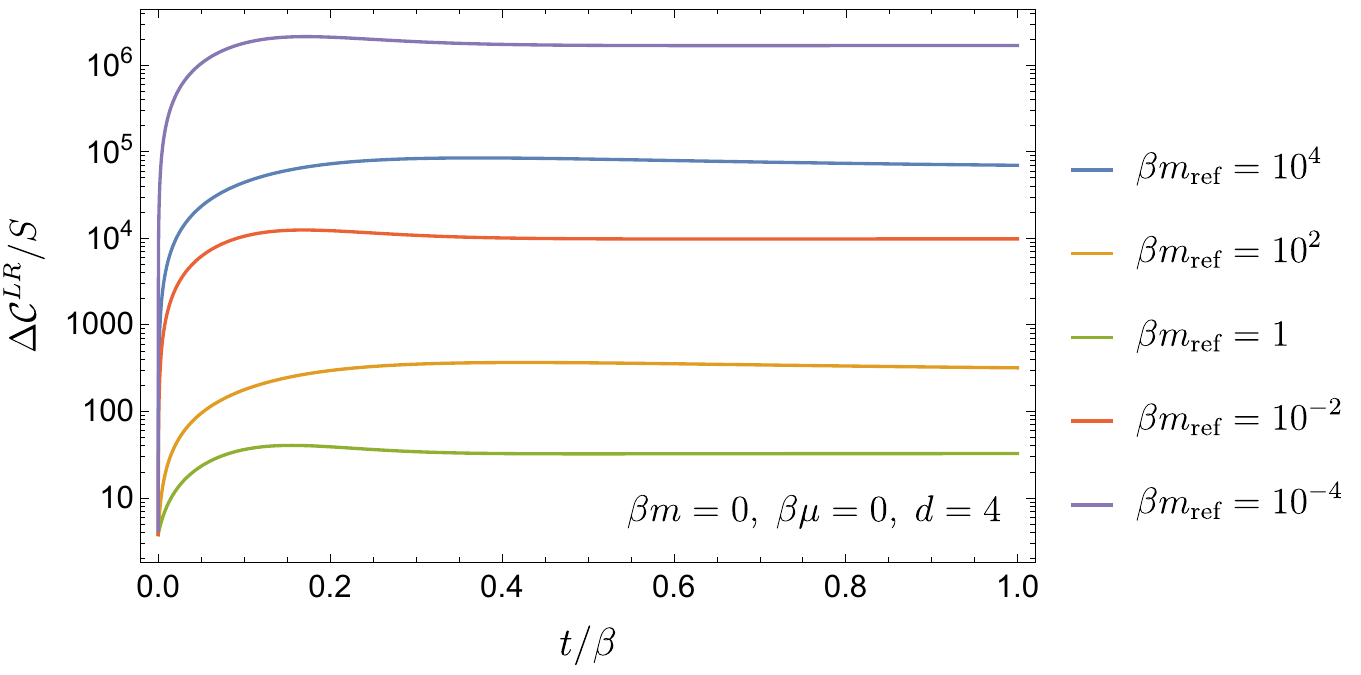}
    \caption{$LR$ basis.}
    \label{fig:whyAmISad}
  \end{subfigure}
  \begin{subfigure}[b]{0.49\textwidth}
    \centering
  \includegraphics[width=\textwidth]{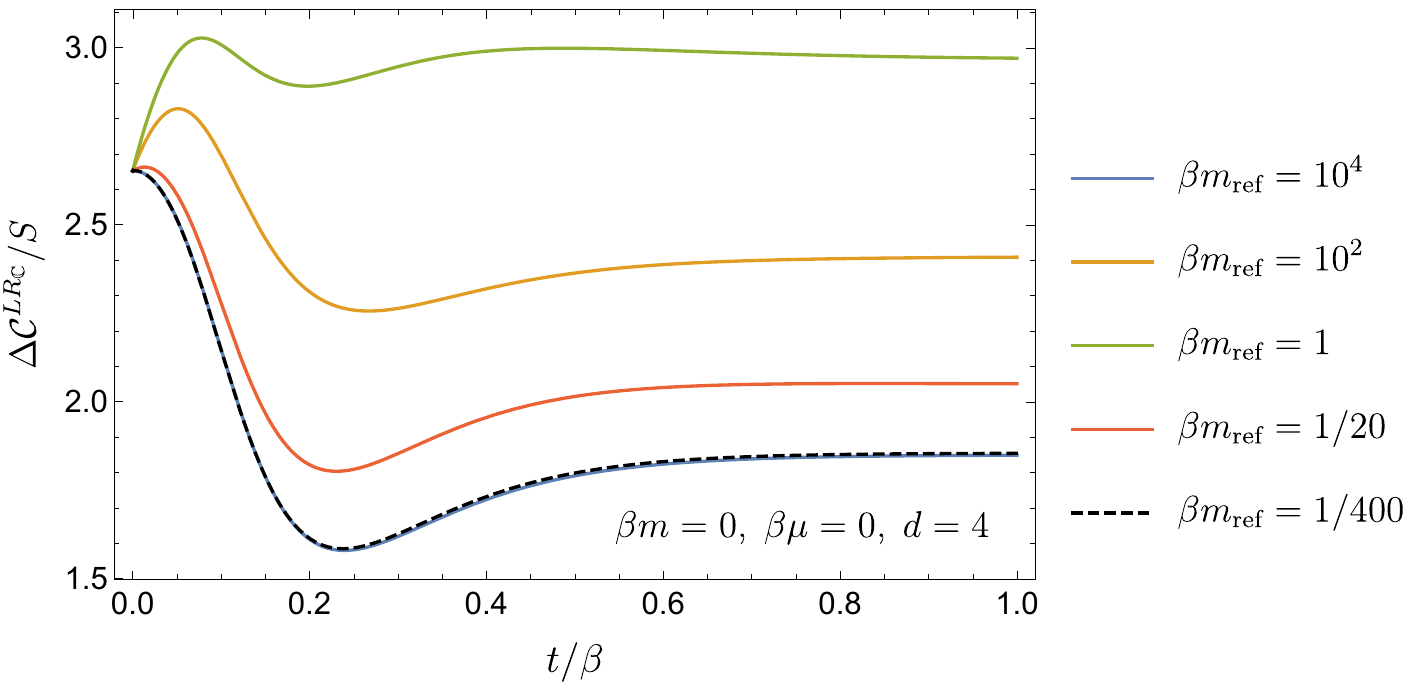}
    \caption{$LR_\Cbb$ basis.}
    \label{fig:IDonTKnow}
  \end{subfigure}
  \caption{Time dependence of the complexity, scaled by the entropy, for the complex scalar cTFD in $d=4$
    with $\beta m=\beta\mu=0$. The curves for various fixed values of $\beta
    m_{\rf}$ are plotted as functions of $t/\beta$.}
  \label{fig:caprese} %\label{fig:risiEBisi}
\end{figure}

\begin{figure}
  \centering
  \begin{subfigure}[b]{0.49\textwidth}
    \includegraphics[width=\textwidth]{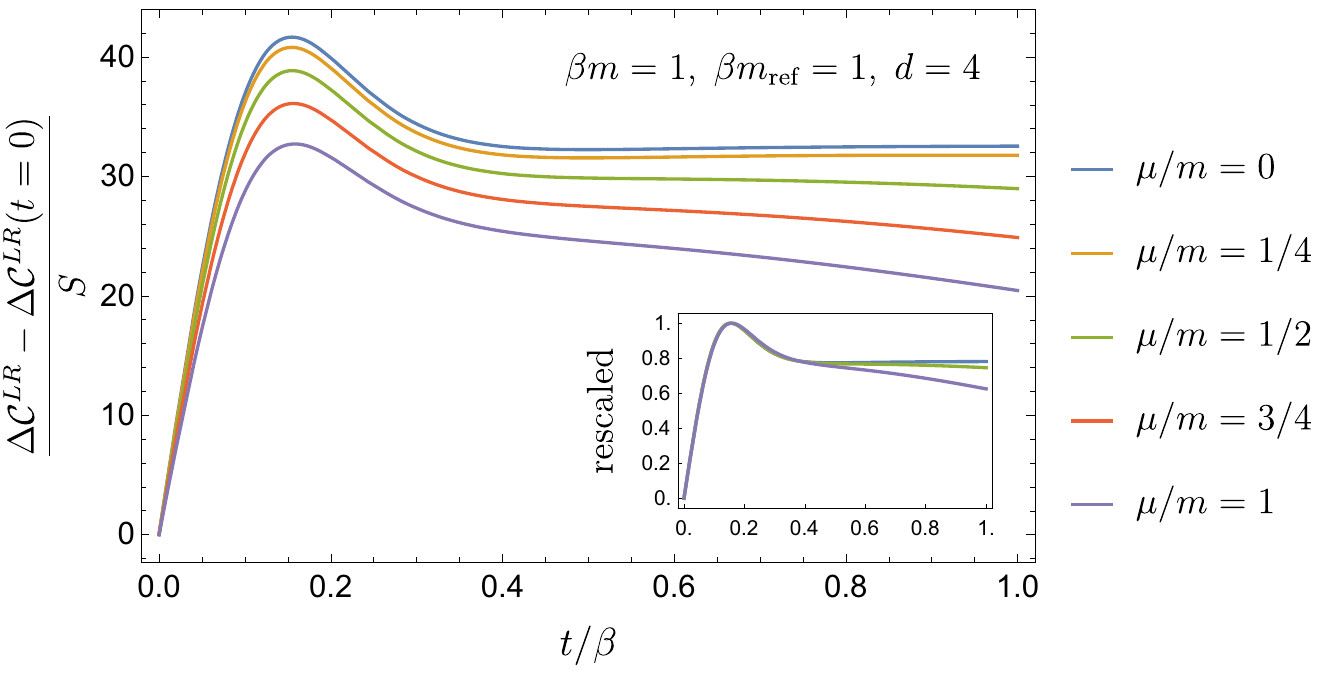}
    \caption{$\beta m=1$.}
    \label{fig:panzenella}
  \end{subfigure}
  \begin{subfigure}[b]{0.49\textwidth}
    \includegraphics[width=\textwidth]{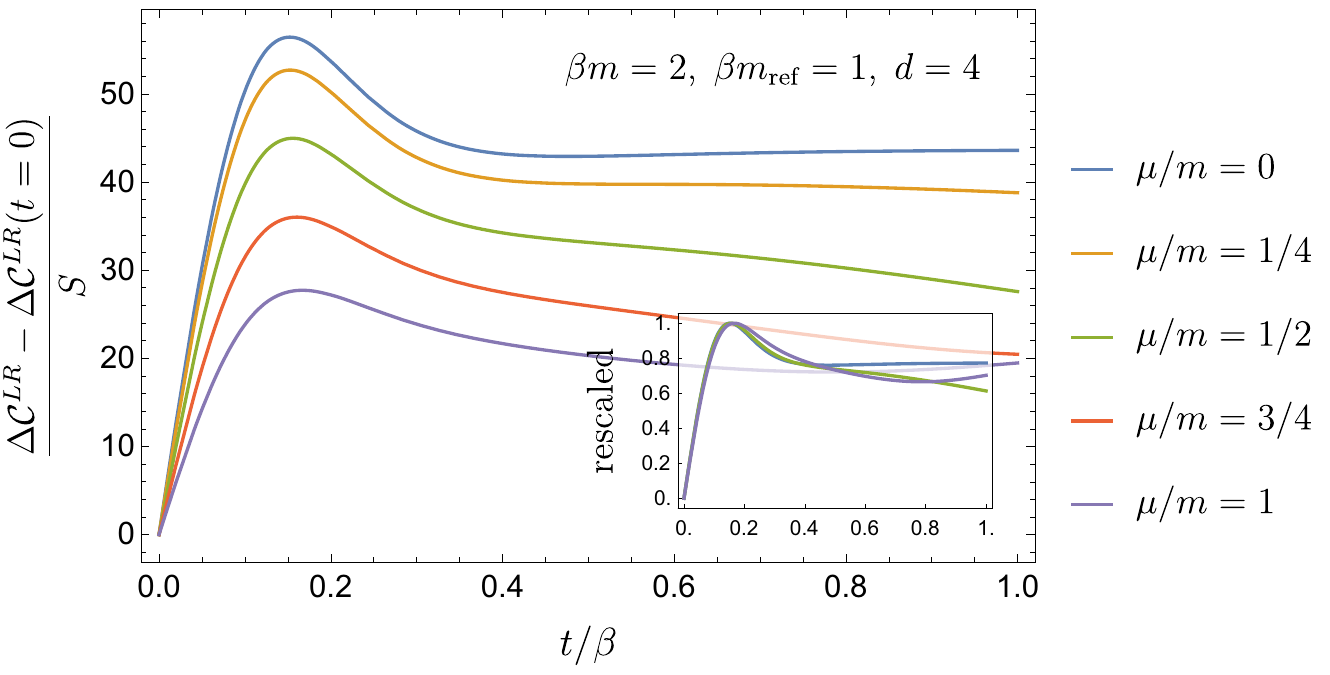}
    \caption{$\beta m=2$.}
    \label{fig:focaccia}
  \end{subfigure}\\\vspace{10pt}
  \begin{subfigure}[b]{0.49\textwidth}
    \includegraphics[width=\textwidth]{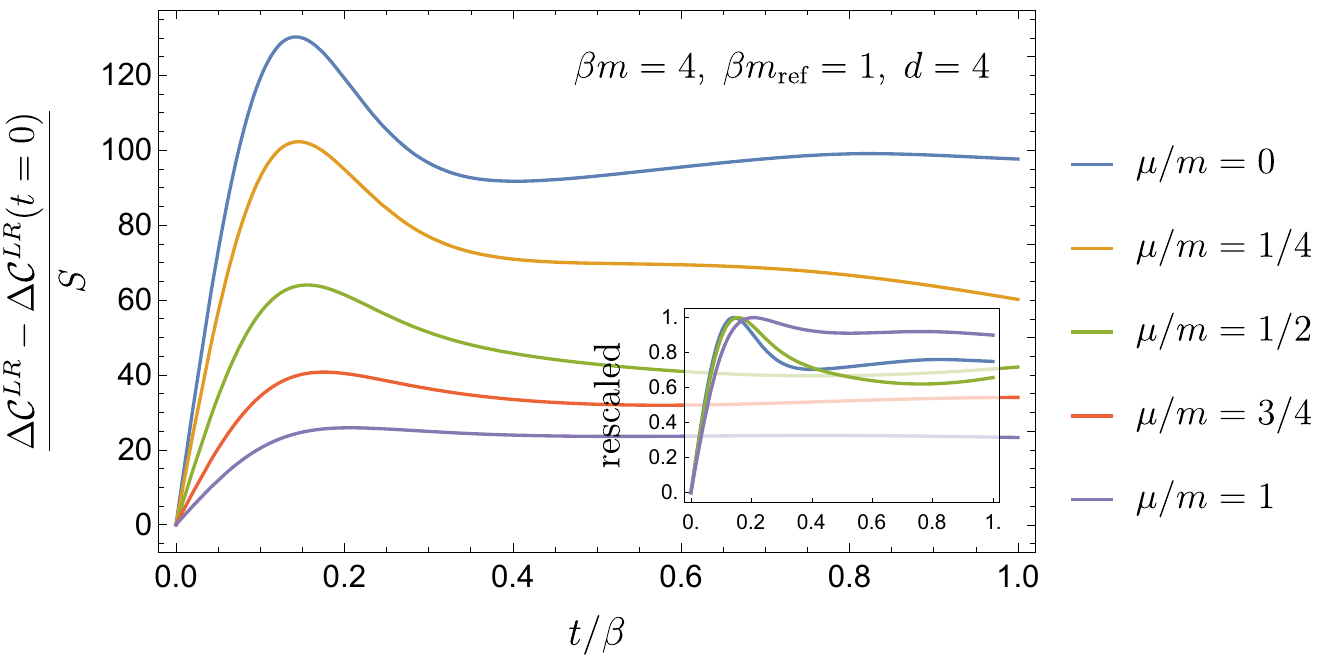}
    \caption{$\beta m=4$.}
    \label{fig:magherita}
  \end{subfigure}
  \begin{subfigure}[b]{0.49\textwidth}
    \includegraphics[width=\textwidth]{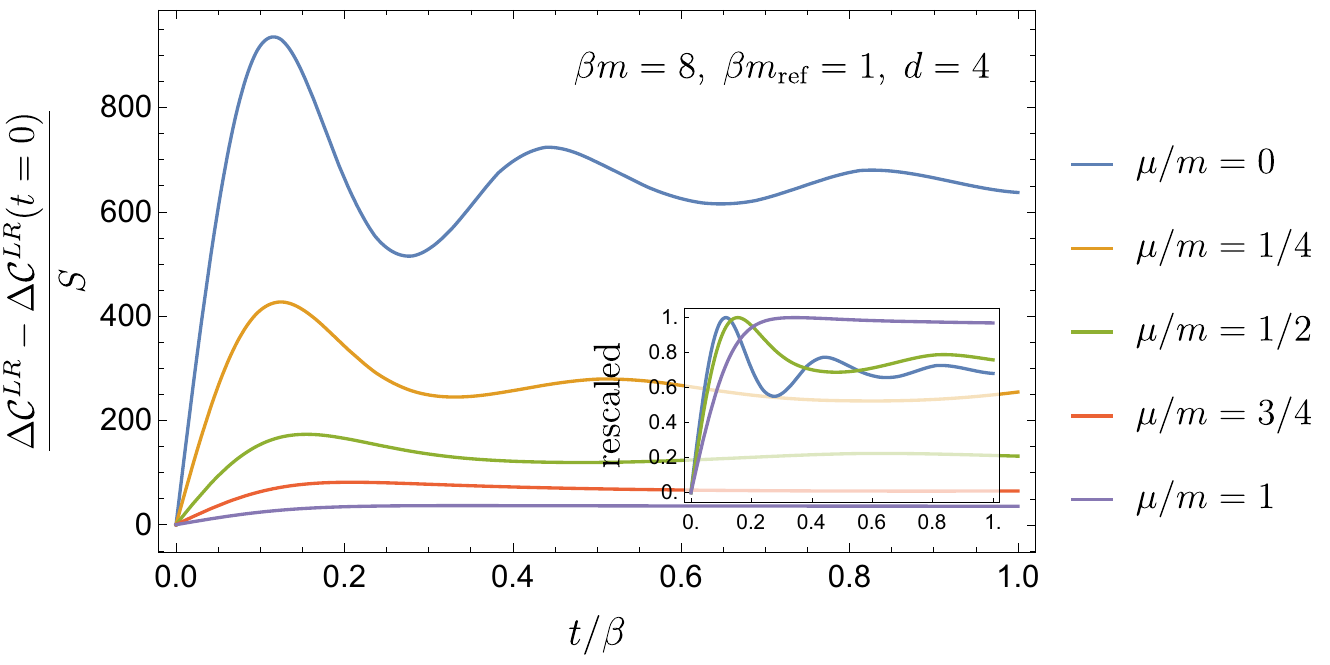}
    \caption{$\beta m=8$.}
    \label{fig:carbonara}
  \end{subfigure}
  \caption{Time dependence of the complexity in the $LR$ basis, scaled by the
    entropy, for the complex scalar cTFD in $d=4$. The subfigures correspond to
    $\beta m_{\rf}=1$ with different fixed values of $\beta m$. Curves for
    different fixed values of $\mu/m$ are plotted as functions of $t/\beta$.}
  \label{fig:buschetta}
\end{figure}

\begin{figure}
  \centering
  \begin{subfigure}[b]{0.95\textwidth}
    \includegraphics[width=\textwidth]{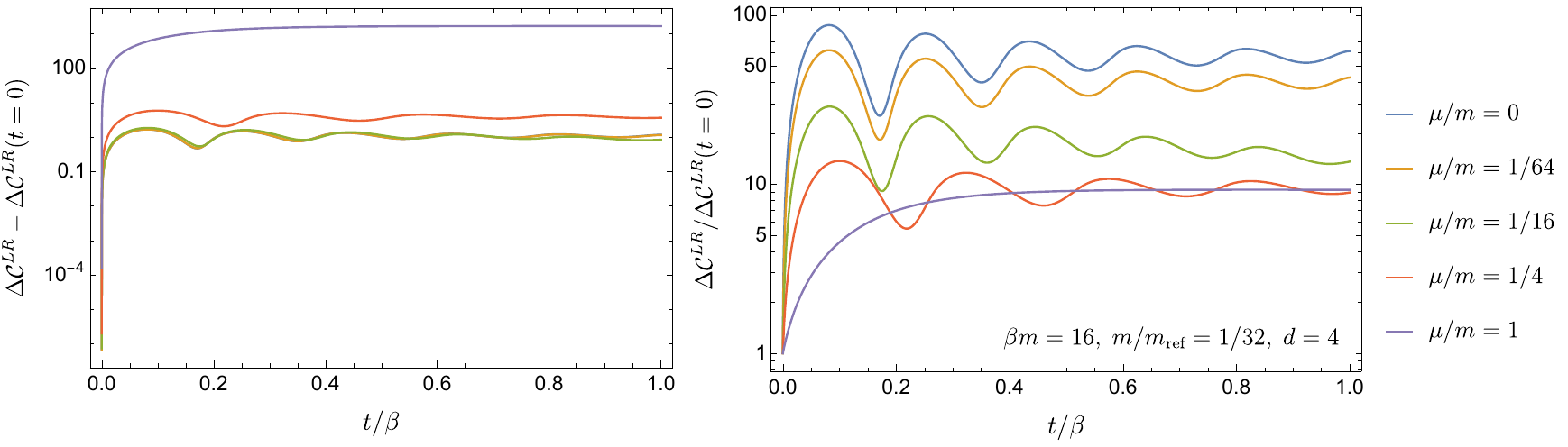}
    \caption{$\beta m=16$, $m/m_{\rf}=1/32$.}
    \label{fig:magherita2}
  \end{subfigure}
  \begin{subfigure}[b]{0.95\textwidth}
    \includegraphics[width=\textwidth]{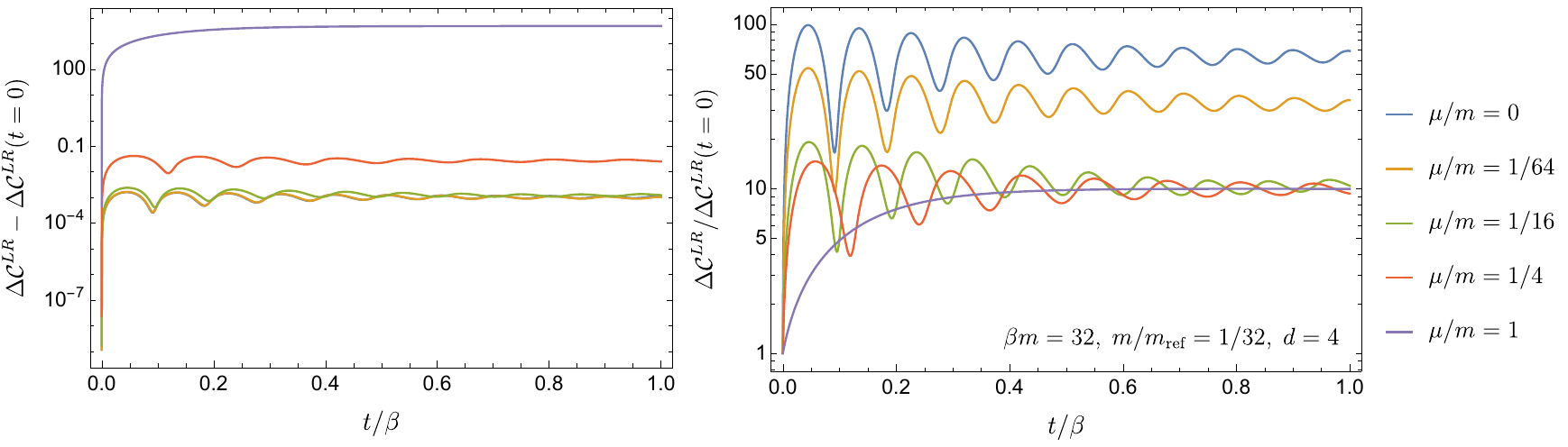}
    \caption{$\beta m=32$, $m/m_{\rf}=1/32$.}
    \label{fig:agliata}
  \end{subfigure}
  \caption{Oscillatory time dependence of the complexity in the $LR$ basis for
    the complex scalar cTFD in $d=4$ for large fixed values of $\beta m$ and
    even larger fixed values of $\beta m_{\rf}$. Curves for different fixed
    $\mu/m$ are plotted as functions of $t/\beta$. Note that the vertical axes
    and values of $\mu/m$ here differ from those in figure \ref{fig:buschetta}.}
  \label{fig:acquacotta}
\end{figure}

\begin{figure}[h]
  \centering
  \begin{subfigure}[b]{\textwidth}
  \end{subfigure}
  \begin{subfigure}[b]{\textwidth}
    \includegraphics[width=\textwidth]{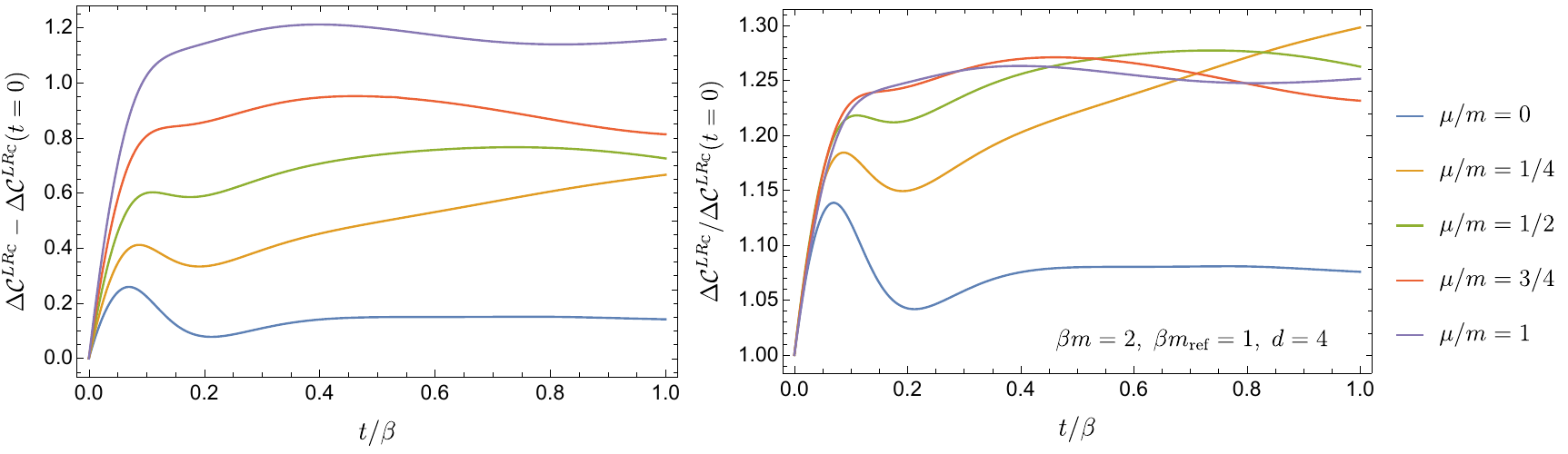}
    \caption{$\beta m=2$.}
    \label{fig:stracciatella}
  \end{subfigure}
  \begin{subfigure}[b]{\textwidth}
    \includegraphics[width=\textwidth]{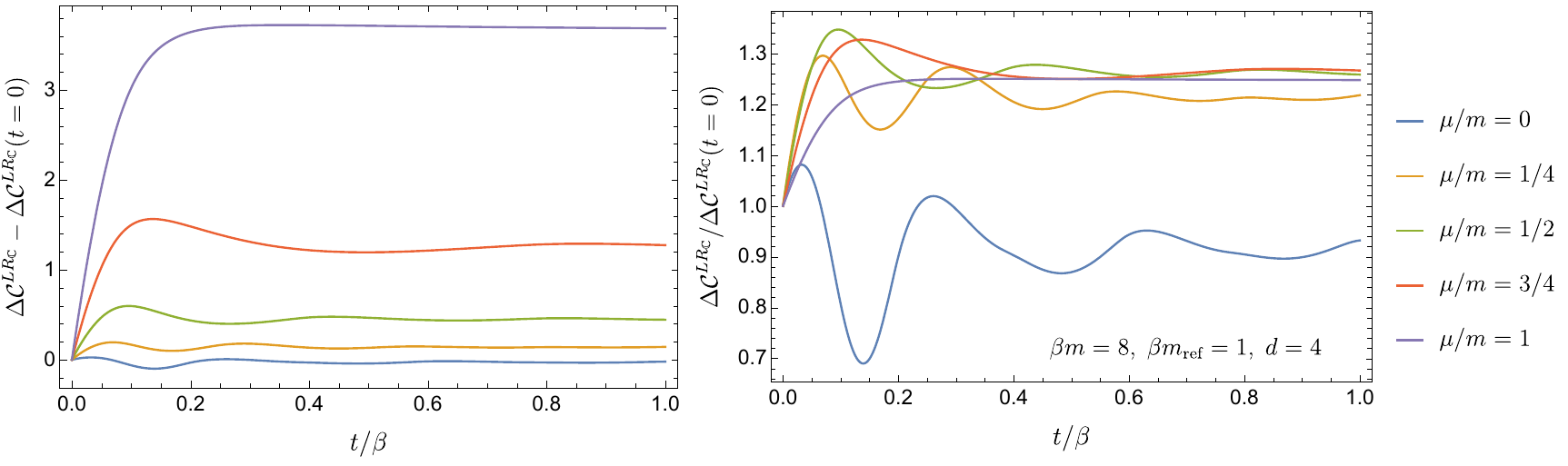}
    \caption{$\beta m=8$.}
    \label{fig:fonduta}
  \end{subfigure}
  \caption{Time dependence of the complexity in the $LR_\Cbb$ basis for the
    complex scalar cTFD in $d=4$ with different fixed values of $\beta m$ and
    with $\beta m_{\rf}=1$. Curves within each plot correspond to different
    fixed values of $\mu/m$ and are plotted as functions of $t/\beta$.}
  \label{fig:Belle}
\end{figure}

\begin{figure}[h]
  \centering
  \begin{subfigure}[b]{0.95\textwidth}
    \includegraphics[width=\textwidth]{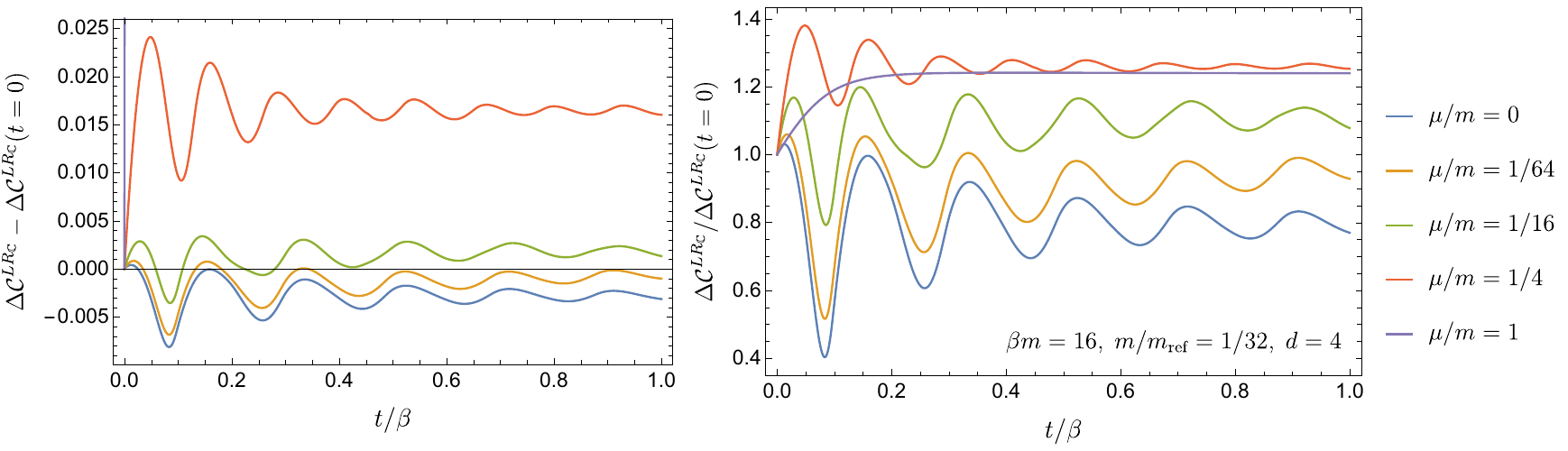}
    \caption{$\beta m=16$, $m/m_{\rf}=1/32$.}
    \label{fig:magherita22}
  \end{subfigure}
  \begin{subfigure}[b]{0.95\textwidth}
    \includegraphics[width=\textwidth]{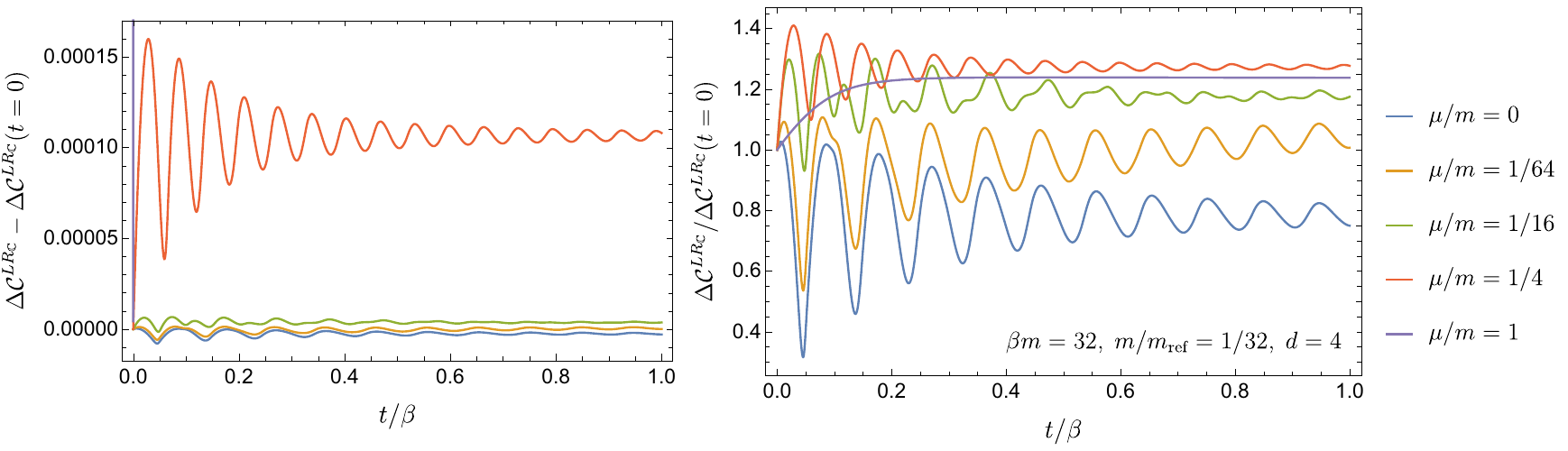}
    \caption{$\beta m=32$, $m/m_{\rf}=1/32$.}
    \label{fig:agliata2}
  \end{subfigure}
  \caption{Oscillatory time dependence of the complexity in the $LR_\Cbb$ basis
    for the cTFD in $d=4$ for large $\beta m$ and even larger $\beta m_{\rf}$.
    Curves for different fixed values of $\mu/m$ are plotted as functions of
    $t/\beta$. Note that the vertical axes and values of $\mu/m$ here differ
    from those in figure \ref{fig:Belle}.}
  \label{fig:acquacotta2}
\end{figure}

Next, we consider the time dependence of complexity for the complex scalar cTFD.
To calculate complexity at arbitrary times, we integrate eq.~\eqref{finalfinal}
over all modes, cf.~eq.~\eqref{eq:flambe}. At general times, it is cumbersome to
write analytic expressions for these relative covariance matrices, so we
immediately resort to numerics. We plot the complexity against time in the
neutral limit (for both the $LR$ and the $LR_\Cbb$ bases) in figure
\ref{fig:caprese} and for general $|\mu|\le m$ in the $LR$ basis in figures
\ref{fig:buschetta}-\ref{fig:acquacotta} and in the $LR_\Cbb$ basis in figures
\ref{fig:Belle}-\ref{fig:acquacotta2}.

Starting with the neutral $\mu=0$ limit in figure \ref{fig:caprese}, we observe
that the complexity in the $LR$ basis (figure \ref{fig:whyAmISad}) does not recover the results
for the time dependence of the uncharged TFD of a real scalar given in
\cite{TFD}. In particular, note that the late time complexity becomes
arbitrarily large as $\beta m_\rf$ is taken to be very small or very large. This
discrepancy with \cite{TFD} is due to the fact that the `$LR$' basis used for
the real scalar there is more analogous to the $LR_\Cbb$ basis of the original
complex operators here. On the other hand, the complexity evaluated in the
$LR_\Cbb$ basis (figure \ref{fig:IDonTKnow}) recovers $\sqrt{2}$ times the time dependence of
the complexity of one uncharged TFD,\footnote{We have consistently used $d=4$
  across all our time-dependence plots. We have, however, separately verified
  that the $d=2$ analogue of the right panel of figure \ref{fig:caprese}, upon
  subtracting off the initial value, matches figure 16 of \cite{TFD} with an
  additional factor of $\frac{1}{\sqrt{2}}$, see subsection \ref{sec:kaboom} and
  footnote \ref{foot:doge}. Recall that the entropy of the complex scalar here
  is double that of the real scalar in \cite{TFD}.} as given in the `$LR$' basis
of \cite{TFD}. In particular, we see that taking extreme values of $\beta m_\rf$
gives a finite limit for the complexity at all times in the $LR_\Cbb$ basis.

In figures \ref{fig:buschetta}-\ref{fig:acquacotta} and
\ref{fig:Belle}-\ref{fig:acquacotta2}, we plot the time dependence of the
complexity in the $LR$ and $LR_\Cbb$ bases respectively for general values of
$|\mu|\le m$ and for various values of $\beta m>0$. Moving between the
subfigures corresponding to different $\beta m$, we see that the complexity
develops oscillations with a frequency proportional to $m$. This is naively to
be expected: in order for the integral over single-mode (vacuum-subtracted)
complexities to be convergent, the contribution of high-frequency modes must
necessarily be suppressed. Hence, we expect the oscillations of the total
(vacuum-subtracted) complexity to result from modes of low frequency, which are bounded from below
by $\omega_{k=0}=m$.

In general, we note that, in the $LR$ basis, the complexity plotted in figures
\ref{fig:whyAmISad} and \ref{fig:buschetta}-\ref{fig:acquacotta}
always initially increases and peaks at a global maximum, never drops below its
initial value, and always saturates to a value fairly close to its global
maximum. In the $LR_\Cbb$ basis, on the other hand, the complexity does not
typically stay above its initial value for all times and indeed sometimes
saturates below its initial value, as shown in figures \ref{fig:IDonTKnow} and \ref{fig:Belle}-\ref{fig:acquacotta2}, in contrast to holographic
complexity.

Similarly to what was found for the uncharged TFD in \cite{TFD}, we observe that
the time dependence of complexity deviates significantly compared to the results
in the holographic systems. In fact, the complexity for our Gaussian cTFD
exhibits damped oscillations around some final value after times of the order of the inverse temperature. This is
perhaps not surprising, since the free systems we consider here are not chaotic.

In the left panels of figures \ref{fig:acquacotta}, \ref{fig:Belle} and \ref{fig:acquacotta2}, we observe that, in
both bases, the time dependence of the (unnormalized) complexity decreases in the low temperature limit (moving between subfigures),\footnote{We expect that this will be the case for values of $\beta m$
  larger than the value for which the unnormalized complexity of formation in the left panels of figures
  \ref{fig:smokedSalmonBagel}-\ref{fig:bologonese} starts decreasing. For
    instance, going from the left panel of figure \ref{fig:stracciatella} to the left panel of \ref{fig:fonduta}, we
    see that only the blue, yellow and green curves appear to
    decrease as $\beta m$ goes from $2$ to $8$, cf.~figure \ref{fig:minestraDiCeci}. Moreover, the change in complexity compared to its initial value in the limiting $m=\mu$ case does not appear to decrease with
    temperature, as shown by the top curves in the left panels of figures
    \ref{fig:acquacotta} and \ref{fig:Belle}, as well as the right panels of
    figure \ref{fig:acquacotta2} (this last case was not included in the left plot because of the very different orders of magnitude and it has to be rescaled with a factor of $\Delta\cmplx^{LR_\Cbb}(t=0)$, cf.~the left panel of figure \ref{fig:salsiccia}, which increases with $\beta m$).} keeping all the other parameters fixed, for all
$|\mu|<m$. This effect
is similar to the one observed in holography where the rate of computation comes
to a halt as the temperature decreases. A precise comparison of the rate at
which computation stops with the decrease of temperature is numerically
challenging and we leave it for the future. Further, moving between the
  curves in each of the right panels, we observe that the amplitude of the
fluctuations in complexity as a function of time, scaled by the initial complexity, increase as the chemical
potential decreases.

%%% Local Variables:
%%% mode: latex
%%% TeX-master: "../ChargedComplexityV7"
%%% End:

\section{Comparison with Holographic Results}\label{sec:holo}
As discussed in section \ref{sec:intro}, there are several conjectures for the
gravitational dual of complexity in holography and it is expected that the
qualitative comparison of free field theory results to those obtained from
gravity will provide hints toward a good definition of complexity on the field
theory side, applicable to the dual field theory of holographic systems.
Therefore, in this section, we perform such a comparison. We start by reviewing
known holographic results for the complexity of cTFDs for QFTs with holographic
duals. We then compare those to our results in the previous section.

In holography, the cTFD state is dual to a charged eternal black hole, see, \eg
\cite{Chamblin:1999tk,Hartnoll:2009sz,Andrade:2013rra,Kundu:2016dyk}. The
complexity in this background was studied in holography in
\cite{Brown2,Carmi:2017jqz} and we provide here a brief summary of those
results.

The complexity of formation in the limit of small chemical potential $\mu/T \ll
1$ was studied for four boundary spacetime dimensions and was found to
obey\footnote{In the result for the CA proposal, we have neglected a term of
  order $(\mu/T)^3$. The coefficient of this term was found in
  \cite{Carmi:2017jqz} to depend on the normalization constant $\alpha$ of the
  null-normals to the boundaries of the WDW patch. Our recent understanding,
  however, is that one should include a counter term \cite{Lehner:2016vdi} to
  remove this parametrization-dependence on the null-normals of the WDW patch.
  It would be interesting to examine the effect of this addition. In the CV
  result, we have neglected a term of order $(\mu/T)^4$. In the setup of
  \cite{Carmi:2017jqz}, the chemical potential was assumed to be positive;
  however, since the definition of the cTFD state is invariant under changing
  $\mu\rightarrow-\mu$ as well as interchanging the left and right sides, we
  know that the result for negative chemical potential should take the same form
  with $\mu \rightarrow |\mu|$ in all the equations below.}
\begin{equation}
  \begin{split}
    \Delta \mC_A & = S \left(c_1+c_2 \left(\frac{\mu}{T}
      \right)^3\ln\left(\frac{\mu}{T}\right) + \ldots\right),\qquad \Delta \mC_V
    = S \left(\tilde c_1+\tilde c_2 \left(\frac{\mu}{T} \right)^2 +
      \ldots\right),
  \end{split}
\end{equation}
where $c_{1,2}$ and $\tilde c_{1,2}$ are order one positive constants, $S$ is
the thermal entropy of each side (or the entanglement entropy between the two
copies) and the subscripts $A$ and $V$ refer to the CA and CV proposals (see
section \ref{sec:intro}), respectively. These expressions reproduce the neutral
results (obtained for neutral eternal black holes) in the limit $\mu\rightarrow
0$. While this statement may seem obvious, checking it in holography is
non-trivial, since neutral and charged eternal black holes have very different
causal structures.

Furthermore, the complexity of formation of charged eternal black holes was
found to be divergent in the opposite limit when the temperature is sent to
zero, \ie for $T/\mu\ll 1$ (using both the CV and the CA proposals)\footnote{In
  these expressions, we have neglected a constant term. For the case of the CA
  complexity, this constant term depends on the normalization constant $\alpha$
  of the null normals. It would be interesting to examine the effect of
  including the counter term restoring reparametrization invariance
  \cite{Lehner:2016vdi} on this constant. Note that this result is not valid for
  the case of spherical black holes with chemical potential below a certain
  critical value. In this case, for a sufficiently small chemical potential,
  pure AdS with a background gauge field is recovered in the zero temperature
  limit rather than an extremal black hole. As a consequence, the entanglement
  entropy and the complexity of formation, using both CV and CA, vanish.}
\begin{equation}
  \begin{split}
    \Delta \mC_{A,V} \sim S \ln\left(\frac{\mu}{T} \right) + \ldots
  \end{split}
\end{equation}
where again the proportionality coefficients in these relations are order one
positive constants. This curious behaviour is a result of an IR divergence due to
the infinitely long throat of the wormhole in the extremal (zero-temperature)
limit. The interpretation suggested in \cite{Carmi:2017jqz} was that of a
``third law of complexity'',\footnote{Note that the third law of thermodynamics
  takes a slightly different form for the free theory compared to the
  holographic system; the value of the entropy approached at zero temperature is
  zero for the free theory, while the entropy approaches a finite constant for
  the zero temperature limit of the holographic system.} namely that cTFD states
at finite chemical potential and zero temperature are infinitely more complex
compared to their finite temperature counterparts and cannot be formed by any
physical process during a finite amount of time. Note that $S$ remains finite as
$T\to 0$ in the holographic setup.

A number of results concerning the time dependence of complexity for charged
eternal black holes are also available \cite{Brown2,Carmi:2017jqz}. Just like
for neutral eternal black holes, also here the complexity exhibits a linear
increase at late times. The authors of \cite{Brown2} have proposed that the late
time rate of change in the CA complexity should obey a modified version of
Lloyd's bound
\begin{equation}
  \frac{d \mC_A}{dt} \leq \frac{2}{\pi}\left[(M-\mu C)-(M-\mu C)|_{\mathrm{gs}}\right],
\end{equation}
where $M$ is the black hole mass and $C$ is its charge and where the subscript
ground state (gs) indicates the thermodynamic quantities associated to the state
minimizing $(M-\mu C)$ for a given value of the chemical potential.\footnote{For
  the case of spherical black holes with small chemical potential, this is the
  vacuum state with a constant gauge field while, for spherical black holes of
  larger the chemical potential or planar and hyperbolic block holes, this is an
  extremal black hole at the same value of $\mu$ \cite{Brown2,Carmi:2017jqz}.}
It was found that large or intermediate charged black holes\footnote{\ie charged
  black holes whose horizon radius is much larger, or of the same order as the
  AdS scale.} violate this bound at late times, while small black holes only
exhibit smaller violations in the approach to late times. In any event, the time
evolution of complexity for charged black holes was found to behave more
regularly than the neutral time evolution using the CA proposal. It was also
found to approach the correct uncharged limit using both the CA and CV
proposals. Furthermore, using both CV and CA, holographic complexity was found
to grow quadratically $\cmplx \sim t^2$ at early times due to the time-reversal
symmetry of the charged black hole geometry.\footnote{The charged case smoothly
  interpolates to the neutral case, where CA interestingly gives a neighbourhood
  around $t=0$ where the rate of change in holographic complexity is exactly
  zero, then discontinuously becomes infinitely negative, only to quickly grow
  to a positive value in a short time
  \cite{Carmi:2017jqz}.\label{foot:willIDieAlone}} Finally, the rate of change
in complexity was found to vanish at all times in the extremal (vanishing
temperature) limit using both the CV and the CA complexity proposals.

Our analysis of the cTFD in the previous section resulted in the following
features which can be tested for consistency and qualitatively compared to the
holographic results. First, all our results obey the symmetry $\mu \rightarrow
-\mu$ of the cTFD state. Furthermore, for the complexity of formation in both
the $LR$ and $LR_\Cbb$ bases as well as for time evolution in the $LR_\Cbb$
basis, we were able to demonstrate that the neutral results are recovered upon
setting the chemical potential to zero, as is the case for holographic systems.
On the other hand, we found that the complexity of formation increases with the
chemical potential, see figures \ref{fig:shepardspie}, \ref{fig:shepardspie2},
\ref{fig:minestraDiCeci} and \ref{fig:salsiccia}. This is unlike what happens in
holography, cf.~figure 26 in \cite{Carmi:2017jqz}. Figures
\ref{fig:shepardspie}, \ref{fig:shepardspie2}, \ref{fig:minestraDiCeci} and
\ref{fig:salsiccia}, further supported by an analytic low-temperature expansion,
also demonstrate that only in the special case $m=|\mu|$ do we possibly have a
``third law of complexity'' for our simple scalar system --- that is, when
$m=|\mu|$, the complexity of formation appears to diverge with decreasing
temperature. Note that, in the case of the complex scalar, we have to keep the
relation $|\mu|\leq m$ and hence we never really approach the conformal limit at
fixed chemical potential, which would be the most relevant limit for comparison
with holography.

We further evaluated the time dependence of complexity and observed that the
maximal (and average) change from the original value of the complexity increases
with the chemical potential. Furthermore, we observe that this change decreases
as the temperature decreases (except possibly for the limiting case $|\mu|=m$).
This can be seen from figures \ref{fig:acquacotta} and \ref{fig:acquacotta2}.
This effect is similar to the holographic observation that the rate of change in
complexity vanishes when we approach the zero-temperature limit, \ie the
computation comes to a halt in this limit.

Similarly to previous works in the neutral case \cite{TFD}, we find that the
time dependence of complexity deviates significantly compared to the results in
the holographic systems. This is perhaps not surprising, since the free systems
we consider do not enjoy the chaotic behaviour expected of holographic systems.
Specifically, we find that complexity exhibits rapid changes followed by damped
oscillations after a time of order the inverse temperature $\beta$, and seems to
converge towards some final value either below or above the initial complexity.
Furthermore, we find that the frequency of the damped oscillations scales
approximately linearly with the field mass $m$.

It is easily verified (at least numerically) that the complexity of the complex
scalar cTFD presented here is time-reversal-invariant, as in the holographic
complexity conjectures. However, the complexity of the complex scalar cTFD does
not have a vanishing rate of change in the early time limit; rather, it appears
to vary linearly as $\sim |t|$, as in the uncharged case \cite{TFD}. We would
like to suggest that the contrast between this and the quadratic growth of
holographic complexity may be due to the orientation of the basis of elementary
gates chosen for evaluating the $F_1$ complexity.

\begin{figure}[h]
  \centering \includegraphics[width=0.6\textwidth]{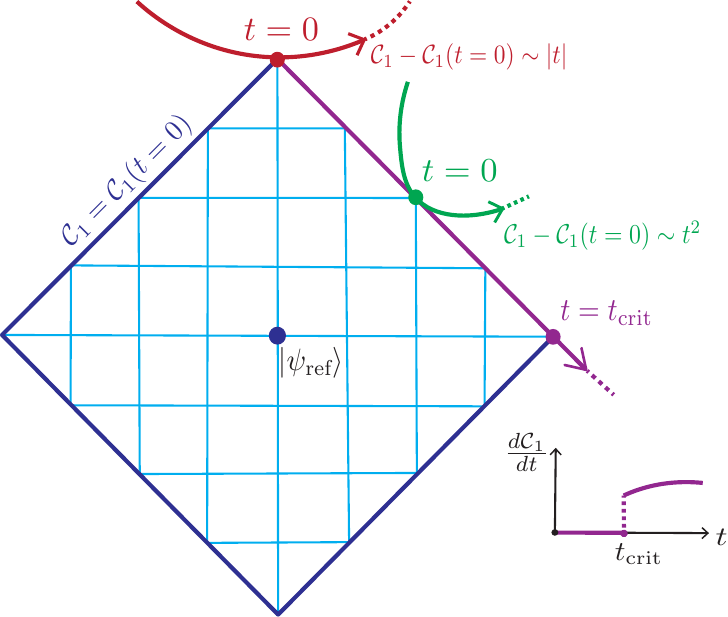}
  \caption{Illustration of the $\cmplx_1$ complexity
    measuring distance from a reference state $|\psi_\rf\rangle$ and various
    trajectories of target states under time evolution. (While points in this
    figure correspond to states, the notion of ``space'' here is borrowed from
    the circuit group manifold.) Here, the action of a set of commuting
    elementary gates induce motion in the north-south and east-west directions
    along paths shown in light blue. The dark blue diamond marks a
    surface of constant $\cmplx_1$ from the reference state. Taking the target
    state to lie on this diamond at the initial time $t=0$, we show three
    possible time-reversal-invariant trajectories: a trajectory touching the
    diamond at $t=0$ non-tangentially, giving a $\sim|t|$ growth in complexity
    (red); a trajectory tangential to and touching the diamond (locally) only at $t=0$,
    producing a quadratic growth in complexity (green); and a trajectory which
    runs along a face of the diamond until a critical time $t_\crit$ when it
    smoothly continues past the boundary of the face, producing a discontinuous
    rate of complexity growth shown in the inset plot (purple).}
  \label{fig:manhatten}
\end{figure}

For visual simplicity, let us consider a subset of elementary gates which
mutually commute. Naively considering these gates alone, the $F_1$ complexity
may be visualized as measuring lengths along grid lines denoting the action of
these gates --- see figure \ref{fig:manhatten}. In this picture, states of fixed
$F_1$ complexity relative to the reference state live on a `diamond' centered on
the reference state with faces\footnote{In the case illustrated in figure
  \ref{fig:manhatten} with two commuting elementary gates which act
  non-trivially on the reference state, the ``faces'' of the diamond are
  one-dimensional segments.} running diagonally with respect to the grid. Our
result $\mathcal{C}_1-\mathcal{C}_1(t=0)\sim |t|$ then suggests
that we have chosen elementary gates, so that the initial time evolution of the
complex scalar cTFD has a component orthogonal to such a diamond. In particular,
requiring `smooth' time evolution and time-reversal invariance around $t=0$,
this analysis would seem to require a trajectory of the cTFD under time
evolution to intersect the diamond in the non-tangential manner shown by the red
curve of figure \ref{fig:manhatten}. On the other hand, the initial quadratic
growth of holographic complexity $\mathcal{C}_1-\mathcal{C}_1(t=0)\sim
  t^2$ (if it is indeed identified with an $F_1$ type complexity)
seems to indicate a time evolution that is initially tangential to the
diamond, resembling the green curve. This picture may also partially provide a mechanism for the
discontinuous rate of change in the CA holographic complexity of neutral black
holes described by footnote \ref{foot:willIDieAlone} --- see purple case in
figure \ref{fig:manhatten}.\footnote{However, note that an explanation for the brief transient period where the complexity decreases, cf.~footnote \ref{foot:willIDieAlone}, is still lacking.}

While we have discussed commuting elementary gates for simplicity, we expect
qualitative features of this discussion to also hold for a complete
non-commuting set of elementary gates; for instance, we expect each surface of
constant complexity to take the shape of a diamond with possibly warped faces and with vertices obtained by the action of
single elementary gates on the reference state. It would be interesting
to check whether an alternative set of gates to those chosen in this paper might
be able to reproduce the quadratic growth exhibited by holographic complexity
--- we leave this for future work.\footnote{Note that this discussion is
  reminiscent of the `first law of complexity' \cite{Bernamonti:2019zyy} where
  variations of the target state tangential to surfaces of equal complexity lead
  to a vanishing first order variation in complexity. In
  \cite{Bernamonti:2019zyy,Bernamonti:2020bcf}, the normal to such a surface is interpreted as the
  `momentum' of the optimal circuits. The leading variation of
  complexity is then of first (second) order
     in a perturbation of
  the target state which is parallel
    (orthogonal) to this momentum.}

%%% Local Variables:
%%% mode: latex
%%% TeX-master: "../ChargedComplexityV6"
%%% End:

\section{Summary and Outlook}\label{sec:summaryoutlook}

In this paper, we presented a systematic framework for studying the complexity in
charged system using the idea of complex phase space operators. We demonstrated
how to use this machinery for the charged thermofield double state of two copies
of a complex scalar field theory. We discovered that, by an appropriately chosen
change of coordinates, the state can be brought to the form of a product of two
uncharged thermofield double states at shifted temperatures and times. We
evaluated, numerically and analytically in certain limits, the complexity of
formation of the cTFD state and demonstrated that this system does not obey, in
general, a third law of complexity. Hence we suggest that this rule might not
hold generally outside the holographic setup. Furthermore, we presented
numerical evaluations of the time dependence of the complexity and explored the effect of
temperature and chemical potential. We found that the rate of change in
complexity vanishes in the limit of vanishing temperature and at finite chemical
potential, similarly to what happens in holographic systems. Finally,
qualitatively drawing upon the early-time linear growth of complexity for the
complex scalar cTFD and the quadratic growth of complexity for holographic cTFDs, we
have speculated on the orientation of the early time evolution of these
states relative to the elementary gates used to compute complexity.

There are a number of open questions which we leave for future studies. One
direction is to conduct an analogous study of the complexity of cTFDs for fermions, which provide a setup more relevant for experimental studies.
In
particular, it would be interesting to explore if, in this setting, one can probe
the conformal limit. Other obvious generalizations include studying the same
problem using different measures of complexity, using \eg different cost functions, the FS metric, or including penalty factors. Furthermore, it would be interesting
to perform an analysis of the entanglement entropy in the cTFD state,
similar to the one that was conducted in \cite{TFD} for the uncharged TFD, and
compare the dependence on the various parameters --- size of the subregion,
chemical potential and temperature --- to that found for holographic systems
\cite{Andrade:2013rra,Kundu:2016dyk}. It would be interesting to explore if the
factorization of each cTFD mode to two uncharged TFDs can be used to define a MERA tensor network
prescription \cite{Vidal_2007,Vidal_2008,vidal2009entanglement} for constructing
cTFD states. In particular, one may consider combining ideas similar to those proposed in the context of the
uncharged TFD state, \eg see comments in \cite{evenbly2015tensor} and appendix E
of \cite{Formation}, with ideas on the incorporation of symmetries in
MERA tensor networks, as in \cite{Singh:2013sda}. Finally, it would be
interesting to evaluate the complexity in the presence of non-abelian charges.

%%% Local Variables:
%%% mode: latex
%%% TeX-master: "../ChargedComplexityV6"
%%% End:

\acknowledgments
\begin{samepage}
We would like to thank Adam Chapman, Lucas Fabian Hackl, Adrian Franco-Rubio, Ben Freivogel, Timothy H.~Hsieh, Rob C. Myers and Giuseppe Policastro for useful correspondence, comments and discussions. The work by HZC was done as part of the Perimeter Scholars International master's program at Perimeter Institute. Research at Perimeter Institute is supported in part by the Government of Canada through the Department of Innovation, Science and Economic Development Canada and by the Province of Ontario through the Ministry of Economic Development, Job Creation and Trade. SC acknowledges funding from the European Research Council (ERC) under starting grant No. 715656 (GenGeoHol) awarded to Diego M. Hofman
and consolidator grant QUANTIVIOL awarded to Ben Freivogel.
\end{samepage}

%\pagebreak

\appendix

\section{Basis of $\ssp(2N,\Rbb)$ for Real/Complex Phase Space Operators}
\label{sec:deadline}
Supplementing the discussion in section \ref{sec:covprel}, we present here
expressions for bases of $\ssp(2N,\Rbb)$ used for real and complex phase space
operators.

Let us begin with real phase space operators \eqref{eq:oink}:
\begin{align}
  \xi_{\mathbb{R}} = (q_1,\ldots,q_N,p_1,\ldots,p_N).
  \label{eq:escitalopram}
\end{align}
For these operators, we will use the same basis of generators as in \cite{TFD}.
Namely, written in terms of the operators $\hat{K}$ from
eq.~\eqref{Qgenerators}, we have
\begin{align}
  \hat{W}_{\Rbb ij}
  =& \frac{1}{2}\{q_i,p_j\},
  &
    (1\le i,j\le N)
  \\
  \hat{V}_{\Rbb ij}
  =& \begin{cases}
    \frac{q_i^2}{\sqrt{2}} & \text{if $i=j$} \\
    q_i q_j & \text{if $i<j$}
  \end{cases} , & (1\le i\le j \le N)
  \\
  \hat{Z}_{\Rbb ij} =& \begin{cases}
    \frac{p_i^2}{\sqrt{2}} & \text{if $i=j$} \\
    p_i p_j & \text{if $i<j$}
  \end{cases}, & (1\le i\le j \le N)
                 \label{eq:bored}
\end{align}
where, $\{q,p\}\equiv qp+pq$ denotes symmetrization. These provide a choice of
elementary gates which each contain as few operators as possible, while being
Hermitian and bilinear in the operators. In terms of the $2N\times 2N$ matrices
$k$, related to the above by eq.~\eqref{Qgenerators}, we have
\begin{align}
  k_{ab}(\hat{W}_{\Rbb ij})
  =& \delta_{ai} \delta_{b,j+N} + \delta_{bi} \delta_{a,j+N},
  &
    (1\le i,j\le N)
  \\
  k_{ab}(\hat{V}_{\Rbb ij})
  =& (\delta_{ai}\delta_{bj} + \delta_{bi}\delta_{aj})\begin{cases}
    \frac{1}{\sqrt{2}} & \text{if $i=j$} \\
    1 & \text{if $i<j$}
  \end{cases} , & (1\le i\le j \le N)
  \\
  k_{ab}(\hat{Z}_{\Rbb ij}) =& (\delta_{a,i+N}\delta_{b,j+N} +
                               \delta_{b,i+N}\delta_{a,j+N}) \begin{cases}
                                 \frac{1}{\sqrt{2}} & \text{if $i=j$} \\
                                 1 & \text{if $i<j$}
                               \end{cases}. & (1\le i\le j \le N)
\end{align}
Here and below, $1\le a,b\le 2N$ are matrix indices. Finally, we may multiply
the above by $\Omega$, as in eq.~\eqref{Kuseful}, to obtain the
corresponding $K$:
\begin{align}
  \tensor{(W_{\Rbb ij})}{^a_b}
  =& -\tensor{\delta}{^a_{i+N}} \delta_{b,j+N} + \delta_{bi} \tensor{\delta}{^a_j},
  &
    (1\le i,j\le N)
  \\
  \tensor{(V_{\Rbb ij})}{^a_b}
  =& -(\tensor{\delta}{^a_{i+N}}\delta_{bj} + \delta_{bi}\tensor{\delta}{^a_{j+N}})\begin{cases}
    \frac{1}{\sqrt{2}} & \text{if $i=j$} \\
    1 & \text{if $i<j$}
  \end{cases} , & (1\le i\le j \le N)
  \\
  \tensor{(Z_{\Rbb ij})}{^a_b} =& (\tensor{\delta}{^a_i}\delta_{b,j+N} +
                                  \delta_{b,i+N}\tensor{\delta}{^a_j}) \begin{cases}
                                    \frac{1}{\sqrt{2}} & \text{if $i=j$} \\
                                    1 & \text{if $i<j$}
                                  \end{cases}. & (1\le i\le j \le N)
\end{align}
In tables \ref{tab:chirp} and \ref{tab:citalopram}, we list as an example the
$\hat{K}$, $k$, and $K$ representations of $W$, $V$, and $Z$ for real phase
space operators in the case of $N=2$, \ie $\ssp(4,\Rbb)$, where we have chosen
$\xi_\Rbb=(q,\bar{q},p,\bar{p})$ as the basis of operators. In $\ssp(8,\Rbb)$,
one repeats the exact same process to obtain the elementary generators written
in terms of the real phase space operators in eq.~\eqref{eq:defLRbasis}:
\begin{equation}
  \xi^{LR}  \equiv \begin{bmatrix}
    q_{L}, & \bar{q}_{L}, & q_{R}, & \bar{q}_{R}, &
    p_{L}, & \bar{p}_{L}, &p_{R}, & \bar{p}_{R}
  \end{bmatrix},
\end{equation}
in which we are interested in the main text.

\begin{table}
  \centering
  \begin{tabular}{|c|cccc|}
    \hline
    & $W_{\Rbb qp}$ & $W_{\Rbb q\bar{p}}$ & $W_{\Rbb \bar{q}p}$ & $W_{\Rbb \bar{q}\bar{p}}$ \\
    \hline
    \rule{-3pt}{\normalbaselineskip}
    $\hat{K}$ & $\frac{qp+pq}{2}$ & $q\bar{p}$ & $\bar{q}p$ & $\frac{\bar{q}\bar{p}+\bar{p}\bar{q}}{2}$ \\
    \rule{-3pt}{1.5\normalbaselineskip}
    $k$ &  $\left[\begin{smallmatrix}
        0 & 0 & 1  & 0 \\
        0 & 0 & 0  & 0 \\
        1 & 0 & 0 & 0 \\
        0 & 0 & 0  & 0 \\
      \end{smallmatrix}\right]$ & $\left[\begin{smallmatrix}
        0 & 0 & 0 & 1  \\
        0 & 0 & 0 & 0  \\
        0 & 0 & 0 & 0 \\
        1 & 0 & 0 & 0  \\
      \end{smallmatrix}\right]$ & $\left[\begin{smallmatrix}
        0 & 0 & 0  & 0 \\
        0 & 0 & 1  & 0 \\
        0 & 1 & 0  & 0 \\
        0 & 0 & 0 & 0 \\
      \end{smallmatrix}\right]$ & $\left[\begin{smallmatrix}
        0 & 0 & 0 & 0  \\
        0 & 0 & 0 & 1  \\
        0 & 0 & 0 & 0  \\
        0 & 1 & 0 & 0 \\
      \end{smallmatrix}\right]$ \\
    \rule{-3pt}{1.5\normalbaselineskip}
    $K$ & $\left[\begin{smallmatrix}
        1 & 0 & 0  & 0 \\
        0 & 0 & 0  & 0 \\
        0 & 0 & -1 & 0 \\
        0 & 0 & 0  & 0 \\
      \end{smallmatrix}\right]$ & $\left[\begin{smallmatrix}
        0 & 0 & 0 & 0  \\
        1 & 0 & 0 & 0  \\
        0 & 0 & 0 & -1 \\
        0 & 0 & 0 & 0  \\
      \end{smallmatrix}\right]$ & $\left[\begin{smallmatrix}
        0 & 1 & 0  & 0 \\
        0 & 0 & 0  & 0 \\
        0 & 0 & 0  & 0 \\
        0 & 0 & -1 & 0 \\
      \end{smallmatrix}\right]$ & $\left[\begin{smallmatrix}
        0 & 0 & 0 & 0  \\
        0 & 1 & 0 & 0  \\
        0 & 0 & 0 & 0  \\
        0 & 0 & 0 & -1 \\
      \end{smallmatrix}\right]$
    \\[0.75\normalbaselineskip]
    \hline
  \end{tabular}
  \caption{Basis of $W$-type generators in $\ssp(4,\Rbb)$ for
    $\xi_\Rbb=(q,\bar{q},p,\bar{p})$.}
  \label{tab:chirp}
\end{table}

\begin{table}
  \centering
  \begin{tabular}{|c|cccccc|}
    \hline
    & $V_{\Rbb qq}$ & $V_{\Rbb q\bar{q}}$ & $V_{\Rbb \bar{q}\bar{q}}$ & $Z_{\Rbb pp}$ & $Z_{\Rbb p\bar{p}}$ & $Z_{\Rbb \bar{p}\bar{p}}$ \\
    \hline
    \rule{-3pt}{\normalbaselineskip}
    $\hat{K}$ & $\frac{q^2}{\sqrt{2}}$ & $q\bar{q}$ & $\frac{\bar{q}^2}{\sqrt{2}}$ & $\frac{p^2}{\sqrt{2}}$ & $p\bar{p}$ & $\frac{\bar{p}^2}{\sqrt{2}}$ \\
    \rule{-3pt}{1.5\normalbaselineskip}
    $k$ & $\left[\begin{smallmatrix}
        \sqrt{2} & 0 & 0 & 0 \\
        0         & 0 & 0 & 0 \\
        0         & 0 & 0 & 0 \\
        0         & 0 & 0 & 0 \\
      \end{smallmatrix}\right]$ & $\left[\begin{smallmatrix}
        0  & 1 & 0 & 0 \\
        1 & 0  & 0 & 0 \\
        0  & 0  & 0 & 0 \\
        0  & 0  & 0 & 0 \\
      \end{smallmatrix}\right]$ & $\left[\begin{smallmatrix}
        0 & 0         & 0 & 0 \\
        0 & \sqrt{2} & 0 & 0 \\
        0 & 0         & 0 & 0 \\
        0 & 0         & 0 & 0 \\
      \end{smallmatrix}\right]$ & $\left[\begin{smallmatrix}
        0 & 0 & 0        & 0 \\
        0 & 0 & 0        & 0 \\
        0 & 0 & \sqrt{2} & 0 \\
        0 & 0 & 0        & 0 \\
      \end{smallmatrix}\right]$ & $\left[\begin{smallmatrix}
        0 & 0 & 0 & 0 \\
        0 & 0 & 0 & 0 \\
        0 & 0 & 0 & 1 \\
        0 & 0 & 1 & 0 \\
      \end{smallmatrix}\right]$ & $\left[\begin{smallmatrix}
        0 & 0 & 0 & 0        \\
        0 & 0 & 0 & 0        \\
        0 & 0 & 0 & 0        \\
        0 & 0 & 0 & \sqrt{2} \\
      \end{smallmatrix}\right]$
    \\
    \rule{-3pt}{1.5\normalbaselineskip}
    $K$ & $\left[\begin{smallmatrix}
        0         & 0 & 0 & 0 \\
        0         & 0 & 0 & 0 \\
        -\sqrt{2} & 0 & 0 & 0 \\
        0         & 0 & 0 & 0 \\
      \end{smallmatrix}\right]$ & $\left[\begin{smallmatrix}
        0  & 0  & 0 & 0 \\
        0  & 0  & 0 & 0 \\
        0  & -1 & 0 & 0 \\
        -1 & 0  & 0 & 0 \\
      \end{smallmatrix}\right]$ & $\left[\begin{smallmatrix}
        0 & 0         & 0 & 0 \\
        0 & 0         & 0 & 0 \\
        0 & 0         & 0 & 0 \\
        0 & -\sqrt{2} & 0 & 0 \\
      \end{smallmatrix}\right]$ & $\left[\begin{smallmatrix}
        0 & 0 & \sqrt{2} & 0 \\
        0 & 0 & 0        & 0 \\
        0 & 0 & 0        & 0 \\
        0 & 0 & 0        & 0 \\
      \end{smallmatrix}\right]$ & $\left[\begin{smallmatrix}
        0 & 0 & 0 & 1 \\
        0 & 0 & 1 & 0 \\
        0 & 0 & 0 & 0 \\
        0 & 0 & 0 & 0 \\
      \end{smallmatrix}\right]$ & $\left[\begin{smallmatrix}
        0 & 0 & 0 & 0        \\
        0 & 0 & 0 & \sqrt{2} \\
        0 & 0 & 0 & 0        \\
        0 & 0 & 0 & 0        \\
      \end{smallmatrix}\right]$
    \\[0.75\normalbaselineskip]
    \hline
  \end{tabular}
  \caption{Basis of $V$- and $Z$-type generators in $\ssp(4,\Rbb)$ for
    $\xi_\Rbb=(q,\bar{q},p,\bar{p})$.}
  \label{tab:citalopram}
\end{table}

Let us now consider the case of complex phase space operators
\eqref{complex:phasespace}:
\begin{align}
  \xi_{\mathbb{C}} = (q_1,q_1^\dagger,\ldots,q_{N/2},q_{N/2}^\dagger,p_1^\dagger,p_1,\ldots,p_{N/2}^\dagger,p_{N/2}).
\end{align}
For these operators, we will use the following basis for $\ssp(2N,\Rbb)$,
written in terms of the operators $\hat{K}$ from eq.~\eqref{Qgenerators}:
\begin{align}
  \tensor{\hat{W}}{_\Cbb^\ell_{ij}}
  =& \frac{1}{\sqrt{2}} \begin{cases}
    i(\{q_i^\dagger, p_j\} - \{q_i,p_j^\dagger\}) & \text{if $\ell=0$} \\
    i(\{q_i^\dagger, p_j^\dagger\} - \{q_i,p_j\}) & \text{if $\ell=1$} \\
    \{q_i^\dagger, p_j\} + \{q_i,p_j^\dagger\} & \text{if $\ell=2$} \\
    \{q_i^\dagger, p_j^\dagger\} + \{q_i,p_j\} & \text{if $\ell=3$} \\
  \end{cases}, & \left(1\le i,j\le \frac{N}{2}\right)
  \\
  \tensor{\hat{V}}{_\Cbb^\ell_{ij}} =& \begin{cases}
    \frac{1}{2} & \text{if $i=j$} \\
    \frac{1}{\sqrt{2}} & \text{if $i<j$}
  \end{cases}
                         \times
                         \begin{cases}
                           i(\{q_i^\dagger, q_j\} - \{q_i,q_j^\dagger\}) & \text{if $\ell=0$ and $i<j$} \\
                           i(\{q_i^\dagger, q_j^\dagger\} - \{q_i,q_j\}) & \text{if $\ell=1$} \\
                           \{q_i^\dagger, q_j\} + \{q_i,q_j^\dagger\} & \text{if $\ell=2$} \\
                           \{q_i^\dagger, q_j^\dagger\} + \{q_i,q_j\} & \text{if $\ell=3$} \\
                         \end{cases}, & \left(1\le i\le j \le \frac{N}{2}\right)
  \\
  \tensor{\hat{Z}}{_\Cbb^\ell_{ij}} =& \begin{cases}
    \frac{1}{2} & \text{if $i=j$} \\
    \frac{1}{\sqrt{2}} & \text{if $i<j$}
  \end{cases}
                         \times
                         \begin{cases}
                           i(\{p_i^\dagger, p_j\} - \{p_i,p_j^\dagger\}) & \text{if $\ell=0$ and $i<j$} \\
                           i(\{p_i^\dagger, p_j^\dagger\} - \{p_i,p_j\}) & \text{if $\ell=1$} \\
                           \{p_i^\dagger, p_j\} + \{p_i,p_j^\dagger\} & \text{if $\ell=2$} \\
                           \{p_i^\dagger, p_j^\dagger\} + \{p_i,p_j\} & \text{if $\ell=3$} \\
                         \end{cases}. & \left(1\le i\le j \le \frac{N}{2}\right)
\end{align}
These provide the simplest possible set of elementary gates (`simplest' in the
sense described below eq.~\eqref{eq:bored}). The corresponding $2N\times 2N$
matrices $k$, related to the above by eq.~\eqref{Qgenerators}, are
\begin{multline}
  k_{ab}(\tensor{\hat{W}}{_\Cbb^\ell_{ij}}) = \frac{1}{\sqrt{2}}\Bigg(
  i^{1-\lfloor \ell/2 \rfloor} \left[ \delta_{a,2i-1}\delta_{b,2j-1+n+(\ell\bmod
      2)} + \delta_{b,2i}\delta_{a,2j+n-(\ell\bmod 2)} \right]
  \\
  +(-i)^{1-\lfloor \ell/2 \rfloor} \left[ \delta_{a,2i}\delta_{b,2j+n-(\ell\bmod
      2)} + \delta_{b,2i-1}\delta_{a,2j-1+n+(\ell\bmod 2)} \right] \Bigg),
  \\
  \left(\text{$1\le i,j\le \frac{N}{2}$ and $0\le \ell \le 3$}\right)
\end{multline}
\begin{multline}
  % \begin{split}
  k_{ab}(\tensor{\hat{V}}{_\Cbb^\ell_{ij}}) =
  \begin{cases}
    \frac{1}{2} & \text{if $i=j$} \\
    \frac{1}{\sqrt{2}} & \text{if $i<j$}
  \end{cases}
  \times \Bigg( i^{1-\lfloor \ell/2 \rfloor} \left[
    \delta_{a,2i-1}\delta_{b,2j-1+(\ell\bmod 2)} +
    \delta_{b,2i}\delta_{a,2j-(\ell\bmod 2)} \right]
  \\
  +(-i)^{1-\lfloor \ell/2 \rfloor} \left[ \delta_{a,2i}\delta_{b,2j-(\ell\bmod
      2)} + \delta_{b,2i-1}\delta_{a,2j-1+(\ell\bmod 2)} \right] \Bigg),
  \\
  \left(1\le i\le j \le \frac{N}{2} \text{ and }
    \begin{cases}
      1\le \ell\le 3 & \text{if $i=j$} \\
      0\le \ell\le 3 & \text{if $i<j$}
    \end{cases}
  \right)
  % \end{split}
\end{multline}
\begin{multline}
  k_{ab}(\tensor{\hat{Z}}{_\Cbb^\ell_{ij}}) \\
  =
  \begin{cases}
    \frac{1}{2} & \text{if $i=j$} \\
    \frac{1}{\sqrt{2}} & \text{if $i<j$}
  \end{cases}
  \times \Bigg( i^{1-\lfloor \ell/2 \rfloor}\left[
    \delta_{a,2i-1+n}\delta_{b,2j-1+n+(\ell\bmod 2)} +
    \delta_{b,2i+n}\delta_{a,2j+n-(\ell\bmod 2)} \right]
  \\
  +(-i)^{1-\lfloor \ell/2 \rfloor}\left[
    \delta_{a,2i+n}\delta_{b,2j+n-(\ell\bmod 2)} +
    \delta_{b,2i-1+n}\delta_{a,2j-1+n+(\ell\bmod 2)} \right] \Bigg).
  \\
  \left(1\le i\le j \le \frac{N}{2} \text{ and }
    \begin{cases}
      1\le \ell\le 3 & \text{if $i=j$} \\
      0\le \ell\le 3 & \text{if $i<j$}
    \end{cases}
  \right)
\end{multline}
Here, $\lfloor x \rfloor$ denotes the floor function (which rounds $x$ down to
an integer) and $(x \mod 2)$ means $x$ modulo $2$ (taking values $0$ and $1$ for
integer $x$). Finally, multiplying the above by $\Omega$ as in
eq.~\eqref{Kuseful}, we find the corresponding matrices $K$:
\begin{multline}
  % \begin{split}
  \tensor{{(\tensor{W}{_\Cbb^\ell_{ij}})}}{^a_b} = \frac{1}{\sqrt{2}}\Bigg(
  i^{1-\lfloor \ell/2 \rfloor} \left[
    -\tensor{\delta}{^a_{2i-1+n}}\delta_{b,2j-1+n+(\ell\bmod 2)} +
    \delta_{b,2i}\tensor{\delta}{^a_{2j-(\ell\bmod 2)}} \right]
  \\
  +(-i)^{1-\lfloor \ell/2 \rfloor} \left[
    -\tensor{\delta}{^a_{2i+n}}\delta_{b,2j+n-(\ell\bmod 2)} +
    \delta_{b,2i-1}\tensor{\delta}{^a_{2j-1+(\ell\bmod 2)}} \right] \Bigg),
  \\
  \left(\text{$1\le i,j\le \frac{N}{2}$ and $0\le \ell \le 3$}\right)
  % \end{split}
\end{multline}
\begin{multline}
  % \begin{split}
  \tensor{{(\tensor{V}{_\Cbb^\ell_{ij}})}}{_{ab}} = -\begin{cases}
    \frac{1}{2} & \text{if $i=j$} \\
    \frac{1}{\sqrt{2}} & \text{if $i<j$}
  \end{cases}
  \times \Bigg( i^{1-\lfloor \ell/2 \rfloor} \left[
    \tensor{\delta}{^a_{2i-1+n}}\delta_{b,2j-1+(\ell\bmod 2)}
    +\delta_{b,2i}\tensor{\delta}{^a_{2j+n-(\ell\bmod 2)}} \right]
  \\
  +(-i)^{1-\lfloor \ell/2 \rfloor} \left[
    \tensor{\delta}{^a_{2i+n}}\delta_{b,2j-(\ell\bmod 2)}
    +\delta_{b,2i-1}\tensor{\delta}{^a_{2j-1+n+(\ell\bmod 2)}} \right] \Bigg),
  \\
  \left(1\le i\le j \le \frac{N}{2} \text{ and }
    \begin{cases}
      1\le \ell\le 3 & \text{if $i=j$} \\
      0\le \ell\le 3 & \text{if $i<j$}
    \end{cases}
  \right)
  % \end{split}
\end{multline}
\begin{multline}
  \tensor{{(\tensor{Z}{_\Cbb^\ell_{ij}})}}{_{ab}} \\
  =
  \begin{cases}
    \frac{1}{2} & \text{if $i=j$} \\
    \frac{1}{\sqrt{2}} & \text{if $i<j$}
  \end{cases}
  \times \Bigg( i^{1-\lfloor \ell/2 \rfloor}\left[
    \tensor{\delta}{^a_{2i-1}}\delta_{b,2j-1+n+(\ell\bmod 2)} +
    \delta_{b,2i+n}\tensor{\delta}{^a_{2j-(\ell\bmod 2)}} \right]
  \\
  +(-i)^{1-\lfloor \ell/2 \rfloor}\left[
    \tensor{\delta}{^a_{2i}}\delta_{b,2j+n-(\ell\bmod 2)} +
    \delta_{b,2i-1+n}\tensor{\delta}{^a_{2j-1+(\ell\bmod 2)}} \right] \Bigg).
  \\
  \left(1\le i\le j \le \frac{N}{2} \text{ and }
    \begin{cases}
      1\le \ell\le 3 & \text{if $i=j$} \\
      0\le \ell\le 3 & \text{if $i<j$}
    \end{cases}
  \right)
\end{multline}
In tables \ref{tab:fluoxetine} and \ref{tab:escitalopram}, we list these
generators in $\ssp(4,\Rbb)$ in the forms $\hat{K}$, $k$, and $K$. Note in this
case that the subscripts $i,j$ above always take the value $1$ and have
therefore been omitted from the table. On the other hand, for the case of
$\ssp(8,\Rbb)$ in which we are primarily interested in the main text, with
phase space operators given by \eqref{eq:defLRCbasis},
\begin{equation}% \label{eq:defLRCbasis}
  \xi^{LR_\Cbb} \equiv
  \begin{bmatrix}
    \tilde q_{L}, & \tilde q_{L}^\dagger, & \tilde q_{R}, & \tilde
    q_{R}^\dagger, & \tilde p_{L}^\dagger, & \tilde p_{L}, & \tilde
    p_{R}^\dagger, & \tilde p_{R}
  \end{bmatrix},
\end{equation}
the subscripts $i,j$ can take values $1,2$ corresponding to the two sets of
complex phase space operators $\{q_L, q_L^\dagger\}$
and $\{q_R,q_R^\dagger\}$ (and their conjugate momenta). The operators listed in
tables \ref{tab:fluoxetine} and \ref{tab:escitalopram} may then be interpreted
as a subset of the full basis of $\ssp(8,\Rbb)$, upon appending subscript $L$ or
$R$ to the phase space operators. All the
matrix generators in this section are orthonormal with respect to the inner
product \eqref{eq:application}.

\begin{table}
  \centering
  \begin{tabular}{|c|cccc|}
    \hline
    & $\tensor{W}{_\Cbb^0}$ & $\tensor{W}{_\Cbb^1}$ & $\tensor{W}{_\Cbb^2}$ & $\tensor{W}{_\Cbb^3}$ \\
    \hline
    \rule{-3pt}{\normalbaselineskip}
    $\hat{K}$ & $\frac{i(-\tilde q\tilde p-\tilde p\tilde q+\tilde
                q^\dagger \tilde p^\dagger + \tilde p^\dagger \tilde
                q^\dagger)}{2\sqrt{2}}$ & $\frac{i\sqrt{2}(-\tilde q\tilde p^\dagger+\tilde
                                          q^\dagger \tilde p)}{2}$ & $\frac{\tilde q\tilde p+\tilde p\tilde q+\tilde q^\dagger \tilde p^\dagger +
                                                                     \tilde p^\dagger \tilde q^\dagger}{2\sqrt{2}}$ & $\frac{\sqrt{2}(\tilde
                                                                                                                      q\tilde p^\dagger+\tilde q^\dagger \tilde p)}{2}$ \\
    \rule{-3pt}{1.5\normalbaselineskip}
    $k$ & $\frac{1}{\sqrt{2}}\left[\begin{smallmatrix}
        0                   & 0                  & i & 0                  \\
        0                   & 0                  & 0                   & -i \\
        -i & 0                  & 0                   & 0                  \\
        0                   & i & 0                   & 0                  \\
      \end{smallmatrix}\right]$ & $\frac{1}{\sqrt{2}}\left[\begin{smallmatrix}
        0                   & 0                  & 0                  & i \\
        0                   & 0                  & -i & 0                   \\
        0                   & i & 0                  & 0                   \\
        -i & 0                  & 0                  & 0                   \\
      \end{smallmatrix}\right]$ & $\frac{1}{\sqrt{2}}\left[\begin{smallmatrix}
        0                  & 0                  & 1 & 0                   \\
        0                  & 0                  & 0                   & 1 \\
        1 & 0                  & 0                   & 0                   \\
        0                  & 1 & 0                   & 0                   \\
      \end{smallmatrix}\right]$ & $\frac{1}{\sqrt{2}}\left[\begin{smallmatrix}
        0                  & 0                  & 0                   & 1 \\
        0                  & 0                  & 1 & 0                   \\
        0                  & 1 & 0                   & 0                   \\
        1 & 0                  & 0                   & 0                   \\
      \end{smallmatrix}\right]$ \\
    \rule{-3pt}{1.5\normalbaselineskip}
    $K$ & $\frac{1}{\sqrt{2}}\left[\begin{smallmatrix}
        -i & 0                  & 0                   & 0                  \\
        0                   & i & 0                   & 0                  \\
        0                   & 0                  & -i & 0                  \\
        0                   & 0                  & 0                   & i \\
      \end{smallmatrix}\right]$ & $\frac{1}{\sqrt{2}}\left[\begin{smallmatrix}
        0                   & i & 0                  & 0                   \\
        -i & 0                  & 0                  & 0                   \\
        0                   & 0                  & 0                  & -i \\
        0                   & 0                  & i & 0                   \\
      \end{smallmatrix}\right]$ & $\frac{1}{\sqrt{2}}\left[\begin{smallmatrix}
        1 & 0                  & 0                   & 0                   \\
        0                  & 1 & 0                   & 0                   \\
        0                  & 0                  & -1 & 0                   \\
        0                  & 0                  & 0                   & -1 \\
      \end{smallmatrix}\right]$ & $\frac{1}{\sqrt{2}}\left[\begin{smallmatrix}
        0                  & 1 & 0                   & 0                   \\
        1 & 0                  & 0                   & 0                   \\
        0                  & 0                  & 0                   & -1 \\
        0                  & 0                  & -1 & 0                   \\
      \end{smallmatrix}\right]$
    \\[0.75\normalbaselineskip]
    \hline
  \end{tabular}
  \caption{Basis of $W$-type generators in $\ssp(4,\Rbb)$ for
    $\xi_\Cbb=(q,q^\dagger,p^\dagger,p)$.}
  \label{tab:fluoxetine}
\end{table}

\begin{table}
  \centering
  \begin{tabular}{|c|cccccc|}
    \hline
    & $\tensor{V}{_\Cbb^1}$ & $\tensor{V}{_\Cbb^2}$ & $\tensor{V}{_\Cbb^3}$ & $\tensor{Z}{_\Cbb^1}$ & $\tensor{Z}{_\Cbb^2}$ & $
                                                                                                                              \tensor{Z}{_\Cbb^3}$ \\
    \hline
    \rule{-3pt}{\normalbaselineskip}
    $\hat{K}$ & $\frac{i(-\tilde q^2+(\tilde
                q^\dagger)^2)}{2}$ & $\tilde q\tilde q^\dagger$ & $\frac{\tilde q^2+(\tilde q^\dagger)^2}{2}$ & $\frac{i(\tilde p^2-(\tilde p^\dagger)^2)}{2}$ & $\tilde p\tilde
                                                                                                                                                                 p^\dagger$ & $\frac{\tilde p^2+(\tilde p^\dagger)^2}{2}$ \\
    \rule{-3pt}{1.5\normalbaselineskip}
    $k$ & $\left[\begin{smallmatrix}
        0 & i & 0 & 0 \\
        -i & 0  & 0 & 0 \\
        0 & 0  & 0 & 0 \\
        0 & 0  & 0 & 0 \\
      \end{smallmatrix}\right]$ & $\left[\begin{smallmatrix}
        1 & 0  & 0 & 0 \\
        0  & 1 & 0 & 0 \\
        0  & 0  & 0 & 0 \\
        0  & 0  & 0 & 0 \\
      \end{smallmatrix}\right]$ & $\left[\begin{smallmatrix}
        0  & 1 & 0 & 0 \\
        1 & 0  & 0 & 0 \\
        0  & 0  & 0 & 0 \\
        0  & 0  & 0 & 0 \\
      \end{smallmatrix}\right]$ & $
                                  \left[\begin{smallmatrix}
                                      0 & 0 & 0  & 0 \\
                                      0 & 0 & 0  & 0 \\
                                      0 & 0 & 0  & i \\
                                      0 & 0 & -i & 0 \\
                                    \end{smallmatrix}\right]$ &
                                                                $\left[\begin{smallmatrix}
                                                                    0 & 0 & 0 & 0 \\
                                                                    0 & 0 & 0 & 0 \\
                                                                    0 & 0 & 1 & 0 \\
                                                                    0 & 0 & 0 & 1 \\
                                                                  \end{smallmatrix}\right]$
    &
      $\left[\begin{smallmatrix}
          0 & 0 & 0 & 0 \\
          0 & 0 & 0 & 0 \\
          0 & 0 & 0 & 1 \\
          0 & 0 & 1 & 0 \\
        \end{smallmatrix}\right]$
    \\
    \rule{-3pt}{1.5\normalbaselineskip}
    $K$ & $\left[\begin{smallmatrix}
        0 & 0  & 0 & 0 \\
        0 & 0  & 0 & 0 \\
        0 & -i & 0 & 0 \\
        i & 0  & 0 & 0 \\
      \end{smallmatrix}\right]$ & $\left[\begin{smallmatrix}
        0  & 0  & 0 & 0 \\
        0  & 0  & 0 & 0 \\
        -1 & 0  & 0 & 0 \\
        0  & -1 & 0 & 0 \\
      \end{smallmatrix}\right]$ & $\left[\begin{smallmatrix}
        0  & 0  & 0 & 0 \\
        0  & 0  & 0 & 0 \\
        0  & -1 & 0 & 0 \\
        -1 & 0  & 0 & 0 \\
      \end{smallmatrix}\right]$ & $
                                  \left[\begin{smallmatrix}
                                      0 & 0 & 0  & i \\
                                      0 & 0 & -i & 0 \\
                                      0 & 0 & 0  & 0 \\
                                      0 & 0 & 0  & 0 \\
                                    \end{smallmatrix}\right]$ &
                                                                $\left[\begin{smallmatrix}
                                                                    0 & 0 & 1 & 0 \\
                                                                    0 & 0 & 0 & 1 \\
                                                                    0 & 0 & 0 & 0 \\
                                                                    0 & 0 & 0 & 0 \\
                                                                  \end{smallmatrix}\right]$
    &
      $\left[\begin{smallmatrix}
          0 & 0 & 0 & 1 \\
          0 & 0 & 1 & 0 \\
          0 & 0 & 0 & 0 \\
          0 & 0 & 0 & 0 \\
        \end{smallmatrix}\right]$
    \\[0.75\normalbaselineskip]
    \hline
  \end{tabular}
  \caption{Basis of $V$- and $Z$-type generators in $\ssp(4,\Rbb)$ for
    $\xi_\Cbb=(q,q^\dagger,p^\dagger,p)$.}
  \label{tab:escitalopram}
\end{table}

%%% Local Variables:
%%% mode: latex
%%% TeX-master: "../ChargedComplexityV6"
%%% End:

\section{Time Evolution of the cTFD State}\label{app:timeevol}
The time evolution in eq.~\eqref{eq:intro:cTFD} might, at first sight, look
strange due to the inclusion of the chemical potential. However, this is easily
understood by coupling the uncharged TFD to an external $U(1)$ gauge field
capturing the effect of the chemical potential
\begin{equation}\label{eq:smoothie}
  A_\mu dx^\mu = \mu dt.
\end{equation}
In holography, this would be interpreted as a turning on a source in the
  boundary field theory dual to a gauge field in the bulk. As a simple example, let us consider the effect of coupling the $U(1)$ gauge
field \eqref{eq:smoothie} to a free complex scalar field
\begin{align}
  \LL
  =& -(D_\mu\phi)^\dagger D^\mu \phi
     -m^2 \phi^\dagger \phi,
  &
    D_\mu
    =& \partial_\mu - iA_\mu,
\end{align}
where we have set the elementary charge to one and the metric is in the mostly
plus convention. Expanding out, we have
\begin{align}
  \LL
  =& -\partial_\mu \phi^\dagger \partial^\mu \phi +i\mu \phi^\dagger \dot{\phi}-i\mu \dot{\phi}^\dagger\phi-(m^2-\mu^2)\phi^\dagger\phi,
\end{align}
which yields the conjugate momenta
\begin{align}
  \pi^\dagger
  =& \frac{\partial \LL}{\partial \dot{\phi}^\dagger}
     = \dot{\phi} -i\mu\phi,
  &
    \pi
    =& \frac{\partial \LL}{\partial \dot{\phi}}
       = \dot{\phi}^\dagger +i\mu\phi^\dagger
\end{align}
and the electric charge density
\begin{align}
  \CC
  =& i(\phi^\dagger\pi^\dagger-\pi\phi).
\end{align}
We thus find that the effect of introducing the coupling to the $U(1)$ gauge
field is to deform the Hamiltonian density by $-\mu \mathscr{C}$:
\begin{align}
  \begin{split}
    \HH =& \dot{\phi}^\dagger \dot{\phi}
    +  \vec \nabla \phi^\dagger \cdot \vec \nabla \phi +(m^2-\mu^2) \phi^\dagger \phi \\
    =& \pi^\dagger \pi +  \vec \nabla \phi^\dagger \cdot \vec \nabla \phi-i\mu \phi^\dagger \pi^\dagger+i\mu\pi\phi + m^2 \phi^\dagger \phi \\
    =& \pi^\dagger \pi+ \vec \nabla \phi^\dagger \cdot \vec \nabla \phi
    +m^2\phi^\dagger \phi - \mu \CC.
  \end{split}
\end{align}
Finally recall from eq.~\eqref{eq:intro:cTFD} that the two copies were selected to have opposite charges such that they are CPT conjugates. The time evolution in eq.~\eqref{eq:intro:cTFD} is obtained by evolving each copy according to $\HH$ above, but with opposite values of the chemical potential, due to (once again) CPT invariance, see discussion below eq.~(2.2) in \cite{Andrade:2013rra}.

%%% Local Variables:
%%% mode: latex
%%% TeX-master: "../ChargedComplexityV6"
%%% End:

\section{Derivation of the Low Temperature Limits}\label{app:lowTlowT}
In this appendix, we present the derivation of the low temperature limits for
the complexity and entropy in eqs.~\eqref{eq:zirkova1}-\eqref{eq:Zooey2}. It
will be helpful in what follows to work in terms of a set of dimensionless variables $x,y,u,\bar{\gamma}$, defined by
\begin{align}
  x \equiv& \beta m,
  &
    y \equiv& \beta \mu,
  &
    u \equiv& \beta k,
  &
    \bar{\gamma}
    \equiv& \frac{1}{\beta m_\rf}.
            \label{eq:quinoa}
\end{align}
In terms of these dimensionless variables, the low temperature limit is given by $x \gg 1$ and $y \gg 1$.
Without loss of generality, we focus on positive chemical potentials $y>0$ (cf.~comment at the end of subsection \ref{sec:argh}). We will see that further assuming $x-y \gg 1$ allows us to derive the low temperature limits for complexity analytically.

First, note that, for large $x$, $\alpha_\Lr$, which is defined in eq.~\eqref{eq:bass} using the modified temperature
\eqref{eq:earlgrey}, is exponentially suppressed for all values of $u$
\begin{align}
  \alpha_\Lr \approx & \exp\left\{-\frac{1}{2}g(u,x,-y)\right\} \ll 1, \qquad(\text{$y\ge 0$ and $x\gg 1$}),
                       \label{eq:poBoy}
\end{align}
where $g$ is defined in eq.~\eqref{eq:hummus}.
For $x-y \gg 1$\footnote{Note that this does not necessarily imply $x\gg y$. For
  example $x=2y$ can satisfy $x-y \gg 1$ but not $x\gg y$.} (\eg when there is
a finite difference between $m$ and $\mu$ while $\beta$ is large), $\alpha_\lR$
is also suppressed for all values of $u$
\begin{align}
  \alpha_\lR \approx & \exp\left\{-\frac{1}{2}g(u,x,y)\right\} \ll 1, \qquad (\text{$x-y\gg 1$}).
                       \label{eq:smokeHouse}
\end{align}
When $y\gg 1$, the suppression in \eqref{eq:poBoy} is far stronger than the
suppression in \eqref{eq:smokeHouse}, so $\alpha_\Lr\ll \alpha_\lR$. Utilizing
the diagonalized form of the relative covariance matrix at $t=0$ obtained in
appendix \ref{sec:cv}, an analytic expression for the generator
\eqref{eq:piano2} of the straight line circuit for the cTFD can be
obtained. Taking $\alpha_\Lr\to 0$ and expanding to linear order in
$\alpha_\lR$, we find, for a single mode of the cTFD,
\begin{align}
  \begin{split}
    \MoveEqLeft[3]\cmplx_1^{LR}(\cTFD \text{ 1 mode}) \\
    \approx\, & 2 \left|\log \frac{\lambda_k}{\lambda_\rf}\right| +4\,
    \alpha_\lR\frac{\max\{\lambda_k^2,\lambda_\rf^2\}}{\lambda_k^2-\lambda_\rf^2}
    \log \frac{\lambda_k}{\lambda_\rf}
    ,\\
    \MoveEqLeft[3]\cmplx_1^{LR_\Cbb}(\cTFD \text{ 1 mode}) \\
    \approx& \sqrt{2} \left|\log \frac{\lambda_k}{\lambda_\rf}\right| + \sqrt{2}
    \,\alpha_\lR \left[
      1+\frac{\lambda_k(1+\lambda_\rf^2)}{\lambda_k^2-\lambda_\rf^2} \log
      \frac{\lambda_k}{\lambda_\rf} \right].
  \end{split}
           &
             (\text{$y\gg 1$ and $x-y\gg 1$})
\end{align}
The complexity of (two copies of) the vacuum is obtained in the
limit\footnote{Note that, when $m>\mu$, the $\beta\to\infty$ limit of the cTFD
  \eqref{eq:chocolate} of each complex scalar mode reduces to the vacuum
  $|0,0\rangle_L |0,0\rangle_R$; in the same limit, $\alpha_{\lR}$ and
  $\alpha_{\Lr}$ vanish, see eqs.\eqref{eq:bass} and \eqref{eq:peppermint}-\eqref{eq:earlgrey}.}
$\alpha_\lR,\alpha_\Lr\to 0$ and is proportional to what was found in
\cite{TFD}:
\begin{align}
  \cmplx_1^{LR}(\vac \text{ 1 mode})
  = & 2\left|\log\frac{\lambda_k}{\lambda_\rf}\right|,
  &
    \cmplx_1^{LR_\Cbb}(\vac \text{ 1 mode})
    = & \sqrt{2}\, \left|\log\frac{\lambda_k}{\lambda_\rf}\right|,
\end{align}
where the proportionality factors of 2 in the $LR$ basis and $\sqrt{2}$ in the
$LR_\Cbb$ basis were explained in subsection \ref{sec:kaboom}.\footnote{In fact,
  the vacuum complexity allows us to explain more explicitly what we meant in
  subsection \ref{sec:kaboom} by saying that the relative factor of $1/\sqrt{2}$ in
  the $LR_\Cbb$ basis is due to the fact that straight line circuit is better
  aligned with the elementary gates of this basis; indeed, the straight line
  circuit for the vacuum is generated by two elementary gates, as opposed to the
  four used in the $LR$ basis. Specifically, in the $LR_\Cbb$ basis, the
  straight line generator $\hat{K}$ for the vacuum is given by
  $-\frac{1}{\sqrt{2}}\log\frac{\lambda_k}{\lambda_\rf}$ times the sum of the
  $\tensor{W}{_\Cbb^2}$ element in table \ref{tab:fluoxetine} written for $L$
  and $R$. In the $LR$ basis, the straight line generator $\hat{K}$ for the
  vacuum is $-\frac{1}{2}\log\frac{\lambda_k}{\lambda_\rf}$ times the sum of the
  $W_{\Rbb q\bar{p}}$ and the $W_{\Rbb \bar{q}p}$ generators in table \ref{tab:chirp}
  written for both $L$ and $R$. \label{foot:doge}} By subtracting these
vacuum contributions, we find the complexity of formation for each mode
\begin{align}
  \begin{split}
    \Delta \cmplx_1^{LR}(\cTFD\text{ 1 mode}) \approx\,& 4 \,\alpha_\lR
    \frac{\max\{\lambda_k^2,\lambda_\rf^2\}}{\lambda_k^2-\lambda_\rf^2} \log
    \frac{\lambda_k}{\lambda_\rf}, \label{eq:bangalore}
    \\
    \Delta \cmplx_1^{LR_\Cbb}(\cTFD\text{ 1 mode}) \approx\,&
    \sqrt{2}\,\alpha_\lR \left[
      1+\frac{\lambda_k(1+\lambda_\rf^2)}{\lambda_k^2-\lambda_\rf^2}\log
      \frac{\lambda_k}{\lambda_\rf}\right].
  \end{split}
                                                            &
                                                              (\text{$y\gg 1$ and $x-y\gg 1$})
\end{align}
To obtain the total complexity of formation of the $\cTFD$ state, we integrate
over all modes (with $u$ given by \eqref{eq:quinoa}):
\begin{align}
  \Delta \cmplx_1(\cTFD)
  = & \frac{\vol}{\beta^{d-1}} \cdot \frac{\Omega_{d-2}}{(2\pi)^{d-1}} \int_0^\infty du\; u^{d-2} \Delta \cmplx_1(\cTFD\text{ 1 mode}).
      \label{eq:jack}
\end{align}
Note that the convergence of the integral is guaranteed by the exponential
suppression of $\Delta\cmplx_1(\cTFD\text{ 1 mode})$ by $\alpha_\lR$. To
continue, we apply the expansion
\begin{align}
  g(u,x,y)
  = & x-y+\frac{u^2}{2x}+x\, \mathcal{O}\left(\frac{u^4}{x^4}\right)\label{eq:naan}
\end{align}
to $\alpha_{\bar L R}$ in eq.~\eqref{eq:smokeHouse} and use it to approximate
eq.~\eqref{eq:jack}. We find that, in the large $x$ limit, the suppression of
$\Delta\cmplx_1(\cTFD\text{ 1 mode})$ due to $\alpha_\lR$ implies that the
integral receives dominant contributions only when $u$ is of order at most
$\vartheta \sqrt{x}$ with $\vartheta$ a large constant smaller than any positive
power of $x$.\footnote{This is since, for $u\gg \sqrt{x}$, the integrand becomes
  exponentially suppressed relative to its value for $u\lesssim \sqrt{x}$.
  Moreover, there is no further significant contribution in the UV, $u \gg x$,
  where the integrand decays exponentially as $e^{-u}$.} Since
\begin{align}
  \lambda = & \lambda_\rf \bar{\gamma} \sqrt{u^2+x^2}
\end{align}
is an approximate constant $\lambda\approx \lambda|_{u=0}=\lambda_\rf
\bar{\gamma} x$ (with $\bar{\gamma}$ given by \eqref{eq:quinoa}) in this region,
we obtain at leading order
\begin{align}
  \begin{split}
    &\hspace{-11pt}\Delta \cmplx_1(\cTFD) \approx\frac{\vol}{\beta^{d-1}} \cdot
    \frac{\Omega_{d-2}}{(2\pi)^{d-1}} \left( \int_0^\infty du\; u^{d-2}
      \alpha_\lR \right)
    \\
    &\hspace{53pt}\times\begin{cases} \frac{ 4\max\{(\bar{\gamma} x)^2,1\} \log
        (\bar{\gamma} x) }{ (\bar{\gamma} x)^2-1 } & \text{$LR$ basis}
      \\
      \sqrt{2}\left[ 1+\frac{ \bar{\gamma}x (\lambda_\rf+\lambda_\rf^{-1})
          \log(\bar{\gamma} x) }{ (\bar{\gamma} x)^2-1 } \right] &
      \text{$LR_\Cbb$ basis}
    \end{cases}.
  \end{split}
    &
      (\text{$y\gg 1$ and $x-y\gg 1$})
      \label{eq:zirkova}
\end{align}
We can evaluate the remaining integral using the approximation
\eqref{eq:smokeHouse} with \eqref{eq:naan}. The integral is simplified to a
Gaussian moment:
\begin{align}
  \int_0^\infty du\; u^{d-2} \alpha_\lR
  \approx\, & \int_0^\infty du\; u^{d-2} e^{-\frac{1}{2}\left(x-y+\frac{u^2}{2x}\right)}
  &
    (x-y\gg 1)\nonumber
  \\
  = \,   & \frac{1}{2}\, \Gamma\left(\frac{d-1}{2}\right) (4x)^{(d-1)/2} e^{-(x-y)/2}.
           \label{eq:absolut}
\end{align}
Upon reverting to the dimensionful physical quantities of the theory (cf.~eq.~\eqref{eq:quinoa}) we obtain the result in equation \eqref{eq:zirkova1}.

We can apply a similar strategy to approximate the integral giving the entropy
\eqref{eq:bourbon} of the cTFD. Inserting \eqref{eq:naan}, then
integrating the logarithmic term by parts, we obtain
\begin{align}
  \begin{split}
    \hspace{-5pt} s(x,y) \approx\,& \frac{\Omega_{d-2}}{(2\pi)^{d-1}}
    \frac{1}{(d-1)x} \int_0^\infty du\; \left(\frac{u^{d-2}}{2}\right) \frac{
      (d+1)u^2 + 2(d-1)x(x-y) }{ e^{\frac{u^2}{2x}+x-y}-1 }\label{eq:whiskey}
    \\
    \approx\,& \frac{\Omega_{d-2}}{(2\pi)^{d-1}}\begin{cases}
      \Gamma\left(\frac{d-1}{2}\right) (2x)^{(d-1)/2} e^{-(x-y)}
      \left(\frac{x-y}{2}\right)
      & \text{if $x-y \gg 1$} \\
      \left(\frac{1}{d-1}\right)
      \Gamma\left(\frac{d+3}{2}\right)\zeta\left(\frac{d+1}{2}\right)
      (2x)^{(d-1)/2} & \text{if $x=y$}
    \end{cases},
  \end{split}
             &
               (x\gg 1)
\end{align}
where, in the case $x=y$, the integral was performed directly; and in the case
$x-y\gg 1$, after selecting the leading contribution. Note that, in the $y\gg 1$
limit, $s(x,y)\gg s(x,-y)$ is the dominant term in the entropy
\eqref{eq:bourbon}; on the other hand, if $y=0$, then the two terms of
\eqref{eq:bourbon} are equal. In the limits $y\gg 1$ and $x-y\gg 1$, we obtain \eqref{eq:bourbon2} upon reverting to the dimensionful physical quantities of the theory (cf.~eq.~\eqref{eq:quinoa}).

Finally, taking the ratio between complexity of formation \eqref{eq:zirkova} and entropy
\eqref{eq:bourbon}, using the approximations \eqref{eq:absolut} and
\eqref{eq:whiskey}, we have
\begin{align}
  \MoveEqLeft[1]\frac{\Delta \cmplx_1(\cTFD)}{S_\cTFD}\nonumber \\
  \approx \, & \frac{2^{(d-1)/2} e^{(x-y)/2}}{x-y}
               \begin{cases}
                 \frac{ 4\max\{(\bar{\gamma} x)^2,1\} \log (\bar{\gamma} x) }{
                   (\bar{\gamma} x)^2-1 } & \text{$LR$ basis}
                 \\
                 \sqrt{2}\left[ 1+\frac{ \bar{\gamma}x
                     (\lambda_\rf+\lambda_\rf^{-1}) \log(\bar{\gamma} x) }{
                     (\bar{\gamma} x)^2-1 } \right] & \text{$LR_\Cbb$ basis}
               \end{cases},  \qquad   (\text{$y\gg 1$ and $x-y\gg 1$}).  \label{eq:Zooey}
\end{align}
This yields eq.~\eqref{eq:Zooey2} when using eq.~\eqref{eq:quinoa} to revert to the dimensionful physical quantities of the theory.

%%% Local Variables:
%%% mode: latex
%%% TeX-master: "../ChargedComplexityV6"
%%% End:

\section{Diagonalization of the Relative Covariance Matrix at $t=0$}
\label{sec:cv}
Below, we present an additional coordinate transformation, used to derive the low
temperature limits in appendix \ref{app:lowTlowT} and section \ref{lowtemplim}.
In this appendix, we have suppressed subscripts $k$ for simplicity of notation,
though we still have in mind that we are treating the $k$-th momentum mode of the complex
scalar (or alternatively, the cTFD of two complex harmonic oscillators).
The reference and target state covariance matrices can be diagonalized by a
transformation $R_{\pm\to\Delta_{\cTFD}}\in \Sp(4,\mathbb{R})$ to a basis which
we shall call the $\Delta_{\cTFD}$ basis
\begin{align}
\begin{split}
  G^{\Delta_{\cTFD}}_{\rf}
  =& R_{\pm\to\Delta_{\cTFD}} \cdot G^\pm_{\rf} \cdot (R_{\pm\to\Delta_{\cTFD}})^\dagger \\
  =& \frac{1}{4\lambda \, \lambda_\rf} \diag\Big[
     (a^\pm+d^\pm) e^{\mp 2(\alpha_{\lR}+\alpha_{\Lr})}, \\
     &(a^\pm-d^\pm) e^{\mp 2(\alpha_{\lR}+\alpha_{\Lr})},
     a^\pm-d^\pm,
     a^\pm+d^\pm
       \Big],
       \label{eq:cheeseCake}
\end{split}\\
  G^{\Delta_{\cTFD}}_{\cTFD}
  =& R_{\pm\to\Delta_{\cTFD}} \cdot G^\pm_{\cTFD} \cdot (R_{\pm\to\Delta_{\cTFD}})^\dagger
     =\mathds{1},
     \label{eq:Moonshine}
\end{align}
where\footnote{Note that, unlike in eqs.~\eqref{xipm} and \eqref{refpm00}, where $\pm$ simply specifies the complete basis, here it
  enumerates two possibilities, \ie $R_{+\,\to\Delta_{\cTFD}}$ transforms the
  operators $\left[ q_{\bar L R}^+, q_{L \bar R}^+,p_{\bar L R}^+, p_{L \bar
      R}^+\right]$ to the new basis, while $R_{-\,\to\Delta_{\cTFD}}$ transforms
  the operators $\left[ q_{\bar L R}^-, q_{L \bar R}^-,p_{\bar L R}^-, p_{L \bar
      R}^-\right]$ to the new basis. Note that the covariance matrices of
    these operators in the reference and cTFD
    states already factorize to $+$ and $-$ coordinate blocks
  $G^+_{\rf},G^-_{\rf}$ and $G^+_{\cTFD},G^-_{\cTFD}$, see
  eqs.~\eqref{eq:tiredAndAlone} and \eqref{refpm00}, and therefore it is
  possible to transform each of the blocks of the covariance matrices separately.} $G^\pm_{\rf}$ is
  given in eq.~\eqref{refpm00},
  $G^\pm_{\cTFD}=G^\pm_\TFD(t_{\lR}=0,\alpha_{\lR})\Oplus
  G^\pm_\TFD(t_{\Lr}=0,\alpha_{\Lr})$ is obtained from
eq.~\eqref{eq:pgsd}, and
\begin{align}
  R_{\pm\to\Delta_{\cTFD}}
  =& (R^{(2)}_{\pm\to\Delta_{\cTFD}} \oplus R^{(2)}_{\pm\to\Delta_{\cTFD}})\cdot (G^\pm_{\cTFD})^{-1/2}
  \notag\\
  R^{(2)}_{\pm\to\Delta_{\cTFD}}
  =& \begin{bmatrix}
    \cos\theta^\pm & -\sin\theta^\pm \\
    \sin\theta^\pm & \cos\theta^\pm
  \end{bmatrix} \in \SO(2),
                       \label{eq:babaganoush}
\end{align}
with
\begin{align}
\begin{split}
a^\pm
  =& \left(e^{\pm 2\alpha_{\lR}}+e^{\pm 2\alpha_{\Lr}}\right)(\lambda^2 + \lambda_\rf^2),
  \\
    b^\pm
  = &\left(e^{\pm 2\alpha_{\lR}}-e^{\pm 2\alpha_{\Lr}}\right)(\lambda^2 + \lambda_\rf^2),
  \\
    c^\pm
  =& 2(\lambda^2-\lambda_\rf^2) e^{\pm(\alpha_{\lR}+\alpha_{\Lr})},
  \\
    d^\pm
    =& \sqrt{(b^\pm)^2+(c^\pm)^2},
    \\
  \theta^\pm
  =& \tan^{-1}\left(\frac{c^\pm}{b^\pm-d^\pm}\right). \label{eq:twoEggBreakfast}
\end{split}
\end{align}
Using eqs.~\eqref{eq:cheeseCake}-\eqref{eq:Moonshine}, it is straightforward to
demonstrate that we in fact have a diagonal relative covariance matrix
\eqref{eq:piano} in this basis
\begin{align}
  \Delta_{\cTFD}^{\Delta_{\cTFD}}
  =& (G_{\rf}^{\Delta_{\cTFD}})^{-1}.
     \label{eq:Willie}
\end{align}

%%% Local Variables:
%%% mode: latex
%%% TeX-master: "../ChargedComplexityV6"
%%% End:

%\input{chapters/apptodo}

\bibliographystyle{JHEP}
\bibliography{references}

\end{document}